\documentclass[%
reprint,
superscriptaddress,
nofootinbib,
 amsmath,amssymb,
 aps,
]{revtex4-2}

\usepackage{graphicx}
\usepackage{dcolumn}
\usepackage{bm}
\usepackage{amsmath}
\usepackage{overpic}
\usepackage{booktabs}%
\usepackage{makecell}
\usepackage{enumerate}
\usepackage[figuresright]{rotating}
\makeatletter

\newcommand{\Rmnum}[1]{\expandafter\@slowromancap\romannumeral #1@}
\makeatother

\newcommand{\mnras}{MNRAS}
\newcommand{\aap}{A\& A}
\newcommand{\apjl}{ApJL}

\newcommand{\araa}{Annu. Rev. Astron. Astrophys}
\usepackage{hyperref}
\usepackage{xcolor}

\begin{document}

\title{A new subclass of gamma-ray burst originating from compact binary merger}

\author{Chen-Wei Wang}
 \affiliation{%
 Key Laboratory of Particle Astrophysics, Institute of High Energy Physics, Chinese Academy of Sciences, Beijing 100049, China}
\affiliation{
 University of Chinese Academy of Sciences, Beijing 100049, China 
}%

\author{Wen-Jun Tan}
\affiliation{%
 Key Laboratory of Particle Astrophysics, Institute of High Energy Physics, Chinese Academy of Sciences, Beijing 100049, China}
\affiliation{
 University of Chinese Academy of Sciences, Beijing 100049, China 
}%

\author{Shao-Lin Xiong}
\email{xiongsl@ihep.ac.cn}
 \affiliation{%
 Key Laboratory of Particle Astrophysics, Institute of High Energy Physics, Chinese Academy of Sciences, Beijing 100049, China}

\author{Shu-Xu Yi}
\email{sxyi@ihep.ac.cn}
 \affiliation{%
 Key Laboratory of Particle Astrophysics, Institute of High Energy Physics, Chinese Academy of Sciences, Beijing 100049, China}

\author{Rahim Moradi}
\email{rmoradi@ihep.ac.cn}
 \affiliation{%
 Key Laboratory of Particle Astrophysics, Institute of High Energy Physics, Chinese Academy of Sciences, Beijing 100049, China}

\author{Bing Li}
 \affiliation{%
 Key Laboratory of Particle Astrophysics, Institute of High Energy Physics, Chinese Academy of Sciences, Beijing 100049, China}

\author{Zhen Zhang}
 \affiliation{%
 Key Laboratory of Particle Astrophysics, Institute of High Energy Physics, Chinese Academy of Sciences, Beijing 100049, China}

\author{Yu Wang}
 \affiliation{%
 ICRANet, Piazza della Repubblica 10, I-65122 Pescara, Italy}
 \affiliation{%
 ICRA, Dipartamento di Fisica, Sapienza Universit\`a  di Roma, Piazzale Aldo Moro 5, I-00185 Rome, Italy}
 \affiliation{%
 INAF, Osservatorio Astronomico d'Abruzzo, Via M. Maggini snc, I-64100, Teramo, Italy}

\author{Yan-Zhi Meng}
 \affiliation{%
 School of Science, Guangxi University of Science and Technology, Liuzhou 545006, China}

\author{Jia-Cong Liu}
 \affiliation{%
 Key Laboratory of Particle Astrophysics, Institute of High Energy Physics, Chinese Academy of Sciences, Beijing 100049, China}
\affiliation{
 University of Chinese Academy of Sciences, Beijing 100049, China 
}%
\author{Yue Wang}
 \affiliation{%
 Key Laboratory of Particle Astrophysics, Institute of High Energy Physics, Chinese Academy of Sciences, Beijing 100049, China}
\affiliation{
 University of Chinese Academy of Sciences, Beijing 100049, China 
}%
\author{Sheng-Lun Xie}
 \affiliation{%
 Key Laboratory of Particle Astrophysics, Institute of High Energy Physics, Chinese Academy of Sciences, Beijing 100049, China}
 \affiliation{Institute of Astrophysics, Central China Normal University, Wuhan 430079, China}
\author{Wang-Chen Xue}
 \affiliation{%
 Key Laboratory of Particle Astrophysics, Institute of High Energy Physics, Chinese Academy of Sciences, Beijing 100049, China}
\affiliation{
 University of Chinese Academy of Sciences, Beijing 100049, China 
}%
\author{Zheng-Hang Yu}
 \affiliation{%
 Key Laboratory of Particle Astrophysics, Institute of High Energy Physics, Chinese Academy of Sciences, Beijing 100049, China}
\affiliation{
 University of Chinese Academy of Sciences, Beijing 100049, China 
}%
\author{Peng Zhang}
 \affiliation{%
 Key Laboratory of Particle Astrophysics, Institute of High Energy Physics, Chinese Academy of Sciences, Beijing 100049, China}
 \affiliation{College of Electronic and Information Engineering, Tongji University, Shanghai 201804, China}
 
\author{Wen-Long Zhang}
 \affiliation{%
 Key Laboratory of Particle Astrophysics, Institute of High Energy Physics, Chinese Academy of Sciences, Beijing 100049, China}
 \affiliation{Qufu Normal University, Qufu 273165, China}

\author{Yan-Qiu Zhang}
 \affiliation{%
 Key Laboratory of Particle Astrophysics, Institute of High Energy Physics, Chinese Academy of Sciences, Beijing 100049, China}
\affiliation{
 University of Chinese Academy of Sciences, Beijing 100049, China 
}%
\author{Chao Zheng}
 \affiliation{%
 Key Laboratory of Particle Astrophysics, Institute of High Energy Physics, Chinese Academy of Sciences, Beijing 100049, China}
\affiliation{
 University of Chinese Academy of Sciences, Beijing 100049, China 
}%

\date{\today}

\begin{abstract}
Type I gamma-ray bursts (GRBs) are believed to originate from compact binary merger usually with duration less than 2 seconds for the main emission. However, recent observations of GRB 211211A and GRB 230307A indicate that some merger-origin GRBs could last much longer. 
Since they show strikingly similar properties (indicating a common mechanism) which are different from the classic ``long"-short burst (e.g. GRB 060614), forming an interesting subclass of type I GRBs, we suggest to name them as type IL GRBs. 
By identifying the first peak of GRB 230307A as a quasi-thermal precursor, we find 
that the prompt emission of type IL GRB is composed of three episodes: 
(1) a precursor followed by a short quiescent (or weak emission) period, 
(2) a long-duration main emission, and 
(3) an extended emission. 
With this burst pattern, a good candidate, GRB 170228A, was found in the \textit{Fermi}/GBM archive data, and subsequent temporal and spectral analyses indeed show that GRB 170228A falls in the same cluster with GRB 211211A and GRB 230307A in many diagnostic figures. 
Thus this burst pattern could be a good reference for rapidly identifying type IL GRB and conducting low-latency follow-up observation. 
We estimated the occurrence rate and discussed the physical origins and implications for the three emission episodes of type IL GRBs. 
Our analysis suggests the pre-merger precursor model, especially the super flare model, is more favored for type IL GRBs. 
More observations in multi-wavelength and multi-messenger are required to deepen our understanding of this subclass of GRBs. 
\end{abstract}
\maketitle


\section{Introduction} \label{introduction}

Gamma-ray bursts (GRBs) are the most violent explosions in the universe. The classification of GRB is an important but difficult task, given the very complicated properties of various GRBs. Traditionally, GRBs can be divided into long GRBs and short GRBs by a simple but effective criteria---$T_{90}$, which corresponds to the duration containing 90\% of photons of a burst \citep{1993Kouveliotou}.
Interestingly, several other properties of GRBs generally follow this classification scheme, such as the relationship between isotropic total energy and the peak energy of the spectrum (Amati relation) \citep{2002A&A...390...81A}, and the relationship between peak luminosity and the peak energy of the spectrum (Yunetoku relation) \citep{2004ApJ...609..935Y}. 
In addition, some temporal properties also in align with the dichotomy of duration, such as the minimum variability timescale \citep{2015ApJ...811...93G}, spectral lag \citep{2015MNRAS.446.1129B} and so on. 
These energetic and time-variability properties indicate that traditionally long bursts and short bursts are generated from essentially different physical origins\cite[e.g.][]{1993Kouveliotou,Woosley1993,Paczynski1998,MacFadyen1999,Eichler1989,Narayan1992}. Finally, the concept of type I GRB and type II GRB have been proposed, considering all these observational properties \citep{Zhang2006type}. Generally speaking, type I GRBs are produced by the merger of compact objects with a shorter duration and harder spectra, while type II GRBs are produced by the collapse of massive stars, which are usually longer and softer.

For both types of GRB, the emission of a GRB can be seen in two stages: the prompt emission and the afterglow emission \citep{2018GRBbook}. Based on intensive studies in decades, the multi-wavelength afterglow is confidently established to arise from the external shock which is generated when the relativistic jet sweeps the medium around the central engine \citep{Zhang2006afterglow,Meszaros1997afterglow,Sari1998afterglow}, while the origin of prompt emission is subject to great debate \citep{2018GRBbook}. Currently, the prompt emission is believed to be produced within the highly relativistic jet by internal shock \citep{Paczynski1994IS,Rees1994IS} or magnetic dissipation \citep{zhang2011ICMART,Zhang2014ICMART}. Photosphere emission are also proposed to interpret the prompt emission for some GRBs \citep{Peer2008photospheric,Meng2019photospheric}. 

Observationally, the prompt emission of GRBs can be further divided into (at least) three episodes: precursor, main emission and extended emission. 
The precursor is the shorter and weaker emission preceding the main emission, usually with a quiescent period between them. 
The extended emission is usually weaker, softer but long-lived compared to the main emission in some GRBs \citep{Norris2006EE}. 
However, all these episodes usually do not occur concurrently in prompt emission, e.g. there is precursor detected only for a small fraction (about 3\% to 20\%) of GRBs \citep{GBM_precursor_PRD,Wang2020_SGRB_precursor}. 

The temporal and spectral characteristics of the precursor can provide many information of the model of progenitor and dissipation process. Especially, for the type I GRBs, the precursor model can be divided into two categories: pre-merger models and post-merger models. For pre-merger models, the precursor can be originated from the interaction between the magnetosphere of the compact stars before the merger or the crust cracking of the compact stars \citep{zhang2022superflare,Hansen2001pre_merger_model,Tsang2012pre_merger_model}. And for post-merger models, the precursor can be generated from the shock breaking-out or the transition of a fireball from being optically thick to optically thin \citep{Meszaros2000post_merger_model}. Studies of precursors in short GRBs show that the majority of precursors of sGRBs have thermal spectra \citep{Wang2020_SGRB_precursor}.

The extended emission can exist in both long GRBs and short GRBs \citep{Zhangxiaolu2020}, and can also have constraints on the central engine, e.g. some studies show the extended emission can be explained as the internal plateau emission of the magnetar central engine for some GRBs. \citep{Rowlinson2010EE,Rowlinson2013EE}.

With the improvement of the GRB monitoring network in recent years, more observations indicate the existence of a special type of GRB called ``long (duration)'' short (original) GRBs. Their duration falls within the typical long GRB region, but other properties resemble type I GRBs or fall between type I and type II GRBs. Well-known examples of such GRBs include GRB 060614, GRB 211227A, GRB 211211A, and GRB 230307A. 
The former two GRBs, namely GRB 060614 and GRB 211227A can be explained as classic short-hard GRBs with soft and long extended emission (sGRBEE), and is considered as the typical ``long'' short GRBs. 
In contrast, the latter two, namely GRB 211211A and GRB 230307A are more complicated and confusing and cannot be classified as classic short-hard GRBs nor typical ``long'' short GRBs.


In this work, we define a new subclass of type I GRB, namely type IL GRB (see section \ref{section2} for details) for them.
Then we performed a comprehensive analysis of two type IL GRB samples, GRB 230307A and GRB 211211A, and found the burst pattern of light curve morphology.
Using the burst pattern as a criterion of type IL GRB, we identified a good candidate of this type of GRB, GRB 170228A, in the historical data from \textit{Fermi}/GBM. 
We further find that type IL GRBs and typical ``long" short GRB show significant differences in the energy budget allocation among main burst and extended emission (see section \ref{section3} for details).

This paper is organized as follows. We present the comprehensive analysis of GRB 230307A and GRB 211211A, the features of type IL GRB and search of type IL GRB in section \ref{section2}. Then constrain of precursor models and progenitors are included in the discussion in section \ref{section3}. Conclusions are given in section \ref{section4}. The data reduction is described in Appendix for details.

All parameter errors in this work are for 68\% confidence level if not otherwise stated. 

\section{Type IL GRB}\label{section2}
In our terminology, the type IL GRB is defined as a sub-class of type I GRB with intrinsic long-duration. The term "intrinsic long-duration" means that the long timescale of these GRBs is not caused by the soft-long extended emission (although they also contribute some), but rather by a hard and long main emission.

In this section, we will first discuss the observational features of two peculiar GRBs, GRB 211211A and GRB 230307A,
and then extract their common patterns of the prompt emission.
With these patterns, searching for type IL GRB in the \textit{Fermi}/GBM catalog will be reported.

\subsection{Two events---GRB 211211A and GRB 230307A}
Our attention is first focused on two similar, peculiar and bright GRBs---GRB 211211A and GRB 230307A. They both have a long duration prompt emission and are associated with kilonovae, which suggest these two long GRBs have a merger-origin and belong to type I GRBs.

GRB 211211A was detected at 2021-12-11T13:09:59.651 (UTC) by \textit{Fermi}/GBM, which is one of the top ten brightest bursts in \textit{Fermi}/GBM burst catalog\citep{Levan2024KN}. The first pulse of light curve, which last for about 0.2s, is well recognized as the precursor of the whole GRB, since it is followed by a short quiescent episode (namely waiting time) and is softer than the later-coming main emission\citep{XS_GRB211211A_QPO}. This precursor is very bright with an equivalent isotropic energy of $E_{\rm{\rm iso}} \sim 7.7\times10^{48}$ erg \citep{XS_GRB211211A_QPO}. The emission after the precursor is clearly divided into two parts, a long-hard main emission and a soft-long extended emission \citep[see e.g.,][]{XS_GRB211211A_QPO, YJ_GRB211211A_nature}.

GRB 230307A is the second brightest GRB ever recorded \citep{2023arXiv230705689S}. On March 7, 2023, at 15:44:06.65 UT ($T_0$), the GECAM-B was triggered in-flight by this exceptionally bright long burst, which is also detected by many other missions such as \textit{Fermi} Gamma-ray Burst Monitor (GBM) \citep{Fermi2023GCN} and Konus-WIND \citep{Konus2023GCN}. The extreme brightness of this burst was firstly reported to the community by GECAM-B with the real time alert data \citep{xiong2023GCN}, and subsequently confirmed by other instruments, leading to a large observation campaign to this event. 

The prompt emission light curve of GRB 230307A is composed of many sharply varying pulses, collectively forming a fast-rise-exponential-decay (FRED) shape with a smooth dip structure \citep{2023arXiv231007205Y} . However, there is one pulse appearing to be more distinct, namely the first pulse (from $T_0$-0.05 s to $T_0$+0.4 s), which is bright and short-lived. The signal decreases rapidly with the increase of photon energy, showing a clear different behavior with other pulses of the prompt emission. A period of low-luminosity but significant emission can be seen between the first pulse and the subsequent pulses. Morphologically, this pulse is very similar to precursor, except with no quiescent period.  
The prompt emission after dip is softer and weaker than the emission before the dip, so it is naturally considered as the extended emission\citep{Peng2024compare}.

The joint analysis of GRB 2303037A is base on GECAM data and \textit{Fermi}/GBM data, while only \textit{Fermi}/GBM data is utilized in the analysis of GRB 211211A. (See Appendix \ref{Data Selection} and Appendix \ref{Data Reduction} for details).
\subsection{Lightcurve pattern of type IL GRB}
\begin{figure*}[http]
\centering
\includegraphics[width=\textwidth]{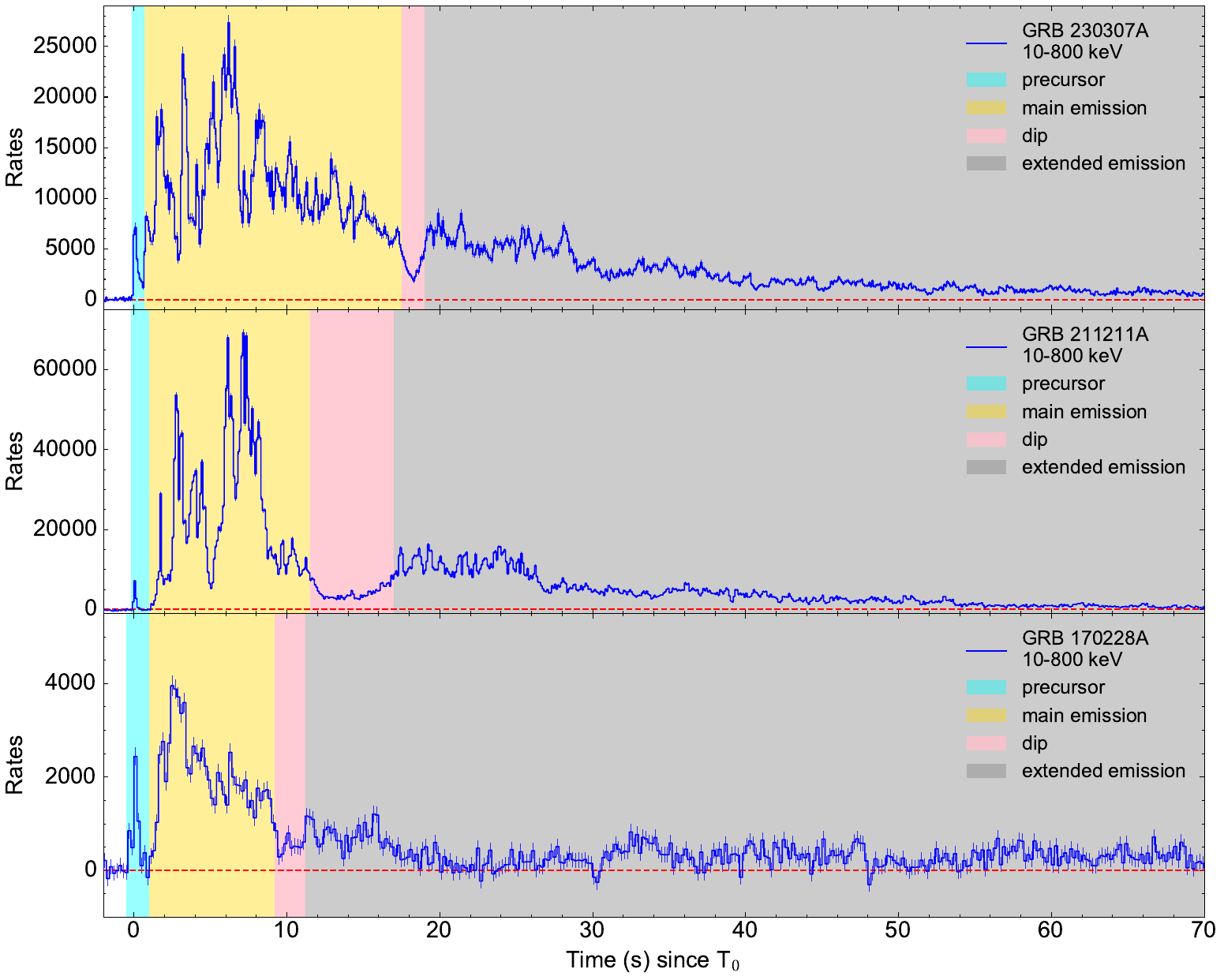}
\caption{\noindent\textbf{The background subtracted lightcurves of GRB 230307A,  GRB 211211A and GRB 170228A. }The light curve of GRB 230307A is obtained from GECAM with GRD01, GRD04, GRD05 summed, the light curve of GRB 211211A is obtained from GBM with n2 and na summed,  the light curve of GRB 170228A is obtained from GBM with n0,n1,n2,n9,na summed. The shaded intervals represent precursor, main emission, dip and extended emission of each light curve structure.}
\label{fig:three_lightcurves}
\end{figure*}

In addition to both having long duration and being associated with kilonova, GRB 230307A and GRB 211211A also exhibit high similarity in their lightcurves. Both GRBs seem to have three phases: precursor, main emission, and extended emission (Fig. \ref{fig:three_lightcurves}).
In this subsection, we will study this common pattern of type IL GRB and reveal the differences in burst characteristics between this type of GRB and other GRBs.

\subsubsection{Identification of precursor in GRB 230307A}
\begin{figure*}[http]
\centering
\includegraphics[width=\textwidth]{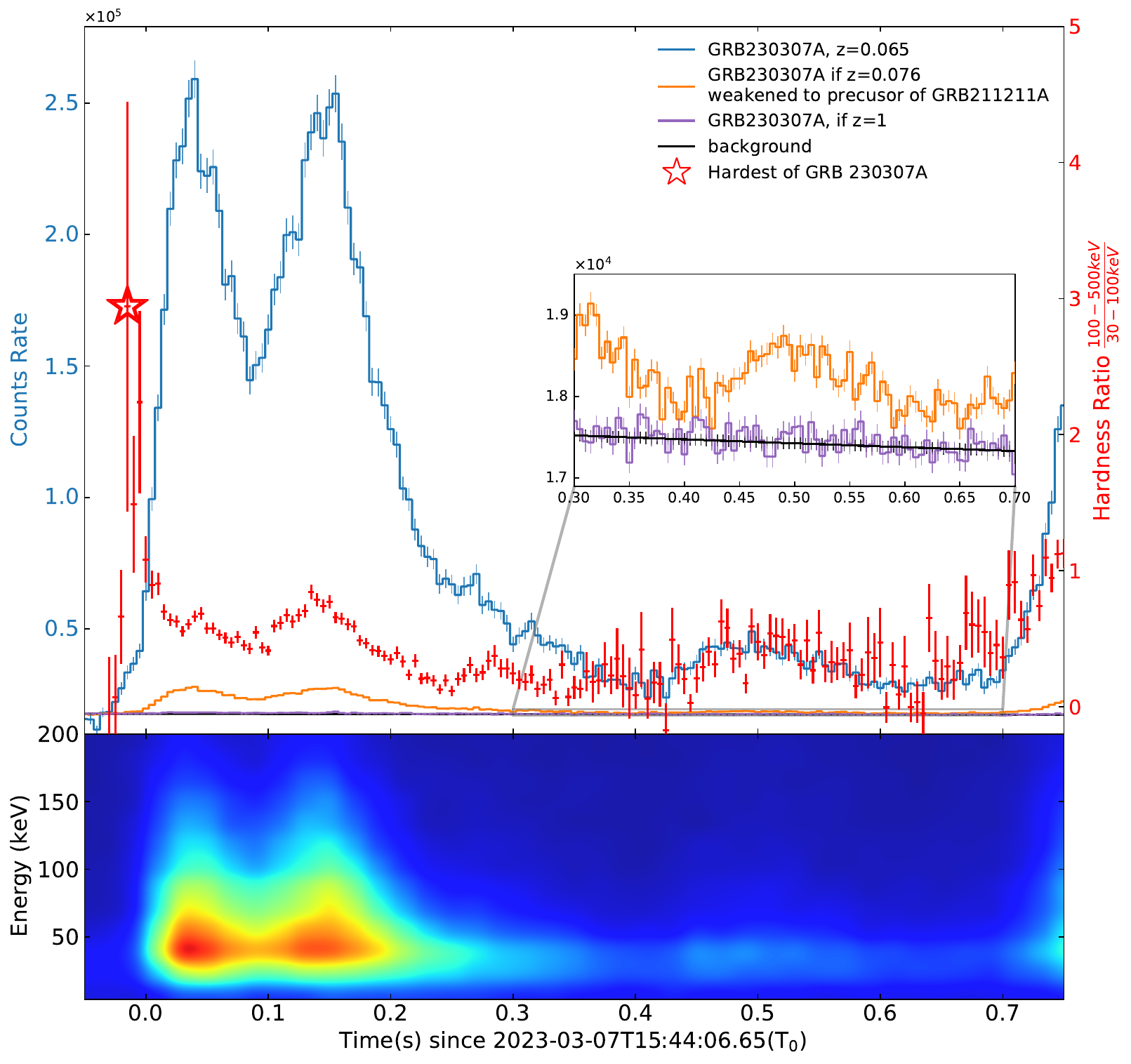}
\caption{\noindent\textbf{The lightcurves and hardness ratio of the precursor in GRB 230307A. }The blue line is the lightcurves of the precursor observed by GECAM-B and GECAM-C. The orange line is the simulated lightcurves of the precursor when move the source to the redshift z=0.076 (the same redshift of GRB 211211A) and weakened to the $E_{\rm iso}$ to $7.7\times10^{48}$\,erg. The violet line is the simulated lightcurves of the precursor when the source is moved to redshift z=1. The red line is the hardness ratio of 100-500\,keV and 30-100\,keV, which shows the hardest point appears shortly after the beginning of the precursor. The following figure shows the evolution of the photon energy distribution.}
\label{fig:precursor_lc}
\end{figure*}

Before drawing any further conclusion, it is necessary to address a potential concern. As shown in Fig.~\ref{fig:precursor_lc}, there is still significant emission between the first pulse (precursor candidate) and subsequent burst of GRB 230307A, which is different from typical precursor behavior. Therefore, the first step is to testify whether the first pulse is a precursor of GRB 230307A. Indeed, we find that many characteristics indicate that the first pulse is significantly different from the main emission in terms of lightcurve, spectra and quiescent time.

\begin{figure*}
\centering
\begin{tabular}{cc}
\begin{overpic}[width=0.41\textwidth]{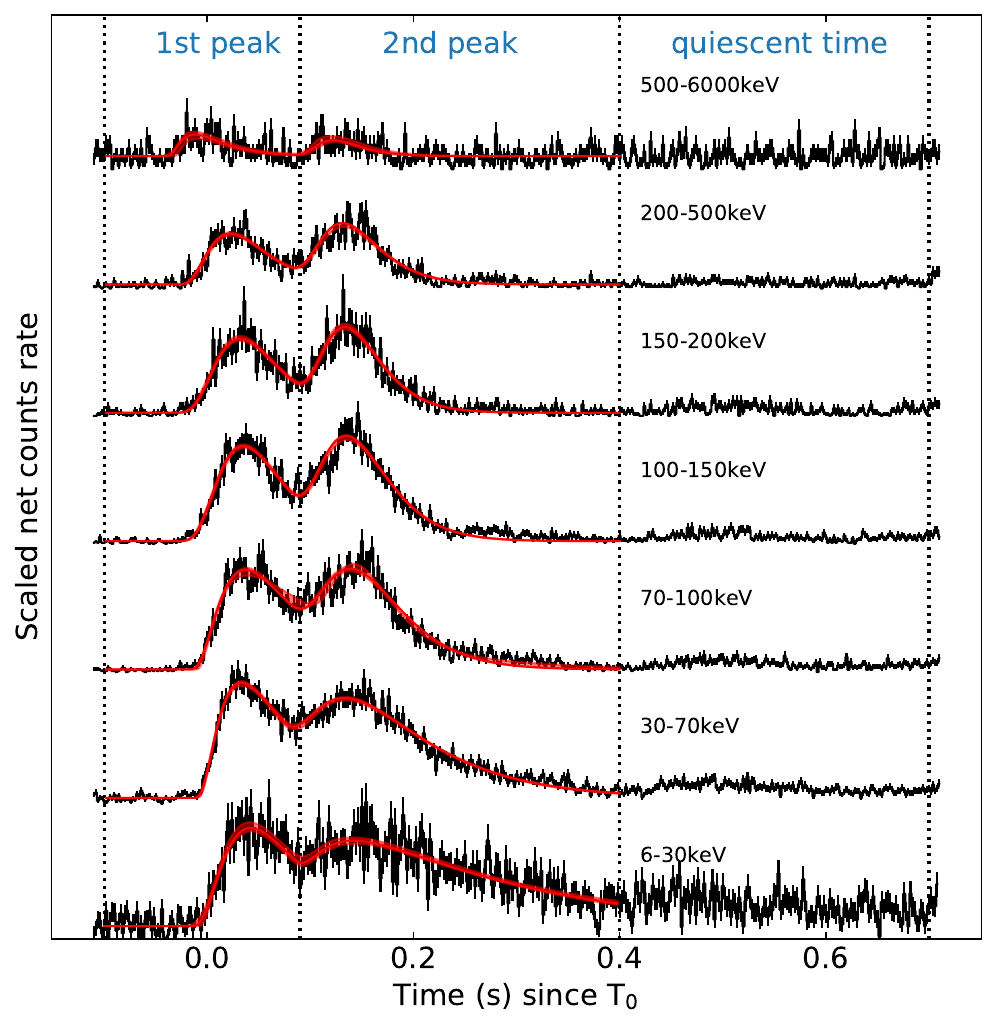}\put(0, 90){\bf a}\end{overpic} &
        \begin{overpic}[width=0.5\textwidth]{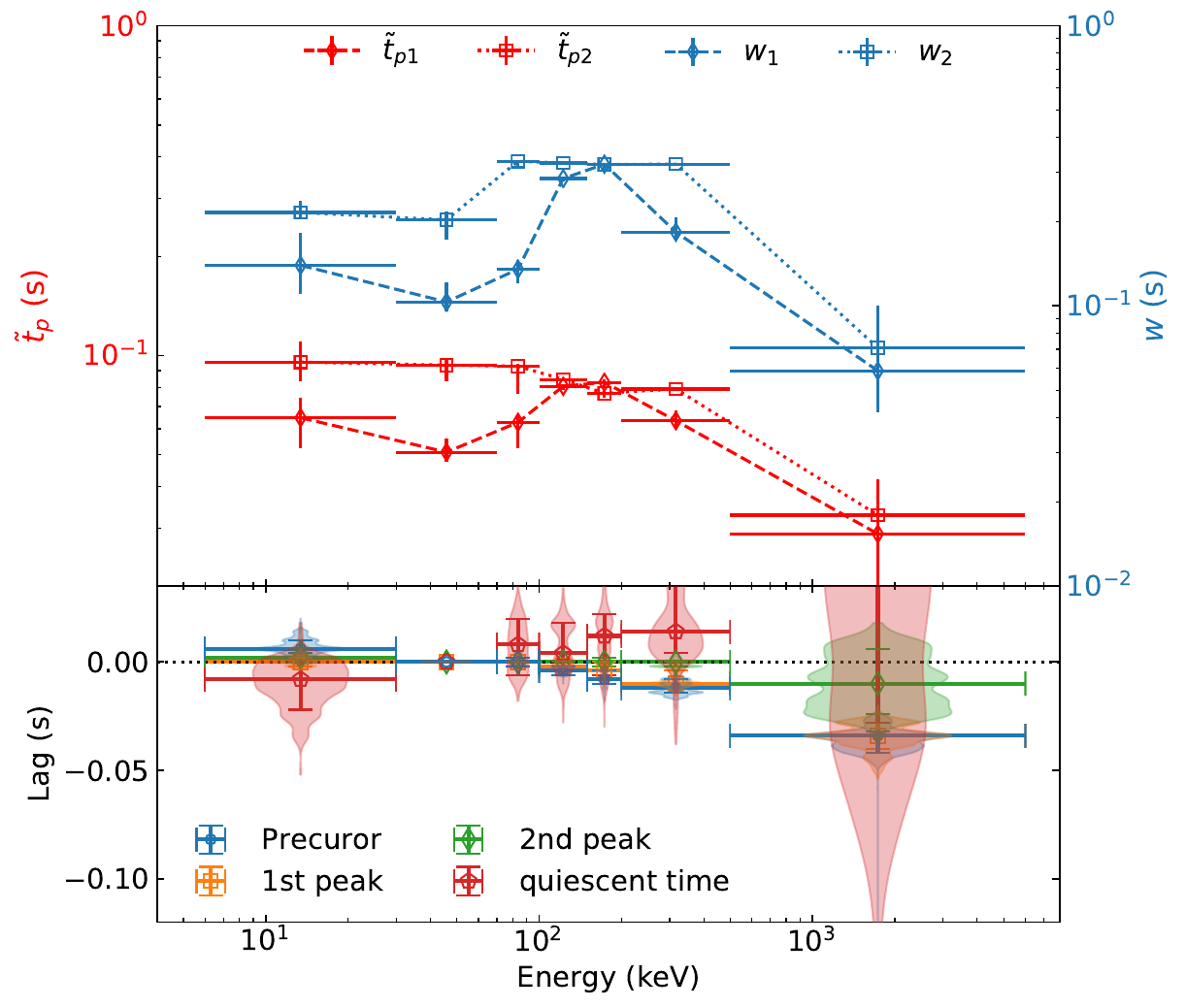}\put(0, 75){\bf b}\end{overpic} \\
\multicolumn{2}{c}{\begin{overpic}[width=0.9\textwidth]{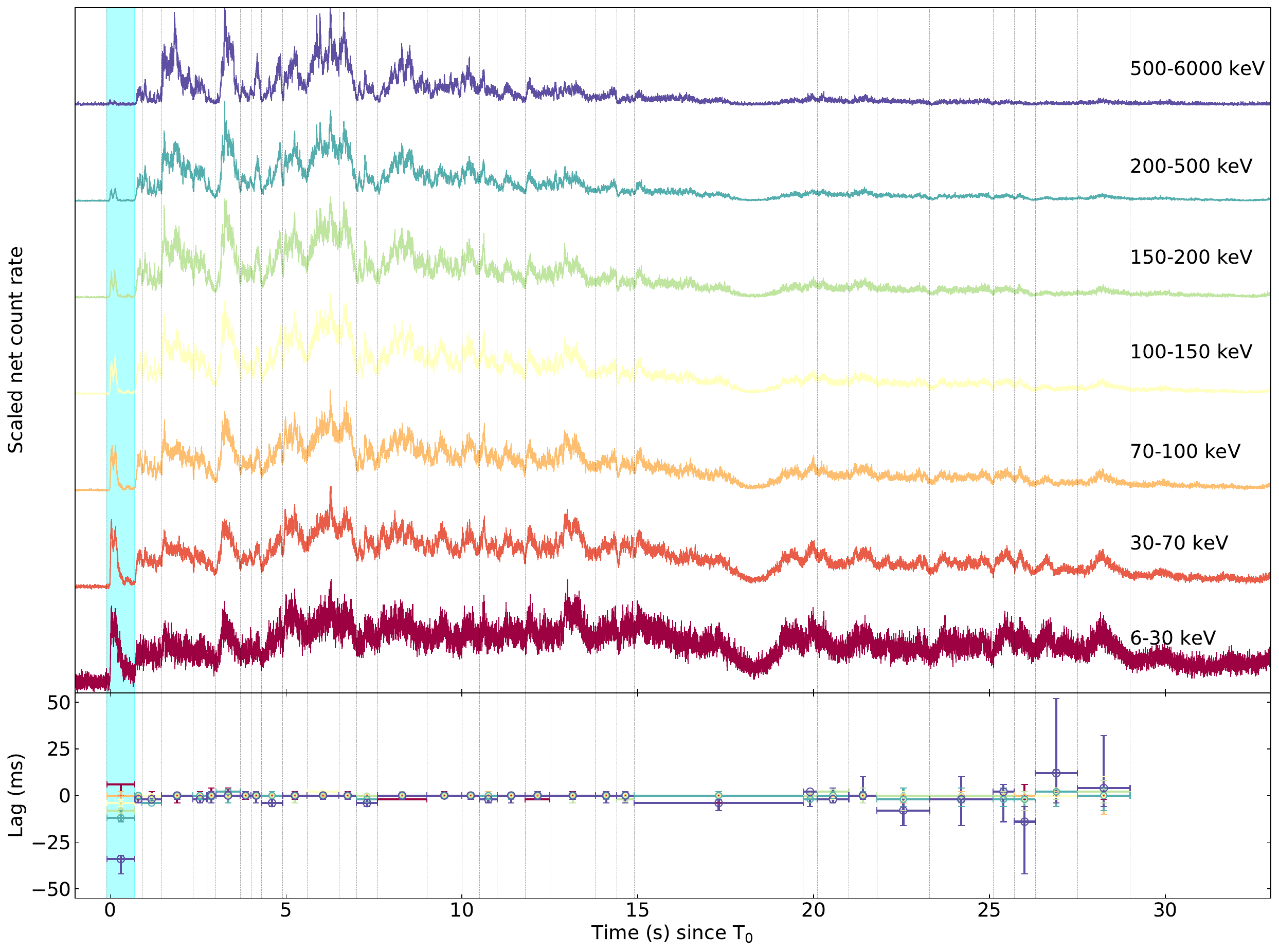}\put(2, 70){\bf c}\end{overpic}} \\
\end{tabular}
\caption{\noindent\textbf{Temporal analysis result of precursor in GRB 230307A. }\textbf{a}, The fitting results of the lightcurves in different energy range. The precursor is divided into three segment, including two peaks and a quiescent time.
\textbf{b}, The evolution of fitted FRED formulation parameters with energy. The results did not show any signs of self-similarity, and there were even no monotonic relationships observed. The bottom panel decipted the spectral for the three segment of the precursor, which shows only the first peak of the precursor have significant spectral lag.
\textbf{c}, The time sliced spectral lag of GRB 230307A. The result shows the spectral lag is only exist in the precursor of GRB 230307A while the spectral lag of other pulses are consistent with zero within the error.
}
\label{fig:precursor_lag}
\end{figure*}

\textbf{Evidence from light-curves}: \\
The most prominent feature is that the first pulse has significant spectral delay compared to the other part of prompt emission, which can be seen in the pulse-resolved spectral lag (Fig.~\ref{fig:precursor_lag}c). The negligible spectra lag in each pulse of the prompt emission is consistent with the result of \cite{2023arXiv231007205Y}. They suggest that these rapid-varying pulses originate from local mini-jet events produced by magnetic reconnection (ICMART event). The time scale of each rapid-varying pulse is related to the size of the local radiation region, and small time scale means a small local radiation region. There would be no significant magnetic field evolution in the local region and therefore no obviously spectral lag, which also results in perfect alignment in time. 
The time scale of the precursor candidate is comparable to or even smaller than other rapid-varying pulses of other part of the prompt emission,  while the spectral lag of the precursor candidate is larger than those pulses, indicating that the precursor candidate may not originate from a local mini-jet.

On the other hand, if the precursor candidate originate from the same ICMART event of the board pulse of the main emission, as the jet continues to expand, the photon observed earlier means that its radiation occurs earlier and has a smaller radiation radius and smaller local radiation region, resulting in smaller spectral lag, which does not match the observation.

Moreover, Yi et al. have \cite{2023arXiv231007205Y} pointed out that these rapid-varying pulses (excluding the precursor candidate) are assembled to a self-similar, ``softer-wider" and ``softer-later" broad FRED pulse, and the self-similarity is also seen in the rapid-varying pulses of the subsequent emission, but not in the precursor candidate (Fig.~\ref{fig:precursor_lag}b) though it can be well fitted by two FRED (Fig.~\ref{fig:precursor_lag}ab) with the parameters in Table\,\ref{tab:lc_fit}, which indicates that the precursor candidate may not be an assembly of indistinguishable mini-jets (e.g. an independent ICMART event). Each FRED component of the precursor candidate even lacks an energy-dependent evolution in width and peak time.

Hence the significant spectral delay and absence of energy-dependent evolution in width and peak time in the precursor candidate suggest that it likely has a different origin from the subsequent emission of GRB 230307A.

\begin{table*}[htbp]
\begin{center}
\caption{The fitting result of the lightcurve of the precursor in GRB 230307A.}\label{tab:lc_fit}%
\begin{tabular*}{\hsize}{@{}@{\extracolsep{\fill}}ccccccccc@{}}
\toprule
Energy range & $\tau_{\rm r}$ & $\tau_{\rm d}$ & t$_s$ & norm & $\tau_{\rm r}$ & $\tau_{\rm d}$ & t$_s$ & norm \\ 
(keV) & (s) & (s) & (s) & (counts$\cdot$s$^{-1}$) & (s) & (s) & (s) & (counts$\cdot$s$^{-1}$) \\ 
\midrule
6-30 & 0.11$^{+0.15}_{-0.06}$ & 0.04$^{+0.02}_{-0.01}$ & -0.03$^{+0.01}_{-0.02}$ & $4.97^{+60.98}_{-3.65}\times10^{5}$ & 0.06$^{+0.03}_{-0.02}$ & 0.15$^{+0.03}_{-0.02}$ & 0.08$^{+0.01}_{-0.01}$ & $5.21^{+2.38}_{-1.47}\times10^{4}$ \\ 
30-70 & 0.06$^{+0.02}_{-0.01}$ & 0.04$^{+0.01}_{-0.01}$ & -0.02$^{+0.00}_{-0.00}$ & $8.62^{+7.79}_{-3.27}\times10^{5}$ & 0.13$^{+0.03}_{-0.03}$ & 0.06$^{+0.00}_{-0.00}$ & 0.06$^{+0.01}_{-0.01}$ & $9.69^{+4.51}_{-3.72}\times10^{5}$ \\ 
70-100 & 0.09$^{+0.03}_{-0.06}$ & 0.04$^{+0.04}_{-0.01}$ & -0.03$^{+0.01}_{-0.00}$ & $8.54^{+8.75}_{-7.17}\times10^{5}$ & 0.30$^{+0.00}_{-0.01}$ & 0.03$^{+0.00}_{-0.01}$ & 0.05$^{+0.02}_{-0.00}$ & $2.07^{+5.93}_{-0.43}\times10^{7}$ \\ 
100-150 & 0.26$^{+0.00}_{-0.00}$ & 0.03$^{+0.00}_{-0.00}$ & -0.05$^{+0.00}_{-0.00}$ & $2.63^{+0.42}_{-0.26}\times10^{7}$ & 0.30$^{+0.00}_{-0.00}$ & 0.02$^{+0.00}_{-0.00}$ & 0.05$^{+0.00}_{-0.00}$ & $5.05^{+0.35}_{-0.51}\times10^{7}$ \\ 
150-200 & 0.30$^{+0.00}_{-0.01}$ & 0.02$^{+0.00}_{-0.00}$ & -0.05$^{+0.00}_{-0.00}$ & $2.51^{+0.57}_{-0.48}\times10^{7}$ & 0.30$^{+0.00}_{-0.01}$ & 0.02$^{+0.00}_{-0.00}$ & 0.06$^{+0.00}_{-0.00}$ & $4.94^{+0.84}_{-1.04}\times10^{7}$ \\ 
200-500 & 0.16$^{+0.01}_{-0.01}$ & 0.03$^{+0.00}_{-0.00}$ & -0.04$^{+0.00}_{-0.00}$ & $2.39^{+0.74}_{-0.53}\times10^{6}$ & 0.30$^{+0.00}_{-0.00}$ & 0.02$^{+0.00}_{-0.00}$ & 0.05$^{+0.00}_{-0.00}$ & $3.70^{+0.25}_{-0.60}\times10^{7}$ \\ 
500-6000 & 0.04$^{+0.03}_{-0.03}$ & 0.03$^{+0.01}_{-0.00}$ & -0.04$^{+0.01}_{-0.01}$ & $3.62^{+5.40}_{-2.79}\times10^{4}$ & 0.16$^{+0.14}_{-0.12}$ & 0.02$^{+0.01}_{-0.00}$ & 0.07$^{+0.02}_{-0.01}$ & $9.83^{+189.68}_{-9.60}\times10^{5}$ \\ 
\botrule
\end{tabular*}
\end{center}
\end{table*}

\textbf{Evidence from Spectra}: 
More direct evidence comes from the result of the spectrum analysis. For the precursor candidate in GRB 230307A, three distinct time segmentation methods are employed to extract the time-resolved and time-integrated spectra, which is defined as S-I, S-II and S-III:\begin{enumerate}[(1)]
\item S-I: the time interval of -0.05-0.7 s, including the precursor candidate and quiescent time period candidate, is utilized as a single time slice for conducting time-integrated spectral analysis;
\item S-II: the time interval of -0.05-0.7 s is divided into 3 time slices, corresponding to the first peak of the precursor candidate, the second peak of the precursor candidate, and the third peak of the quiescent time period candidate, respectively;
\item S-III: the time interval of -0.05-0.4 s is divided into 7 time slices to study the spectral evolution.
\end{enumerate} 

\begin{figure*}
\centering
\begin{tabular}{cc}
\begin{overpic}[width=0.45\textwidth]{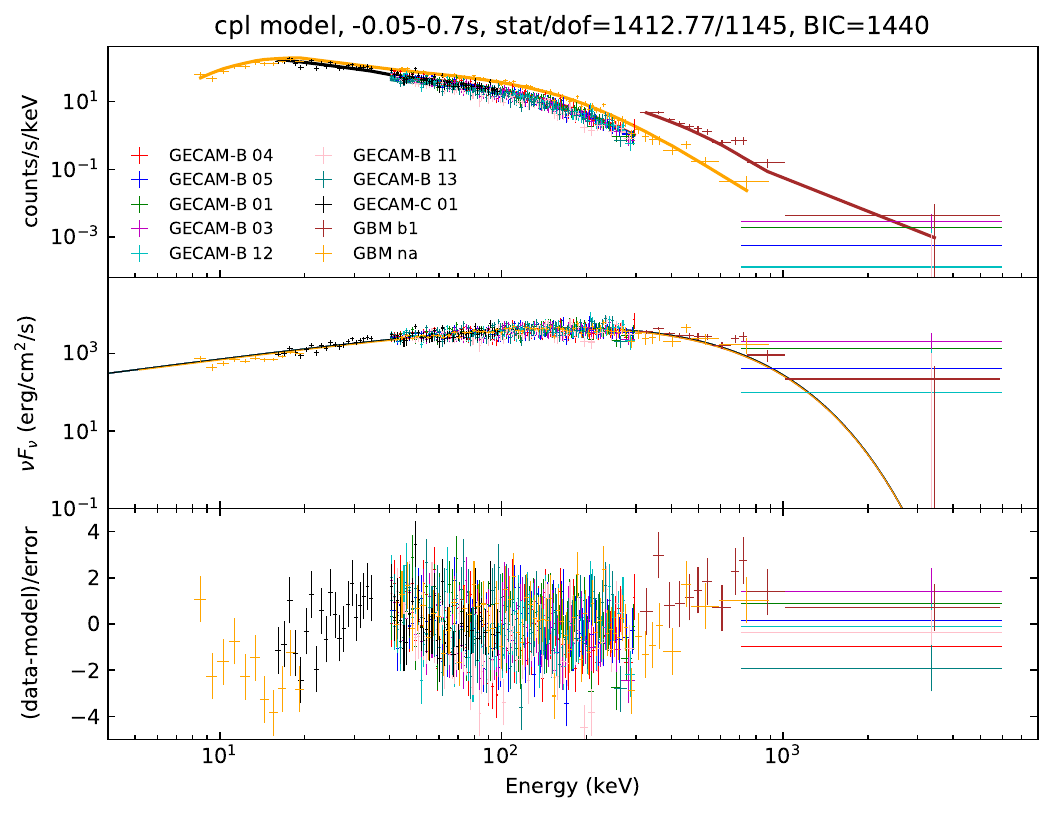}\put(0, 70){\bf a}\end{overpic} &
        \begin{overpic}[width=0.45\textwidth]{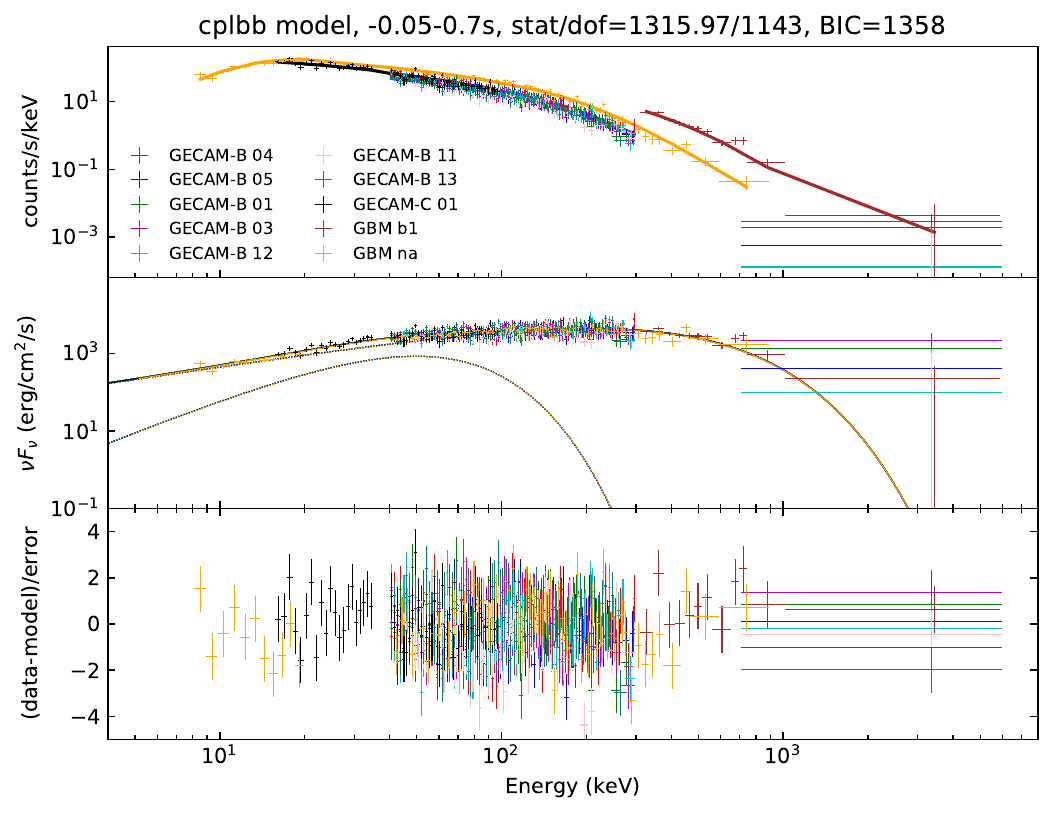}\put(0, 70){\bf b}\end{overpic} \\
\begin{overpic}[width=0.45\textwidth]{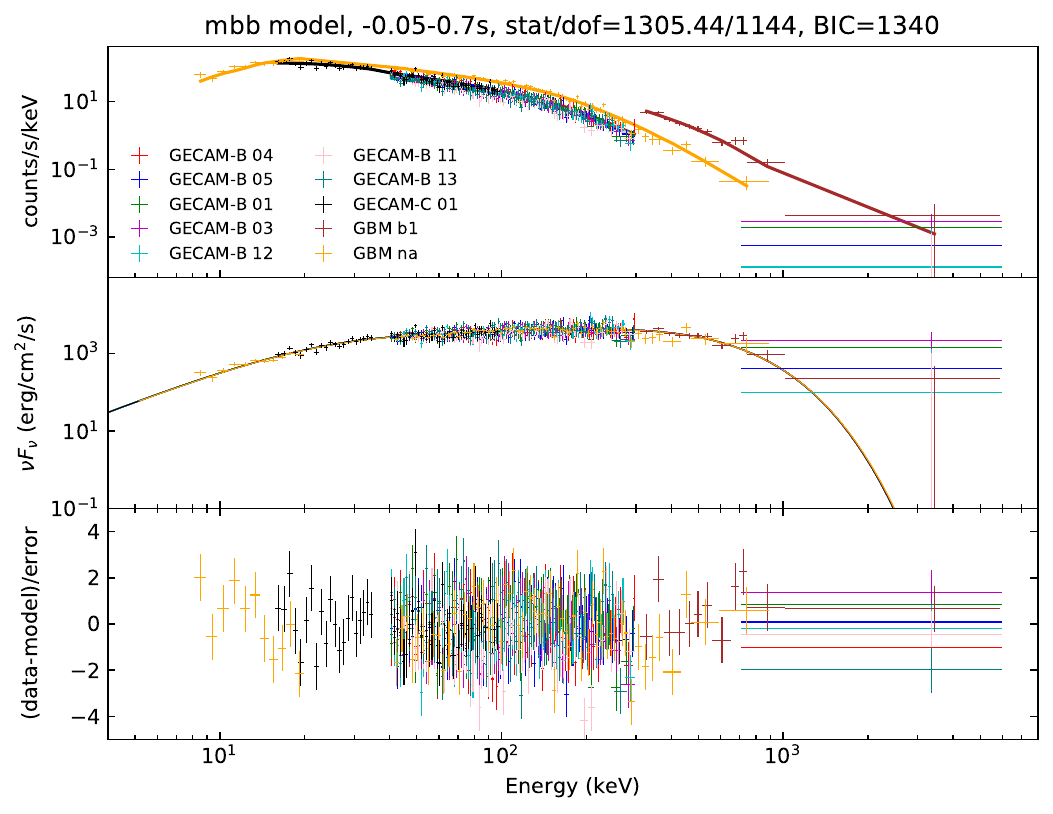}\put(0, 70){\bf c}\end{overpic} &
        \begin{overpic}[width=0.45\textwidth]{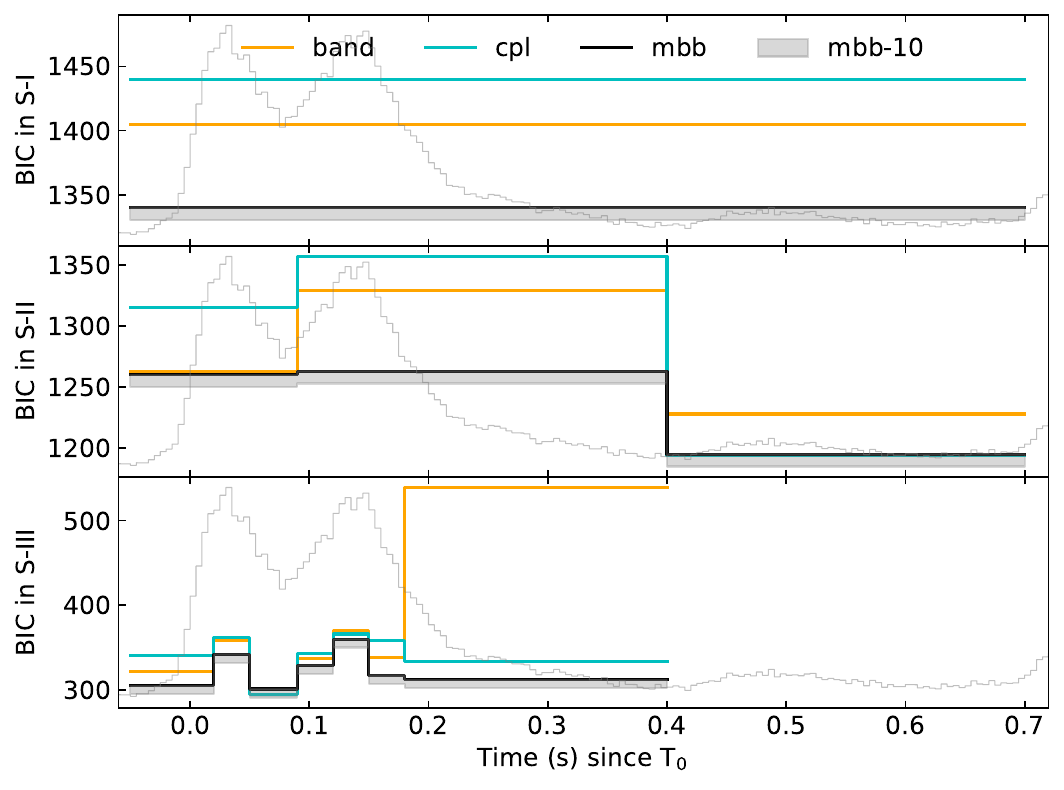}\put(0, 70){\bf d}\end{overpic} \\
\begin{overpic}[width=0.4\textwidth]{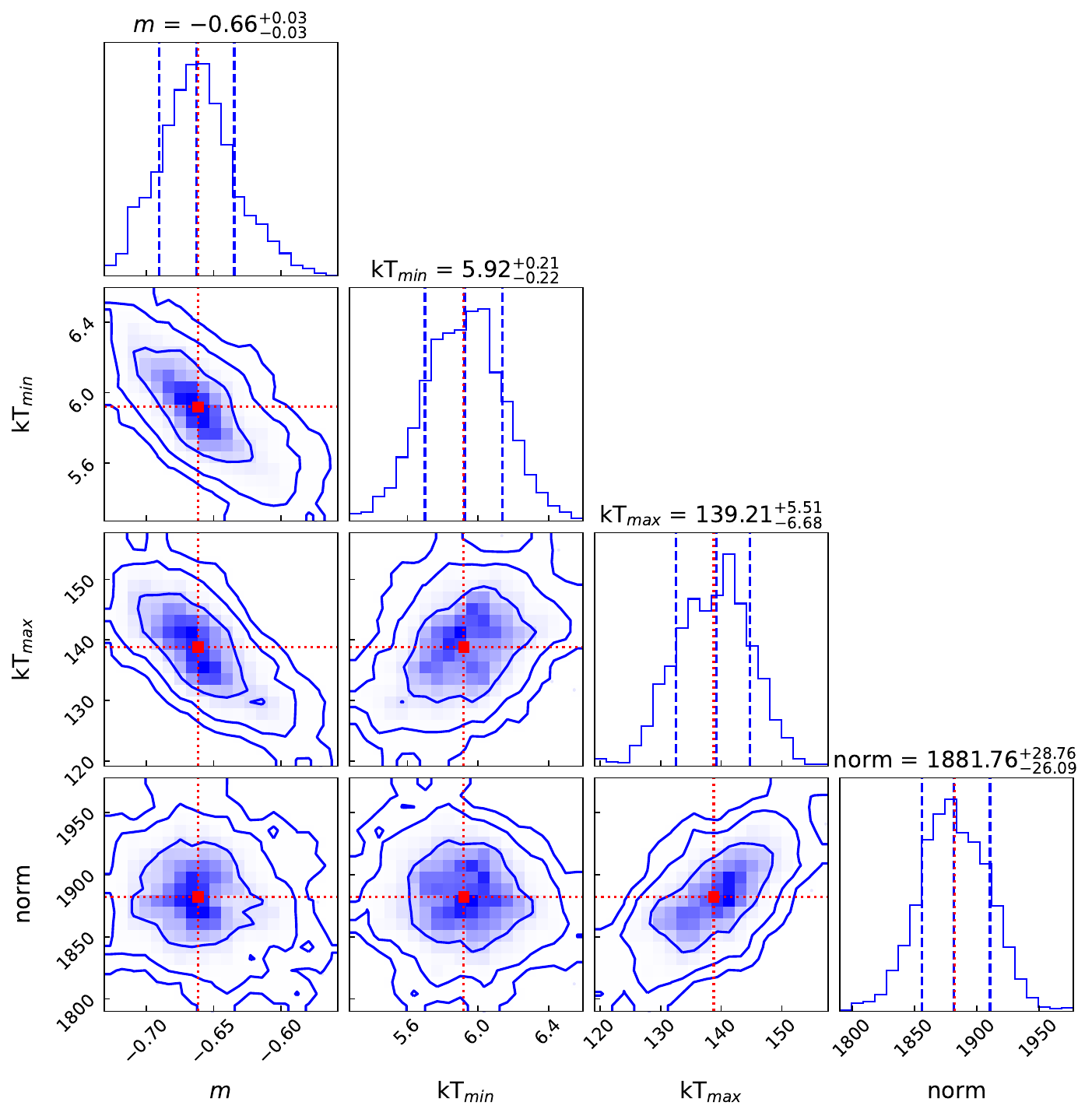}\put(0, 90){\bf e}\end{overpic} &
        \begin{overpic}[width=0.4\textwidth]{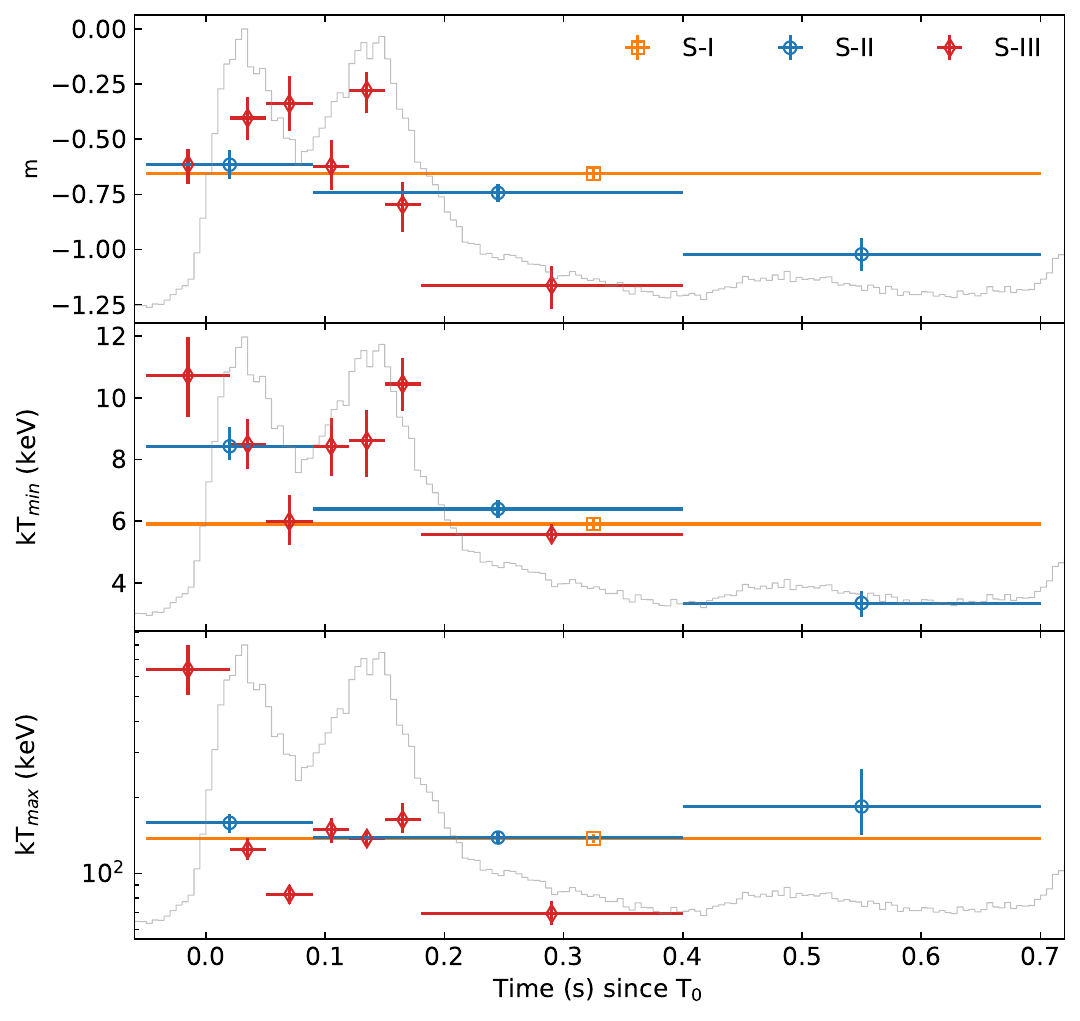}\put(0, 90){\bf f}\end{overpic} \\
\end{tabular}
\caption{\noindent\textbf{Spectra analysis result of precursor of GRB 230307A. }\textbf{a}, Fit results of the CPL model for the S-I spectrum. The structure in fit residuals show a break at low energy. 
\textbf{b}, Fit results of the CPL model with a blackbody component for the S-I spectrum. There is no apparent structure in the fit residuals, and the BIC is lower than that of the CPL model, supporting the presence of a thermal component in the spectrum. 
\textbf{c}, Fit results of the mBB model for the S-I spectrum. There is no apparent structure in the fit residuals with a smaller BIC than that of the CPL model and CPL+bb model. 
\textbf{d}, The BIC of different model of S-I, S-II and S-III spectra in the precursor. The results indicates that throughout the entire precursor, the best model of spectra is almost always mBB. 
\textbf{e}, The corner of the fit results of the mBB model for the S-I spectrum. 
\textbf{f}, The spectral evolution as reflected by the best-fit parameter of the mBB model. In the three stages of precursor (S-II in Table~\ref{tab:spec_fit}), the kT$_{\rm {max}}$ remains almost unchanged, while both kT$_{\rm {min}}$ and m gradually decrease over time. In finer time bins (S-III in Table~\ref{tab:spec_fit}), more complex changes can be observed. For the first spike, both kT$_{\rm {max}}$ and kT$_{\rm {min}}$ show a decrease over time, while in the second spike, they demonstrate a gradual increase. 
}
\label{precursor_spec}
\end{figure*}

The fit result of S-I is shown in the Fig.~\ref{precursor_spec} a,b,c. The structure in the residuals of Fig.~\ref{precursor_spec} indicate that CPL does not describe S-I well. The comparison between  Fig.~\ref{precursor_spec}a and Fig.~\ref{precursor_spec}b shows that the addition of a thermal component has greatly improved the spectral fitting results, as proved by both BIC and residual structures, suggesting the presence of a thermal component in the S-I spectrum. Alternatively, this could be a quasi-thermal spectrum. The CPL+bb model is highly similar in shape to the bandcut model \citep{Oganesyan2018bandcut}, so we also tried the bandcut model, obtaining a better fit. The fitting results of the bandcut model are also better than those of the single CPL and even better than the CPL+bb model, but the low-energy spectral $\alpha_1$ in Eqs.~\ref{equ:bandcut_Model} is positive, which also suggests the presence of some thermal components. 

A commonly used semi-empirical quasi-thermal model in GRB spectra analysis is mBB, which could fit the S-I spectrum well (Fig.~\ref{precursor_spec}c). Combining BIC, fitting residuals, and parameter reasonableness, mBB is considered to be the best model for the S-I spectrum. Additionally, the results of BIC indicate that the mBB model is more favored during the entire period of the precursor candidate (including the quiescent period candidate) as depicted in Fig.~\ref{precursor_spec}d. 

Therefore, the spectrum analysis indicates that the precursor candidate has a quasi-thermal origin, which is significantly different from the main emission spectrum of GRB 230307A. The quasi-thermal spectrum of the precursor candidate is consistent with the presence of thermal components in most short GRB precursors, supporting the idea that the first pulse is a precursor.

\textbf{Evidence from Quiescent (weak emission) Period}: 
The strongest evidence comes from the identification of the quiescent period. The general four criteria of precursor are: 
\begin{enumerate}[(1)]
\item precursor is the the first pulse of the lightcurves;
\item the peak count rate of precursor is lower than that of the main pulse;
\item the count rate during the quiescent period or waiting time, i.e., the time interval between the precursor and the main pulse, is consistent with the background level. We note that this does not mean no emission at all. Instead, there could be weak emission that is too weak to be detected.
\item significant enough to be identified as signal.
\end{enumerate}

A specific precursor, satisfying these four criteria, has been found in GRB 211211A \citep{XS_GRB211211A_QPO}, which is also produced by merger \citep{YJ_GRB211211A_nature,Troja_KN_11A}. As for GRB 230307A, the criterion (1), (2) and (4) are clearly satisfied. The debate about whether ``the first pulse is a precursor" stems from the absence of a quiescent period between this pulse and subsequent emission, with a high signal-to-noise ratio ($\textit{SNR}$) of 71 from $T_0$+0.4\,s to $T_0$+0.7\,s. 

Previous research has shown that the distribution of quiescent time of long GRBs could be well described by the sum of two Gaussian component \citep{GBM_precursor_PRD}. The conventional or ``empirical" quiescent period could be due to low emission which is comparable to background level, rather than a true shutdown of the central engine. Later scenario has been observed in the brightest GRB 221009A \citep{SXY_precursor_09A}, which is a long type-II burst. Considering that GRB 230307A, although much fainter than GRB 221009A, is the second brightest GRB, it is necessary to investigate whether emission between the precursor candidate and the subsequent emission can still be observed when the brightness decreases. 

We conducted three tests to weaken the signal, as depicted in Fig.~\ref{fig:precursor_lc} and Table~\ref{tab:weaken_test_table}. If we weaken the $E_{\rm iso}$ of the precursor candidate ($\sim$1.3$\times$10$^{50}$ ergs) to the level of the precursor of GRB 211211A ($\sim$7.7$\times$10$^{48}$ ergs), and simultaneously put the burst at the same location as GRB 211211A (from z=0.065 to z=0.076), the SNR of all GECAM GRDs is 3.03 and the SNR of a single GECAM GRD is 1.55. If the energy of this precursor candidate is maintained and it is simply moved to a redshift of z=0.5, the SNR of all GECAM GRDs has already decreased to 0.73, and 0.37 for the SNR of a single GECAM GRD. When this source is moved to a larger redshift of z=1, the SNR  of all GECAM GRDs and a single GRD will further decrease to 0.13 and 0.07 respectively. From this perspective, the quiescent time after the precursor of GRB 211211A could be a kind of ``tip of iceberg" and the GRB 230307A with extreme brightness provides an opportunity to unveil the weak emission during the so-called quiescent period. We note that the high luminosity of the precursor with multi-peaks profile can have constrains on the models of precursor.

\begin{table*}[htbp]
\caption{\centering{Weaken test result of the quiescent period in GRB 230307A.}}
\begin{tabular*}{\hsize}{@{}@{\extracolsep{\fill}}cccccc@{}}
\toprule
test ID & redshift & $E_{\rm iso}$ (erg) & SNR of 25 GRDs & SNR of single GRD\footnote{GRD04 in GECAM-B and GRD01 in GECAM-C.} & SNR of single NaI\footnote{NaI detector na of \textit{Fermi}-GBM.}\\
\hline
1 & 0.076 & $7.7\times10^{48}$\footnote{$E_{\rm iso}$ of precursor in GRB 211211A.}  & 3.03 & 1.55 & 2.50\\
2 & 0.5 & $1.3\times10^{50}$\footnote{$E_{\rm iso}$ of precursor in GRB 230307A.}  & 0.73 & 0.37 & 0.62\\
3 & 1 & $1.3\times10^{50}$  & 0.13 & 0.07 & 0.11\\
\botrule
\end{tabular*}
\label{tab:weaken_test_table}
\end{table*}

\begin{table*}[htbp]
\caption{\centering{Timescale of different phases of three GRBs}}
\begin{tabular*}{\hsize}{@{}@{\extracolsep{\fill}}ccccc@{}}
\toprule
GRB name& $T_{90}$(s) & $T_{\rm pre}$(s) & $T_{wt}$(s)\\
\hline
GRB 230307A & 41.52 & 0.40 & 0.30 \\
GRB 211211A & 43.18 & 0.20 & 0.93 \\
GRB 170228A & 60.20 & 0.65 & 1.20 \\
\botrule
\end{tabular*}
\label{tab:timescale_of_GRBs}
\end{table*}

\subsubsection{Three-episode burst pattern}\label{Burst Pattern}
By comparing GRB 211211A and GRB 230307A, we can notice that both of these GRBs exhibit three distinct episodes in their prompt emission, which is one of the most important features of type IL GRB. But these are not sufficient to fully characterize the burst behavior of type IL GRB. More surprising features are hidden within the precursor.

We calculated the precursor duration, the waiting time between precursor and main emission ($T_{wt}$), and the total duration ($T_{90}$) of GRB 230307A and GRB 211211A (Fig.~\ref{fig:classification}, Table.~\ref{tab:timescale_of_GRBs}). In general, for those GRBs which have precursor, long GRBs tend to have a longer waiting time, while short GRBs have a shorter waiting time. What makes these two GRBs peculiar is that they last a long time but the waiting time is relatively short. As depicted in the $T_{90}$-$T_{wt}$ distribution map (Fig.~\ref{fig:classification}c), the $T_{90}$ of these two GRBs are closer to the Type II GRB region while the waiting time of them are closer to the Type I GRB region, which results in them being far from both classic type I and type II GRBs.

This is a very prominent and important supplementary feature of the type IL GRB, which can serve as a crucial reference for type IL GRB searching and provide constraints for the unified model of type IL GRBs.

Hence, the burst pattern of type IL GRB can be summarized as:
\begin{enumerate}[(1)]
\item there must be an evident precursor;
\item the waiting time $T_{wt}$ should be relatively short;
\item a relatively long $T_{90}$ of the prompt emission,  including precursor, main emission and extended emission;
\item having an extended emission, which is separated from the main emission by a dip-like structure.
\end{enumerate}

The second description about the waiting time is crucial among these descriptions. As we know there are also some type II GRBs exhibiting precursor, main emission, and extended emission, such as GRB 160625B \citep{Zhang160625B} and GRB 221009A \citep{HXMT_GECAM_221009A}, with a $T_{wt}$ over hundreds seconds, thus do not follow the burst pattern of type IL. 

The accuracy and validity of this burst pattern will be further confirmed in the search results of the Section \ref{Burst Sample}.

\subsection{Burst Sample} \label{Burst Sample}
To increase the sample size and estimate the occurrence rate of type IL GRB, we conducted a search in the archive data of \textit{Fermi}/GBM from 2014 to 2024. We demanded that the GRBs should satisfy the burst pattern of type IL GRB described in Section~\ref{Burst Pattern}. As a result, we found one good candidate, GRB 170228A, which is very similar to GRB 230307A and GRB 211211A in shape of light curve and follows the burst pattern strictly. 

\begin{figure*}
\centering
\begin{tabular}{c}
\begin{overpic}[width=0.5\textwidth]{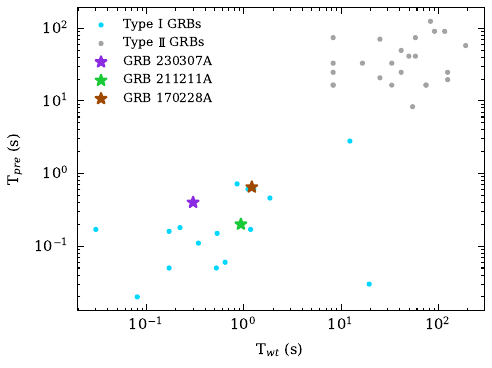}\put(0, 70){\bf a}\end{overpic} \\
        \begin{overpic}[width=0.5\textwidth]{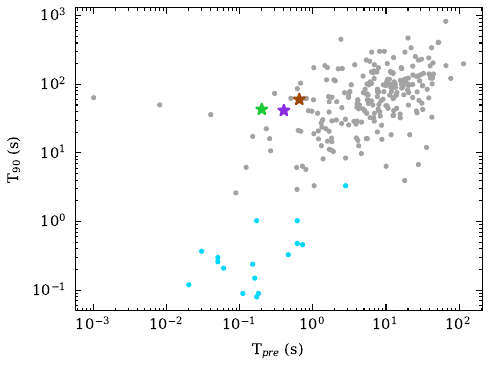}\put(0, 70){\bf b}\end{overpic} \\
        \begin{overpic}[width=0.5\textwidth]{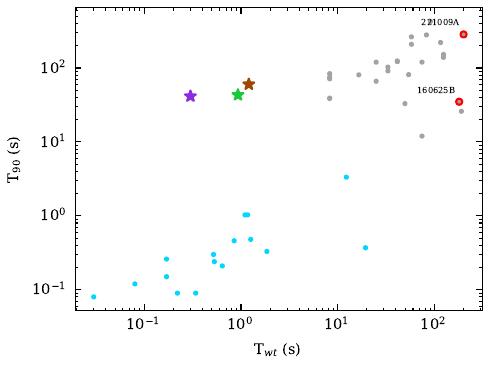}\put(0, 70){\bf c}\end{overpic} \\
\end{tabular}
\caption{\noindent\textbf{The temporal properties of type IL GRB samples.} \textbf{a}, The $T_{90}$ vs T$_{pre}$ diagram. \textbf{b}, The T$_{pre}$ vs $T_{wt}$ diagram. \textbf{c}, The $T_{90}$ vs $T_{wt}$ diagram. In \textbf{a}--\textbf{c}, type I and type II GRBs are represented by cyan and gray solid circles, respectively, GRB 230307A, GRB 211211A and GRB 170228A are highlighted by the star in color of purple, green and brown, respectively. In \textbf{c},two typical type II GRBs with extended emission, GRB 221009A and GRB 160625B are put in red circle. }
\label{fig:duration}
\end{figure*}

GRB 170228A was detected at 2017-02-28T19:03:00.17 (UTC) by \textit{Fermi}/GBM. Unfortunately, there is no publicly available follow-up observation for this GRB, so the redshift is unmeasured. Like the GRB 230307A and GRB 211211A, GRB 170228A can also be divided into three episodes: the precursor lasts for about 0.6\,s, which is relatively significant over the background, and is followed by a bright main emission which lasts about 9\,s, and an extended emission up for about 50\,s. The $T_{90}$ of GRB 170228A is up to about 60s. Additionally, as one can see, GRB 170228A locates in the same region as GRB 211211A and GRB 230307A and formed a separate cluster of data, making themselves one peculiar classification in $T_{90}$-$T_{wt}$ distribution map (Fig.~\ref{fig:duration}c).

\begin{figure*}
\centering
\begin{tabular}{cc}
\begin{overpic}[width=0.45\textwidth]{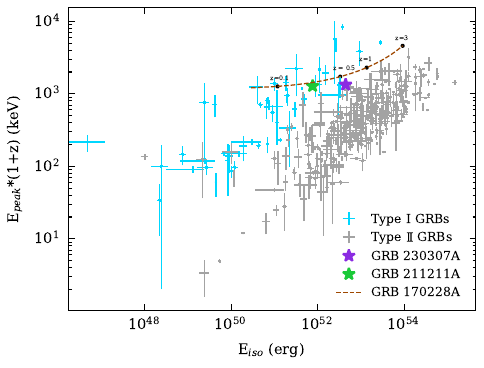}\put(0, 70){\bf a}\end{overpic} &
        \begin{overpic}[width=0.45\textwidth]{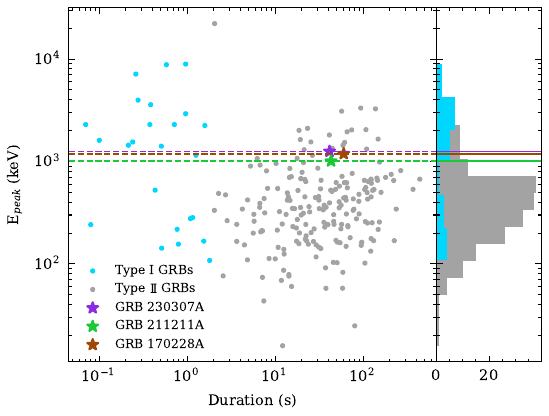}\put(0, 70){\bf b}\end{overpic} \\
\begin{overpic}[width=0.45\textwidth]{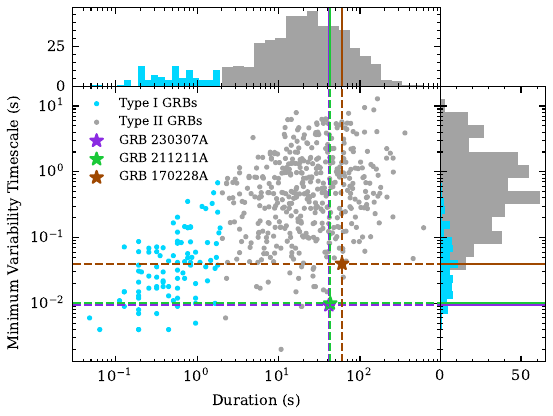}\put(0, 70){\bf c}\end{overpic} &
        \begin{overpic}[width=0.45\textwidth]{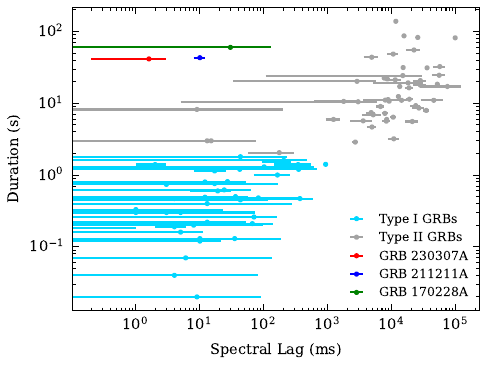}\put(0, 70){\bf d}\end{overpic} \\
\begin{overpic}[width=0.45\textwidth]{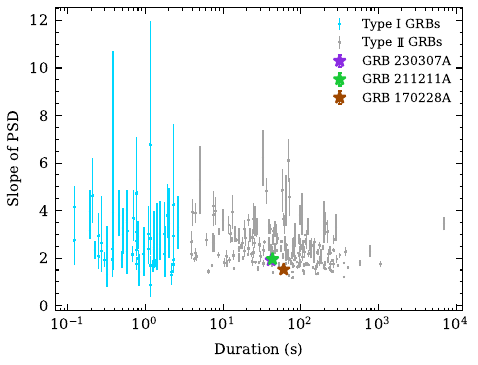}\put(0, 80){\bf e}\end{overpic} &

\end{tabular}
\caption{\noindent\textbf{Position of type IL GRB samples in various GRB classification schemes.} \textbf{a}, The E$_{p,z}$ and E$_{\rm iso}$ correlation diagram. GRB 230307A and GRB 211211A are highlighted by the star in color of purple and green, respectively. The brown dashed line represent the evolution of the  E$_{p,z}$ and E$_{\rm iso}$ of GRB 170228A as the redshift changes from 0.05 to 4. 
\textbf{b}, The peak energy versus duration diagram. \textbf{c}, The minimum variability timescale versus duration diagram. \textbf{d}, The duration versus lag diagram. \textbf{e}, The slope of PSD versus duration diagram. In \textbf{a}--\textbf{e}, type I and type II GRBs are represented by cyan and gray solid circles, respectively. All error bars on data points represent their 1$\sigma$ confidence level. In \textbf{b}--\textbf{e},GRB 230307A, GRB 211211A and GRB 170228A are highlighted by the star in color of purple, green and brown, respectively.}
\label{fig:classification}
\end{figure*}

To further prove the similarity between GRB 170228A, GRB 211211A and GRB 230307A, more analysis were conducted for other properties of GRB 170228A, including spectrum and temporal properties. The time-integrated spectrum of GRB 170228A can be well described by a single CPL model, which is shown in Fig.~\ref{fig:28A_spec}. As shown in the Amati relation (Fig.~\ref{fig:classification}a), GRB 170228A is more close to type I GRB region, just like GRB 230307A and GRB 211211A. Additionally, both spectral lag and MVT fall in the type I GRB region even if GRB 170228A is a long-duration GRB. All these features suggest that GRB 170228A is a special “short” (merger origin) GRB.

As for other properties of these three GRBs, not only did they fall in type I GRB region, but they also resemble each other,including $E_{peak}$ (Fig.~\ref{fig:classification}b), slope of PSD (Fig.~\ref{fig:classification}e), which indicates the similarity in the origin of these GRBs. The successful identification of GRB 170228A as type IL GRB by the morphology of the light curves also suggests the speciality of this burst pattern.

\section{Discussion}\label{section3}
Based on the three type IL GRB samples and the successful application of burst pattern in searching type IL GRB, here we discuss the physical implications of the burst pattern.

\subsection{Precursor}
The temporal and spectral behavior of precursors are essential components of the burst pattern that provide crucial information for understanding type IL GRBs.

The first characteristic of precursors of type IL GRB is their high luminosity. For the two GRBs in the sample with known redshifts (GRB 211211A and GRB 230307A), their precursors are too bright compared with other GRBs, especially the precursor of GRB 230307A, whose $E_{\rm iso}$ is as high as $\sim1.3\times10^{50}$\,erg. This leads to many models having energy budget issues and being ruled out \citep{2023Dichiara}.

The second characteristic of precursors of type IL GRB is their short $T_{wt}$ compared to the long $T_{90}$. If $T_{90}$ is considered more fundamental, the short $T_{wt}$ of type IL GRB must have an unified explanation. While if the $T_{wt}$ can directly constrain the progenitor system, then the explanation of why all type IL GRBs have long duration is needed.

The third characteristic of precursors in type IL GRB is their multi-peak structure. Compared to the simpler precursor structures of other GRBs, the precursors of type IL GRB typically exhibit more complex structural features. To be specific, the precursor of GRB 211211A shows multiple peaks, displaying Quasi-Periodic Oscillation (QPO) features \citep{XS_GRB211211A_QPO}; the precursor of GRB 230307A has two distinct peaks, while the precursor of GRB 170228A also shows multiple peaks. Consequently, models capable of producing multiple FRED-like precursors are more favored in explaining type IL GRBs.

\subsubsection{Pre-merger Model}

There are many branches of theoretical models suggesting that the precursor of merger-type GRB originates from the interaction between the compact star before merger, including the magnetosphere interaction between two neutron stars \citep[see e.g.,][and references therein]{Hansen2001pre_merger_model, 2010ApJ...723.1711T}, and the resonant shattering flare (RSF) of NS crusts before coalescence \citep{Tsang2012pre_merger_model, 10.1093/mnras/stac1645}. The high luminosity of the precursor requires the model to have a high energy budget.

The typical energy budget of RSF is $\sim10^{46} - 10^{47}$\,erg which is from the shattering of the NS crust driven by a crust-core interface mode and enough to generate the luminous precursor of GRB 230307A. The RSF model is used to explain the precursor of GRB 230307A in some research, and due to the extremely high luminosity this requires a relatively high gamma-ray efficiency, otherwise the lower limit on the surface magnetic field of the neutron star is even higher than $\sim1.2\times10^{15}$\,G \citep{2023Dichiara}. However the spectrum of RSF is near-thermal only for weak magnetic field while the emission will be non-thermal when the magnetic field is strong. 

It is worth noting that, if there is a magnetar in a binary neutron star system, a superflare (SF) with energy hundred times larger than giant flares of magnetar will have the opportunity to be released just seconds or sub-seconds before the coalescence from a catastrophic global crust destruction when the tidal-induced deformation surpass the maximum that the magnetar’s crust can sustain \citep{zhang2022superflare}. The SF model can give a good explanation of the high energy budget ($\sim7.7\times10^{48}$\,erg) and Quasi-Periodic Oscillation (QPO) ($\sim$22\,Hz) in precursor of GRB 211211A \citep{zhang2022superflare}, and it can provide a much lager energy budget than RSF.

The luminosity of superflare extracted from both the core and crust can be as high as
\begin{equation}   
L_{\rm max}\sim10^{51}{\rm erg}\cdot {\rm s}^{-1} \left(\frac{v}{c}\right) \left(\frac{B_{\rm eff}}{10^{15}{\rm G}}\right)^2 \left(\frac{R_{\Sigma}}{10\rm km}\right)^2
\label{equ:Lmax_Model}
\end{equation}
where $\frac{v}{c}$ is the velocity of the perturbations, $R_{\Sigma}$ is the linear scale of the crust involved in deconstruction, $B_{\rm eff}$ denotes the enhanced magnetic-field strength at the radius of r $\sim$ $R_{\Sigma}$ after the crust disruption which can be the same order as the magnetic-field strength at the outer boundary of the NS core \citep{Lander2013NS} and be 1–2 orders of magnitude higher than that at the magnetar surface before destruction \citep{Henriksson2013NS}. If the perturbations propagate in the same speed as Alfvén waves, $\frac{v}{c}$$\gtrsim$ 0.1, the energy budget is sufficient to produce the precursor of GRB 230307A.

The SF model predicts that if the double neutron star (DNS) merger is consist of two magnetars, there could be two precursor peaks, which is highly appealing for GRB 230307A because most model cannot produce such a complex precursor lightcurve. Additionally, the time interval between the beginning of the SF and the coalescence (corresponding to the duration of the precursor) is positively correlated with its magnetic field strength (corresponding to the energy budget) in SF model, which is consistent with the observation of GRB 211211A and GRB 230307A.

Another prediction of the SF model is the presence of QPO in precursor, with a positive relationship between the QPO frequency and the total energy, which is not observed in the precursor of GRB 230307A, this may be due to the high level of red noise. 

A notable difference between the precursor of GRB 230307A and the precursor of GRB 211211A is that the best-fit spectrum model for the former is a quasi-thermal mBB, while the best-fit model for the latter is a non-thermal CPL. Specifically, in the precursor energy spectrum of GRB 230307A, there is a clear low-energy break, which is absent in GRB 211211A. 

\begin{figure*}
\centering
\begin{tabular}{cc}
\begin{overpic}[width=0.45\textwidth]{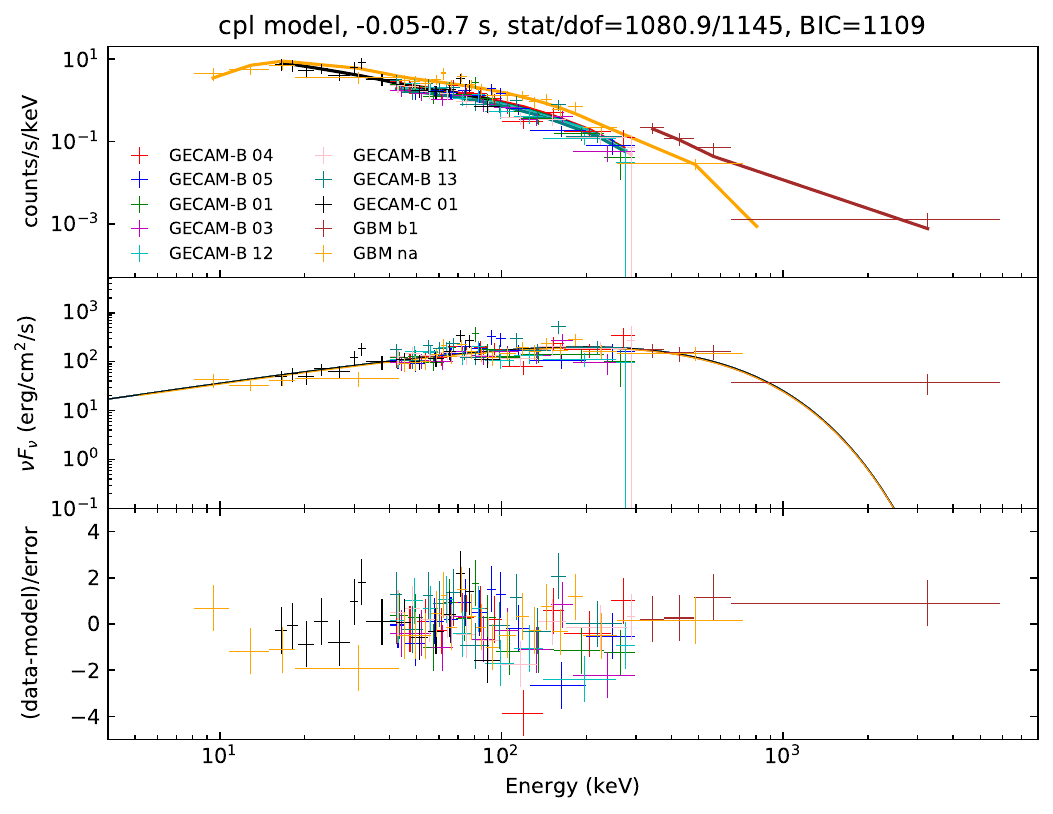}\put(0, 70){\bf a}\end{overpic} &
        \begin{overpic}[width=0.45\textwidth]{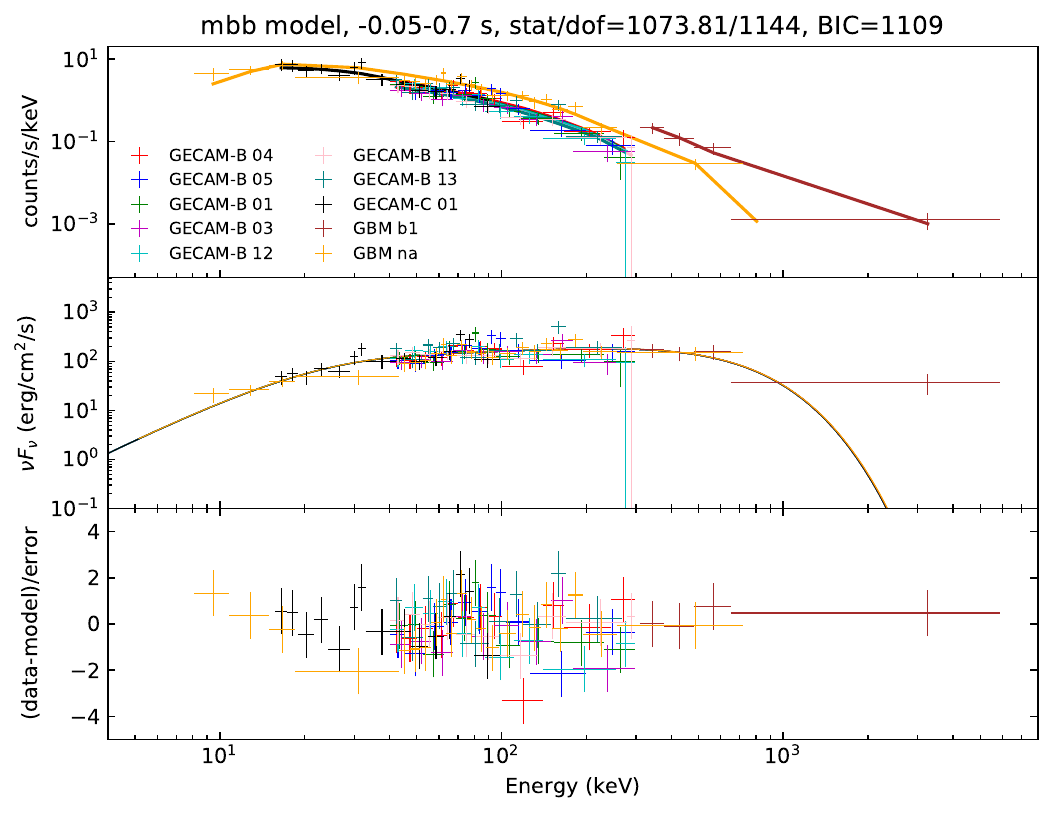}\put(0, 70){\bf b}\end{overpic} \\
\begin{overpic}[width=0.45\textwidth]{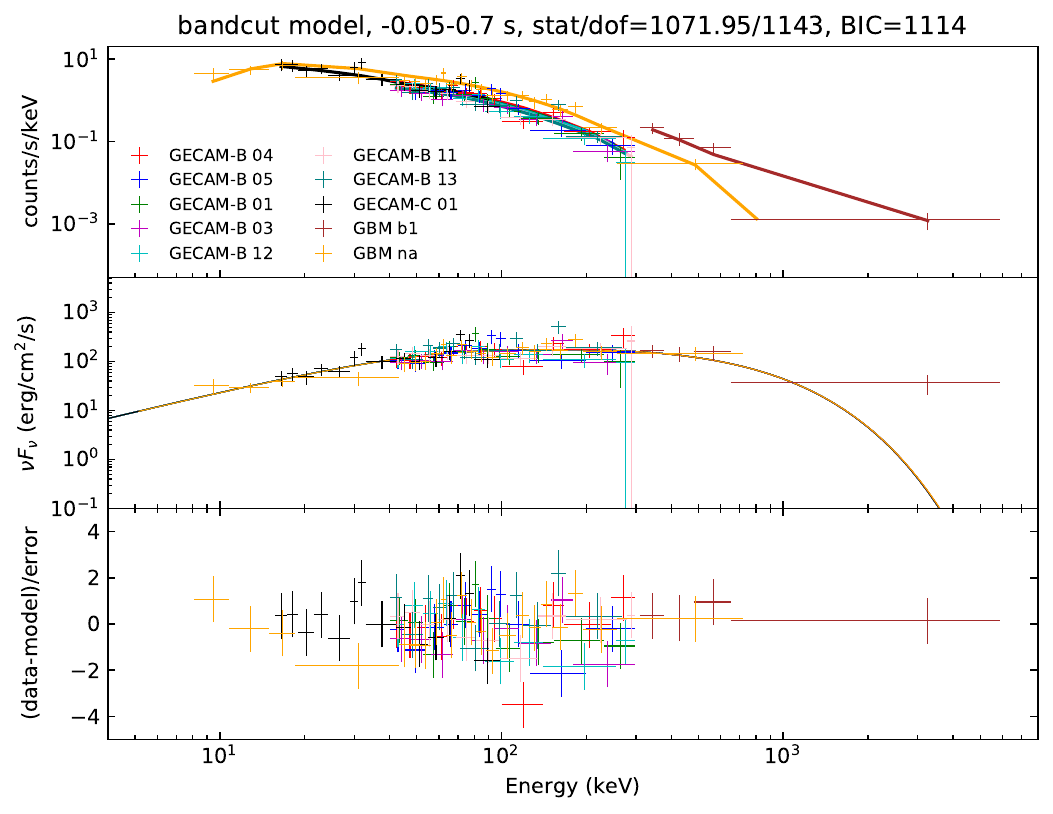}\put(0, 90){\bf c}\end{overpic} &
        \begin{overpic}[width=0.45\textwidth]{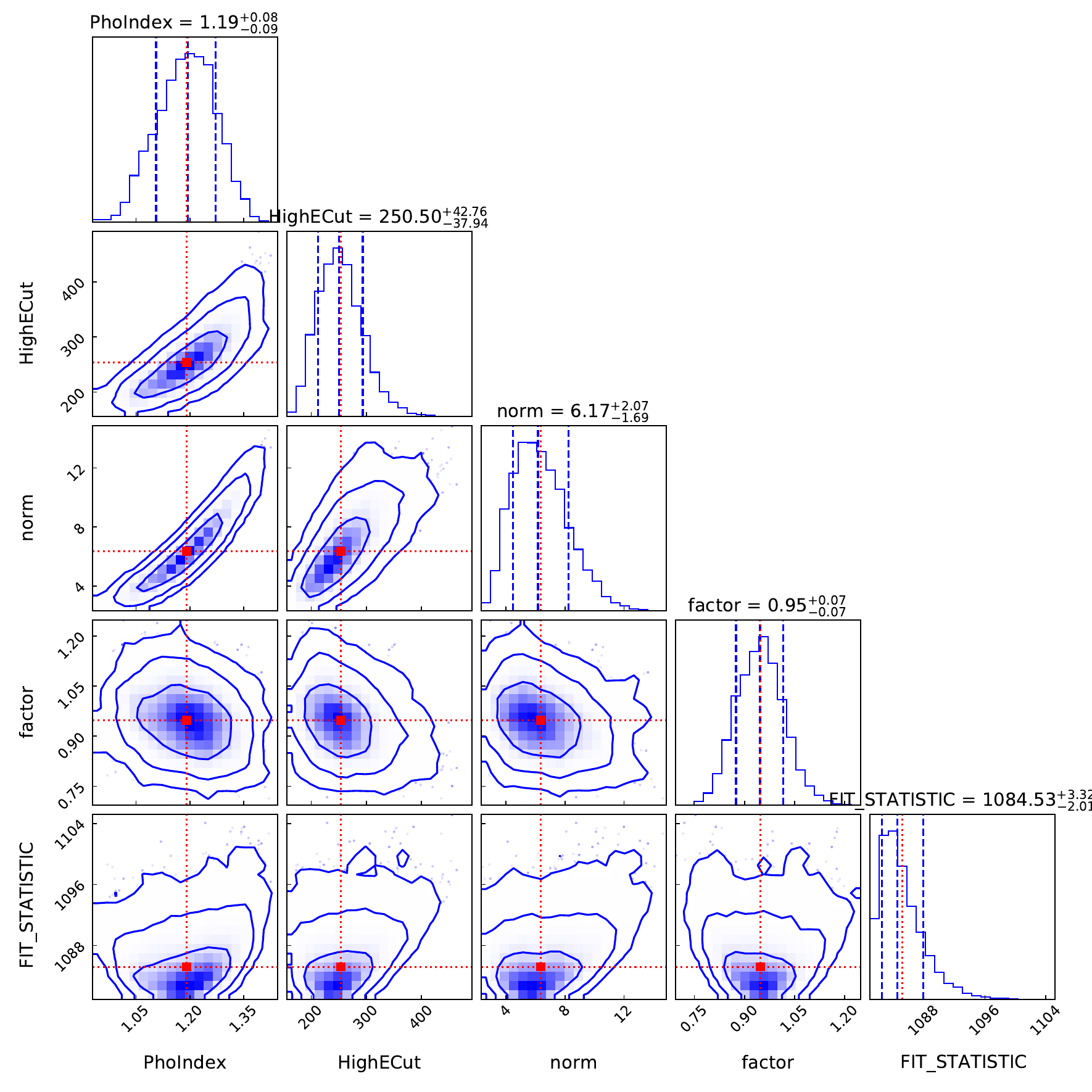}\put(0, 90){\bf d}\end{overpic}
\end{tabular}
\caption{\noindent\textbf{The comparison of the spectra of precursor in GRB 230307A and GRB 211211A. }\textbf{a-c}, Fit results of different model for the simulated quasi-thermal precursor spectrum. The simulated spectrum is obtained using the fakeit function in pyxspec by assuming a mBB model and response matrix. The norm parameter of the assuming mBB model is the best fit of S-I spectrum and reduced by a factor, which is the same as the weaken test 1. 
\textbf{d}, The corner of the fit results of the CPL model for the simulated quasi-thermal precursor spectrum. 
The fit results show that even if the spectrum of GRB 211211A is a quasi-thermal spectrum similar to GRB 230307A, due to the low photon count, the low-energy break cannot be statistically identified, and the observed spectrum will degrade to a non-thermal CPL model. Therefore, the unique non-thermal spectrum of the precursor of GRB 211211A may not necessarily be due to its non-thermal origin, but could also be a result of insufficient statistics to reveal its quasi-thermal characteristics. And the same origin of the precursor of GRB 230307A and GRB 211211A cannot be ruled out based on their energy spectra.}
\label{weaken_spec}
\end{figure*}

To investigate this difference, we decreased the norm of the best-fit mBB model of the S-I spectrum of GRB 230307A precursor by the same factor of weaken test 1 in Table~\ref{tab:weaken_test_table}. Using the \textit{fakeit} \footnote{\href{https://heasarc.gsfc.nasa.gov/xanadu/xspec/python/html/extended.html\#fakeit}{https://heasarc.gsfc.nasa.gov/xanadu/xspec/python/html\\/extended.html\#fakeit}} provided by PyXspec, we simulated the spectra of the precursor of GRB 230307A observed by GECAM and \textit{Fermi}/GBM when $E_{\rm iso}$ was reduced to the same level as the precursor of GRB 211211A and moved the source to the same redshift as GRB 211211A. We then fitted the simulated spectra, and the results are shown in Fig.~\ref{weaken_spec}. The low-energy break is no longer identifiable in the spectrum, and the CPL model is the best-fit model for the weakened spectra, because the statistics of photons are no longer sufficient to display the low-energy break in the data. This suggests that, although the precursor of GRB 211211A exhibits a non-thermal CPL spectrum, it does not mean that it necessarily has a non-thermal origin. It is still possible that the luminosity is insufficient to reveal the characteristics of a quasi-thermal spectrum. Hence the precursor of GRB 230307A and GRB 211211A could have the same thermal origin.

It is believed that non-thermal emission is generated in both SF model and RSF model. This is one of the evidence that suggests the precursor of GRB 211211A originated before the merger, but this is inconsistent with the observation of quasi-thermal spectrum of GRB 230307A. If the magnetic field is very strong then the spectrum of the precursor generated by RSF should be non-thermal \citep{Tsang2012pre_merger_model}, and SF model predicted that if the released energy propagates along or interacts with magnetic field lines, the resultant emission will be non-thermal. However, RSF and SF mainly describe the energy supply mechanism. During radiation process, synchrotron radiation spectra may still be thermalized due to scattering, resulting in the observed quasi-thermal spectra.

A potential issue with magnetars being involved in compact star merger is that the magnetar cannot maintain its strong magnetic field until the merger occurs due to different processes \citep{Goldreich1992Magnetic}, such as Hall effect \citep{Hollerbach2002Halldriftohmic}, Ohmic dissipation \citep{Hollerbach2002Halldriftohmic} and ambipolar diffusion \citep{Igoshev2023Ambipolar}.
Therefore, the strong magnetic field of the magnetar may have already decayed to the normal magnetic field of a typical NS when the merger occurred. But some research suggest the magnetic field can be amplified to magnetar-level ($\sim10^{16}$\,G) due to the instabilities and turbulence arise from the tidal interaction of its companion in the internal fluid \citep{Giacomazzo2015amplify} and will solve this problem \citep[see also][and references therein]{2024arXiv240409251W}.

\subsubsection{Post-merger Model}
Various models suggest that different processes can lead to a precursor after the coalescence, including emission from the photosphere, and from the jet breaking out, etc \citep{10.1046/j.1365-8711.2002.05875.x, 2010ApJ...723.1711T, 2016ApJ...819..120L, NAKAR20201}. Most models expect to obtain a thermal or quasi-thermal spectrum. 

If the main emission of GRB 230307A is produced by internal shock and the precursor is produced by the transition of the fireball from optically thick to optically thin or by the interaction of the jet with the progenitor (such as jet breaking out), the time variability of the main emission $\Delta t$ should be larger than the time variability of the precursor $\delta t$, because the radius of internal shock ($R_{\rm IS}\sim2\Gamma^2c\Delta t$) should be larger than the photosphere radius ($R_{\rm PS}\sim2\Gamma^2c\delta t$). While the $R_{\rm IS}-R_{\rm PS}\sim2\Gamma^2cT_{wt}$, thus $\Delta t$ should larger than $T_{wt}$ \citep{Lazzati2005Precursor}. But the MVT of the main emission of GRB 230307A is $\sim24\,{\rm ms}$ for GECAM data and $\sim15\,{\rm ms}$ for \textit{Fermi}/GBM data \citep{2023Dichiara} which is much shorter than the $T_{wt}$ ($\sim300\,{\rm ms}$ for GECAM data) of GRB 230307A. Hence, the observations cannot support the model of the main emission of GRB 230307A being produced by internal shocks when the precursor is produced by the fireball becoming optically thin \citep[see also][]{2010ApJ...723.1711T}.

Another interesting topic is whether this precursor can be naturally generated within the framework of the Internal-Collision-Induced Magnetic Reconnection and Turbulence \citep[ICMART][]{zhang2011ICMART}, especially since some studies have indicated that GRB 230307A probably has a Poynting-flux-dominated jet \citep{2023arXiv231007205Y,lv2024ICMART}. In the scenario of ICMART, the quiescent time can be explained as the random waiting time of different mini-jets, and a precursor-like structure can be found in the simulation results of ICMART \citep{2023arXiv231007205Y}. However, it is expected that there will be no difference in the spectral lag between the fast pulse of precursor emission and the fast pulse of main emission.

Due to limitations in statistics, it is hard to perform a time-resolved spectrum analysis with finer time slices, but the evolution of the spectrum can be investigated using hardness ratios. The hardness ratio between 30-100 keV and 100-500 keV is shown in the Fig.~\ref{fig:precursor_lc}. At the onset of the burst, there is a very rapid increase in the hardness ratio, reaching the maximum value for the entire burst in the corresponding energy band. Then, it decreases rapidly as the rising edge of the first spike of the precursor is not yet finished. The two peaks of the precursor correspond to two hardening events in the hardness ratio. During the decay of the second spike, there is also a brief increase in the hardness ratio. Such a hardness ratio change does not follow the ``hard to soft evolution" pattern (which is usually followed by ICMART events) nor the ``intensity tracking" pattern (which is usually followed by Internal shock events).

Therefore, further and more detailed investigation is needed to test whether the ICMART model can explain the observed characteristics of the precursor in GRB 230307A and also in GRB 211211A, if we believe they have a similar origin.

\subsection{Main emission and extended emission}

Using the large BATSE sample with time-tagged event (TTE) data, \citet{Norris2006EE} have demonstrated that, for bright bursts where significant measurements are possible, the extended emission (EE) of short GRBs, which were present in $\sim 1/3$ of their main sample,  exhibits negligible spectral lag. This characteristic is consistent with the main emission of short GRBs and thus offers a method to distinguish whether the persistent emission following the initial hard peak is the extended emission of a short GRB or the low-intensity pulses of a long GRB. Obviously, the very small lag of the extended emission in the three type IL GRBs discussed in this paper clearly confirms this evidence.



Furthermore, the fact that the spectral lag of the extended emission is as negligible as that of the main emission; as shown in Fig.~\ref{fig:precursor_lag} for GRB 230307A and analysis by \citet{Peng2024Comparative} for GRB 211211A, indicates that the extended emission is probably produced in the same region as the main emission. Otherwise, the spectral lag of the extended emission would be larger due to the curvature effect at a large radius \citep{Norris2006EE} .


Considering the duration of the extended emission and the interval between the extended emission and the main emission, it is expected that the main emission and extended emission in type I GRBs are generated by different processes \citep[see, e.g.,][for a similar discussion about sGRBEE]{Metzger08, 10.1111/j.1365-2966.2011.19810.x}. Similar to sGRBEEs, where the extended emission is probably powered by late-time activity of the central engine, a successful model of type I GRBs should take this into account \citep[see, e.g.,][and references therein]{10.1111/j.1365-2966.2011.19810.x, 2022ApJ...939..106J}.



For the typical sGRBEE, magnetars are one of the most popular potential central engine candidates \citep{Metzger08, 10.1111/j.1365-2966.2011.19810.x, 2013MNRAS.431.1745G}. The main emission in sGRBEE is often explained by the accretion process after merger, while the relatively long and soft extended emission can be explained by several models involving magnetar in some activities, such as relativistic wind that extracts the rotational energy of the magnetar during its cooling and spin-down \citep{Metzger08}, the process of fall-back accretion onto a newborn magnetar \citep{Gibson17} and delayed energy injection after merger. The interval between the main emission and the extended emission can be explained by the cooling time of the newly formed magnetar or the time required for the matter that are ejected into highly eccentric orbits during the NS’s tidal disruption to fall-back accretion \citep{2013MNRAS.431.1745G}. 

A robust model for type IL GRB should address some more observation features, including a consistently small spectral lag and a long duration of main emission. The small yet relatively constant spectral lag observed in both the main emission and extended emission suggests that they should be generated in the same region, even by different mechanisms. So those models in which the extended emission site is at large distance from the main emission site would be disfavored.

More importantly, for those sGRBEE models, the characteristic time-scale for accretion is about several seconds, which corresponds to the duration of main emission in sGRBEE. But the type IL GRBs in our paper exhibit longer duration of main emission (see discussion below), which may need some explanation other than pure accretion process.

\begin{figure*}
\centering
\begin{tabular}{cc}
\begin{overpic}[width=0.45\textwidth]{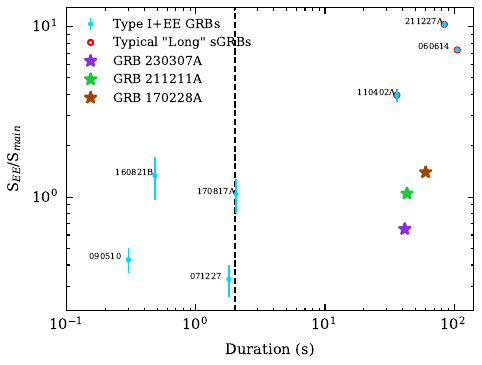}\put(0, 70){\bf a}\end{overpic} &
        \begin{overpic}[width=0.45\textwidth]{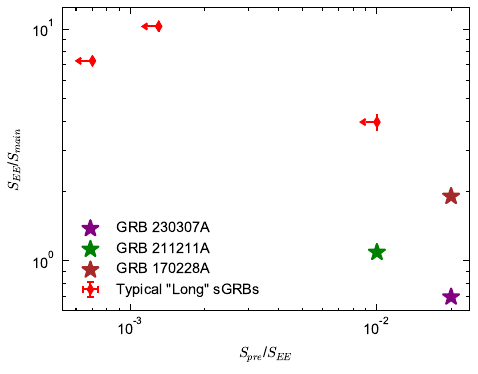}\put(0, 70){\bf b}\end{overpic}

\end{tabular}
\caption{\noindent\textbf{Comparison of type IL GRB with typical ``long" sGRB.} 
\textbf{a}, The ratio of photon fluence from 15\,keV to 100\,keV between the extended emission and the main emission ($S_{EE}/S_{main}$) versus duration diagram. The red hollow circles represent the typicle ``long'' short GRBs, GRB 060614, GRB 211227A and GRB 110402A. 
\textbf{b}, ($S_{EE}/S_{main}$) versus the ratio of photon fluence between precursor and extended emission $S_{pre}/S_{EE}$. The leftward arrows represent the 3$\sigma$ upper limits of $S_{pre}/S_{EE}$ of typical ``long" sGRB without obvious precursors.
The type IL GRB 230307A, GRB 211211A and GRB 170228A are highlighted by the star in color of purple, green and brown, respectively. }
\label{fig:ratio}
\end{figure*}

If only the extended emission is considered, the difference between type IL GRB and classical ``long" short GRB (including GRB 060614 and GRB 211227A) cannot be discerned, as they exhibit similar behavior. Therefore, we 
define the ratio of photon fluence between long-soft extended emission and main emission ($S_{\rm EE}/S_{\rm main}$). As depicted in Fig.~\ref{fig:ratio}a, $S_{\rm EE}/S_{\rm main}$ of type IL GRB are close to unity, which means the photon fluence of the extended emission is comparable to that of the main emission, and in some cases, the latter may even exceed the former. The property of $S_{\rm EE}/S_{\rm main}$ significantly distinguish type IL GRBs from classical ``long" short GRBs, as $S_{\rm EE}/S_{\rm main}$ of these ``long" short GRBs are relatively large. In other words, probably more energy is allocated to the main emission rather than the extended emission in type IL GRB. 

After taking into account the photon fluence of the precursor ($S_{\rm pre}$), more differences between type IL GRB and typical ``long" sGRB are revealed in the diagram of $S_{\rm EE}/S_{\rm main}$ versus $S_{\rm pre}/S_{EE}$ (Fig.~\ref{fig:ratio}b). Since there is no precursor detection in typical ``long" sGRB, we calculate the 3 sigma upper limits of the precursor fluence and then determine the $S_{\rm pre}/S_{\rm EE}$ of them. We can notice that $S_{\rm pre}/S_{\rm EE}$ of type IL GRB is much larger that of typical ``long" sGRB, which is actually a more quantitative description of the presence of precursors in type IL GRB.
But $S_{\rm pre}/S_{\rm main}$ (can be derived by $S_{\rm EE}/S_{\rm main} \cdot S_{\rm pre}/S_{\rm EE}$) of type IL GRB and typical ``long" sGRB have the same distribution, suggesting the energy allocation of precursor and main emission of type IL GRB and typical ``long" sGRB may be similar, and brighter main emission correspond to brighter precursor. Based on these two conclusions, we find it it difficult to consistently argue that type IL GRB and typical ``long" sGRB have the same origin. 
First, if we consider that non-detection of precursor in typical ``long" sGRB is due to the low brightness of main emission, it is inconsistent with the presence of a bright precursor with weak main emission of GRB 170228A.  Alternatively, we could consider that the precursor and main emission originate from the same jet observed off-axis, but analysis of GRB 211211A and GRB 230307A suggest that post-merger model are disfavored in type IL.

Hence, all these evidence suggest that type IL GRB and typical ``long" sGRB are distinct groups of type I GRB.
Additionally, these features could provide strong constraints for establishing a unified model to describe the origin of these type IL GRBs, which means that physical models must explain why the energy budget distribution between the these emission episodes of type IL GRBs is so different from typical ``long" sGRBs and type II GRBs with extended emission.


Among all types of GRBs, the long duration of main emission may be explained as the continuous activity of the central engine, such as the free-fall timescale ($t_{\rm ff}$) of the star for the accretion-powered or spindown-powered GRB engine \citep{YJ_GRB211211A_nature}. 
For the former, it is the general scenario of type II GRB, with $t_{\rm ff}\sim\left( \frac{3\pi}{32G\rho} \right)^{1/2}$, where $\rho$ is the mean density of the accreted matter \citep{2018GRBbook}. 
For the latter, the spin-down of the millisecond magnetar may get into play. 
The long duration of main emission can also be explained as the duration of the energy dissipative process, such as the scenario of internal shock happening in a large radius or the scenario of ICMART \citep{2023arXiv231007205Y}. We note that both of them could be in effective.

In any case, the observation do suggest that there seems to be close relation between precursor, main emission and extended emission for type IL GRBs. Moreover, we note that there seems to be a dip structure in the light curve between these two episodes for all three type IL GRBs in this paper.

\subsection{Occurrence Rate}
Over the past decade (from 2014-01-01T00:00:00 to 2024-01-01T00:00:00), \textit{Fermi}/GBM has detected over 2000 GRBs, among of which about 400 have a $T_{90}<2$\,s. These three type IL GRBs are observed during this period, and the occurrence rate can be roughly estimated. 

If the long duration of type IL GRB is generated by the continuous activity of the central engine and assuming they have similar progenitor (such as NS-WD\citep{YJ_GRB211211A_nature}, NS-BH\citep{Meng2024NSBH}, or even NS and massive star binary systems), the proportion of merger systems capable of producing type IL GRB among merger systems capable of producing type I GRBs should be less than about 1$\%$.
Interestingly, we note that this value is roughly compatible to the proportion of magnetars in NSs ($\lesssim1\%$) \citep{Kaspi2017Magnetars}. This, to some extent, supports the plausibility of the SF model in explaining the precursors of GRB 230307A and GRB 211211A.

The three episodes of type IL GRB could originate from different processes, e.g. the precursor could be magnetar flares with possible isotropic radiation, while the main emission and extended emission both originate from relativistic jets but with different opening angles. This could result in the main emission being undetectable in off-axis observations, with only the precursor or both the precursor and extended emission being observable. 
This could result in observational results resembling a typical sGRB or appearing as a normal lGRB with a precursor that has a short waiting time. Although spectral lags can rule out this scenario, our current search does not consider spectral lag.
Also, for some very faint type IL GRBs, only the (peak regions) of the main emission could be detected, resembling a short GRB, while the precursor or extended emission may not be identifiable.

But we need to emphasize that it is required that all three episodes of burst pattern are very prominent in our search. Although this criterion ensures the quality of the sample, it could also omit some samples in which some of the three episodes are not detectable owing to low brightness. Thus this ratio of 1$\%$ is probably underestimated.

\section{Summary}\label{section4}
By utilizing the observations data of the GECAM-B, GECAM-C, and \textit{Fermi}/GBM, a comprehensive analysis is conducted for the first pulse of GRB 230307A (from $T_0$-0.05 s to $T_0$+0.7 s). We found that many characteristics indicate that the first pulse of GRB 230307A has significant differences from the main emission. The spectral results indicate that the first pulse is quasi-thermal, which is consistent with most precursors of merger-origin GRBs. Additionally, we proved that when the brightness of GRB 230307A is reduced, there would be a quiet period between the first pulse and the main emission. These evidences strongly suggest that the first pulse of GRB 230307A is a precursor.

Hence we find that light curves of these two special long GRBs associated with kilonovae, GRB 211211A and GRB 230307A, consist of three episodes: precursor, main emission and extended emission. 
Compared to sGRBEEs and classic long-short GRBs, these two GRBs have significant differences. Specifically, the main emission of sGRBEEs is typically hard-short, but the main emissions of these two events are intrinsic long. 
Therefore, we classified these two events as a subclass of type I GRB, namely type IL, where L means intrinsically long in duration of the main emission.

Moreover, we noticed that the $T_{wt}$ of type IL GRB fall in a different position on the $T_{90}$-$T_{wt}$ distribution compared to type I and type II GRB, which is the most prominent feature to identify this kind of burst pattern. 
Combining the above characteristics, we have summarized the burst pattern of type IL GRB. 
More importantly, with this burst pattern, we have identified a good type IL GRB candidate, GRB 170228A, in the archived data of the \textit{Fermi}/GBM. 

It is surprising that these three GRBs with similar pattern also share the same characteristics, such as spectral lag, MVT, and the position in the Amati relation (being closer to type I GRBs), except for the duration being typical of long GRBs. 
Considering that both GRB 211211A and GRB 230307A were firmly observed to be associated with kilonovae, indicating a special compact star merger origin (such as neutron star and white dwarf mergers), we boldly speculate that GRB 170228A also has a similar merger origin.

Unfortunately, as a seemingly ordinary event of moderate brightness, GRB 170228A did not gain enough attention and there is no publicly available follow-up observation for this GRB. In fact, it is also the same situation for most GRBs. With the limited instruments available, it is difficult to quickly identify whether a GRB is special and worthy of further follow-up observations. Now, the burst pattern of type IL GRB found in this work provides a good reference for rapidly identifying type IL GRB and conducting low-latency follow-up observations, thereby to increase the sample size of this kind of very interesting bursts.

Additionally, this three-episode burst pattern provides important clues on the physical processes of this kind of burst.
The precursor has two prominent features: high luminosity and short $T_{wt}$. 
The high luminosity of the precursor in type IL GRBs requires a high energy budget, which can rule out some models with limited energy resources. And the short $T_{wt}$ also provide some constraints on the radiation radius for the post-merger models. Therefore, considering all features together, we find that pre-merger model, especially the superflare model, is more favored to explain the precursor of type IL GRB.
The $S_{EE}/S_{main}$ is another important diagnostic parameter of type IL GRB, which revealed the difference of energy allocation among these two episodes between type IL GRB and sGRBEE. This implies that type IL GRB may have a special central engine which has an unusual activity history.

We estimated the lower limit of the occurrence rate of type IL GRB to be roughly 1$\%$, which is somewhat consistent with the observed ratio of magnetars among neutron stars. This seems to lend some supports to the hypothesis that type IL GRB may originate from magnetars. However, the origin and underlying physics of Type IL GRB need further studies, especially multi-wavelength and multi-messenger observations.


\acknowledgments
Chen-Wei Wang and Wen-Jun Tan contributed equally to this work.
Chen-Wei Wang thanks the insightful discussion with Xue-Ying Shao on the potential physical implication of the precursor. This work is supported by the National Key R\&D Program of China (2021YFA0718500). 
Shao-Lin Xiong acknowledges the support by the National Natural Science Foundation of China (Grant No. 12273042). 
Shu-Xu Yi acknowledges support from the Chinese Academy of Sciences (grant Nos. E329A3M1 and E3545KU2). 
Rahim Moradi acknowledges support from the Chinese Academy of Sciences (E32984U810).
The GECAM (Huairou-1) mission is supported by the Strategic Priority Research Program on Space Science (Grant No. XDA15360000) of Chinese Academy of Sciences. We appreciate the public data and software of Fermi/GBM.

\appendix

\section{Data Selection} \label{Data Selection}
\subsection{GECAM}
GECAM (Gravitational wave high-energy Electromagnetic Counterpart All-sky Monitor) constellation is composed of four instruments, GECAM-A/B (December 2020)\citep{xiao_GECAMB_time_calibration}, GECAM-C (July 2022)\citep{Zhang2023HEBS} and GECAM-D (March 2024)\citep{Wang2024GTM}, dedicated to monitor all-sky gamma-ray transients.
GECAM-A and GECAM-B feature a dome-shaped array of 25 Gamma-ray detectors (GRD) and 8 Charged particle detectors (CPD) while GECAM-C has 12 GRDs and 2 CPDs. Most of GRDs operate in two readout channels: high gain (HG) and low gain (LG). Thanks to the dedicated design of instrument, neither GECAM-B nor GECAM-C suffered from data saturation during the whole burst of GRB 230307A despite of its extreme brightness, and have high quality data for temporal and spectral analysis. 

For GRB 230307A, the data reduction process is similar with Sun et al. \cite{2023arXiv230705689S} and Yi et al. \cite{2023arXiv231007205Y}.
For temporal analysis, the energy range is 30-300\,keV for GECAM-B HG, 0.3-6\,MeV for GECAM-B LG and 6-30\,keV for GECAM-C HG. No data from GECAM-C LG is used in the temporal analysis. To investigate the data in detail as much as possible, all the data of 25 detectors of GECAM-B and GRD01-06 of GECAM-C are selected for temporal analysis.
For spectral analysis, the energy range is 40-300\,keV for GECAM-B HG, 0.7-6\,MeV for GECAM-B LG and 15-100\,keV (except the 35-40\,keV) for GECAM-C HG. No data from GECAM-C LG is used in the spectral analysis. The GRDs (GRD01, 03-05, 11-13 of GECAM-B and GRD01 of GECAM-C) whose incident angle is smaller than 50 $^\circ$ are selected of spectral analysis. 

\subsection{\textit{Fermi}/GBM}
As one of the two instruments onboard the Fermi Gamma-ray Space Telescope, the Gamma-ray Burst Monitor (\textit{Fermi}/GBM) is composed of 14 detectors with different orientations: 12 Sodium Iodide (NaI) detectors (labelled from n0 to nb) covering the energy range of about 8-1000 keV, and 2 Bismuth Germanate (BGO) detectors (labelled as b0 and b1) covering energies about 0.2-40 MeV \citep{GBM_overview, GBM_calibration}.

Since GRB 230307A is extremely bright, GBM suffers pulse pile-up and data loss during the prompt emission\footnote{ \href{https://fermi.gsfc.nasa.gov/ssc/data/analysis/grb230307a.html}{https://fermi.gsfc.nasa.gov/ssc/data/analysis/grb230307a.html}} unfortunately and the extreme brightness may amplify the inconsistencies between different GBM detectors. But the data is still valid for the first pulse of GRB 230307A with a ordinary brightness of normal GRBs. 
For temporal analysis of precursor in GRB 230307A, the energy range is 6-500\,keV for NaI detectors.
For spectral analysis of precursor in GRB 230307A, the energy range is 8-900\,keV for NaI detectors (except 20-40\,keV) and 0.3-6\,MeV for BGO detector based on the Time-Tagged Events (TTE) data. 

For GRB 211211A, TTE data are used for temporal analysis. We use the GBM n2 and na detectors with energy bands of 10-800 keV for the temporal analysis.

For GRB 170228A, TTE data are used for both temporal and spectral analysis. We use the GBM na detectors with incident angle smaller than 60 $^\circ$, which are the n0, n1, n2, n9 and na, with energy bands of 10-800 keV for the temporal analysis. In order to improve statistics in spectral analysis, we choose 5 NaI detectors with incident angle smaller than 50$^{\circ}$, which are n1, n2, n9 and na, with energy bands of 8-900 keV (except 20-40\,keV), and a BGO detector B1 (0.3-20 MeV).

We use the software $GBMDataTools$ \footnote{ \href{https://fermi.gsfc.nasa.gov/ssc/data/analysis/gbm/}{https://fermi.gsfc.nasa.gov/ssc/data/analysis/gbm/}} to extract the net light curves, full spectral, background spectral and response file for three GRBs.

\section{Data Reduction} \label{Data Reduction}
\subsection{Temporal analysis}
Considering signal of the first pulse of GRB 230307A decreases rapidly with energy increased, the GECAM data are divided into seven energy bands (1st to 7th channels), namely 6-30 keV, where the data are from GECAM-C; 30-70 keV, 70-100 keV, 100-150 keV, 150-200 keV, 200-500 keV, 500-6000 keV, where the data are from GECAM-B. The GBM data are divided into channel from 1st to 6th. The optical path difference between GECAM-B to GECAM-C and GCEAM-B to GBM has been corrected. The light curves of GECAM are binned into two different time binsize: 2 ms and 5 ms, including raw light curves, background light curves and net light curves, while the light curves of GBM are binned into 5 ms. 

A raw light curve without background subtraction of the precursor candidate in the full energy band of GECAM (6-6000 keV) is shown in Fig.~\ref{fig:precursor_lc} (blue line). The profile of the precursor candidate can be described as a pulse with two distinguishable FRED-like spikes and a weak emission superimposed on the tail of the pulse.

The 2 ms binned light curves are used in calculating the spectral lag of GECAM data and fitting of the precursor of GRB 230307A. The 5ms binned light curves are used in calculating the hardness ratio and the spectral lag comparison between GECAM and GBM for GRB 230307A.

Two FRED shape individual spikes can be distinguished in the precursor candidate. Follow the analysis as Yi et al. \cite{2023arXiv231007205Y}, the two FRED are fitted by Eq.~\ref{equ:Norris05} as shown in Fig.~\ref{fig:precursor_lc}. 
\begin{equation}
L(t) = \frac{A_1}{\exp(\frac{\tau_{\rm{r1}}}{t-t_{s1}}+\frac{t-t_{s1}}{\tau_{\rm{d1}}})} + \frac{A_2}{\exp(\frac{\tau_{\rm{r2}}}{t-t_{s2}}+\frac{t-t_{s2}}{\tau_{\rm{d2}}})},
\label{equ:Norris05}
\end{equation}
where $t_{\rm{s1}}$, $t_{\rm{s2}}$ are the starting instance of the two spikes, $\tau_{\rm{r1}}$, $\tau_{\rm{r2}}$, $\tau_{\rm{d1}}$ and $\tau_{\rm{d2}}$ are the rising and decaying time scales for the two pulses, respectively. The fitting results are listed in Table~\ref{tab:lc_fit}. The peak time $t_{\rm{p}}$ and the width $w$ of the pulse are therefore defined as: 
$t_{\rm{p}}=t_{\rm{s}}+\sqrt{\tau_{\rm{r}}\tau_{\rm{d}}}$ and $w=\tau_{\rm{r}}+\tau_{\rm{d}}$. However there is no clear energy dependence in both $w$ and $\tilde{t}_{\rm{p}}\equiv t_{\rm{p}}-t_{\rm{s}}$, as observed in prompt emission, as depicted in Fig.~\ref{fig:precursor_lag}b.

The minimum variability timescale (MVT) is defined as the shortest timescale with significant variation that exceeds statistical noise in the GRB light curve. It can be identified as the rise time of the shortest pulse in the whole GRB profile. Generally speaking, type II GRBs have longer MVT than type II GRBs do. To determine the MVT, we use the Bayesian block algorithm \citep{2013ApJ...764..167S} on the entire light curve within the 10--1000 keV energy range to identify the shortest block which represent one single pulse, and take the rise time of this pulse as the MVT. We find that the MVT of GRB 170228A is about 40 ms, which mostly lie in the type I GRBs region (Fig.~\ref{fig:classification}b) rather than type II GRBs region in the distribution of the MVTs.

Spectral lag refers to the time delay between the soft-band and hard-band light curves. Spectral lags in type II GRBs are usually more evidently than type I GRBs, which may indicate the difference in their spectral evolution. We use the cross-correlation function \citep{1997ApJ...486..928B} to determine the observed lag between light curves with energy bands 100-150 and 200-250 keV, and the lag of GRB 170228A is $30_{-90}^{+100}$ ms. Note that the spectral lag of GRB 170228A is relatively small (Fig.~\ref{fig:classification}c), which indicate a probable type I GRB regime.

The study of power spectral density (PSD) continuum properties can help explore the dynamic property of the GRB jet. Here we use a power law plus a constant (PLC) to determine the slope of PSD of these three GRBs in energy bands 50-150 keV \citep{GuidorziPSD}. All PSDs of three GRBs are calculated within energy band 50-150 keV. We note that the slope of PSD of these three GRBS are quite similar (Fig.~\ref{fig:classification}d), indicating there are similar properties in three GRBs.
\begin{equation}
S_{plc}({\upsilon}) = A{\upsilon}^{-\alpha} + b
\end{equation}

Where $\alpha$ is the power-law index (i.e. slope).

Finally, to determine the time interval of each episode of these three GRBs, we use the Bayesian block to divide the whole profile of the light curve in energy bands 10-800 keV, and obtain the required time interval, including the duration of precursor (T$_{pre}$) and waiting time ($T_{wt}$). Each timescale of these three GRBs are listed in Table~\ref{tab:timescale_of_GRBs}.

\subsection{Spectrum analysis}
For each of above time slices, the spectral fitting is conducted using the Pyxspec software \citep{pyxspec} with PGSTAT.
Five types of photon spectrum model are adopted in the fitting. 

\begin{figure*}
\centering
\begin{tabular}{cc}
\begin{overpic}[width=0.45\textwidth]{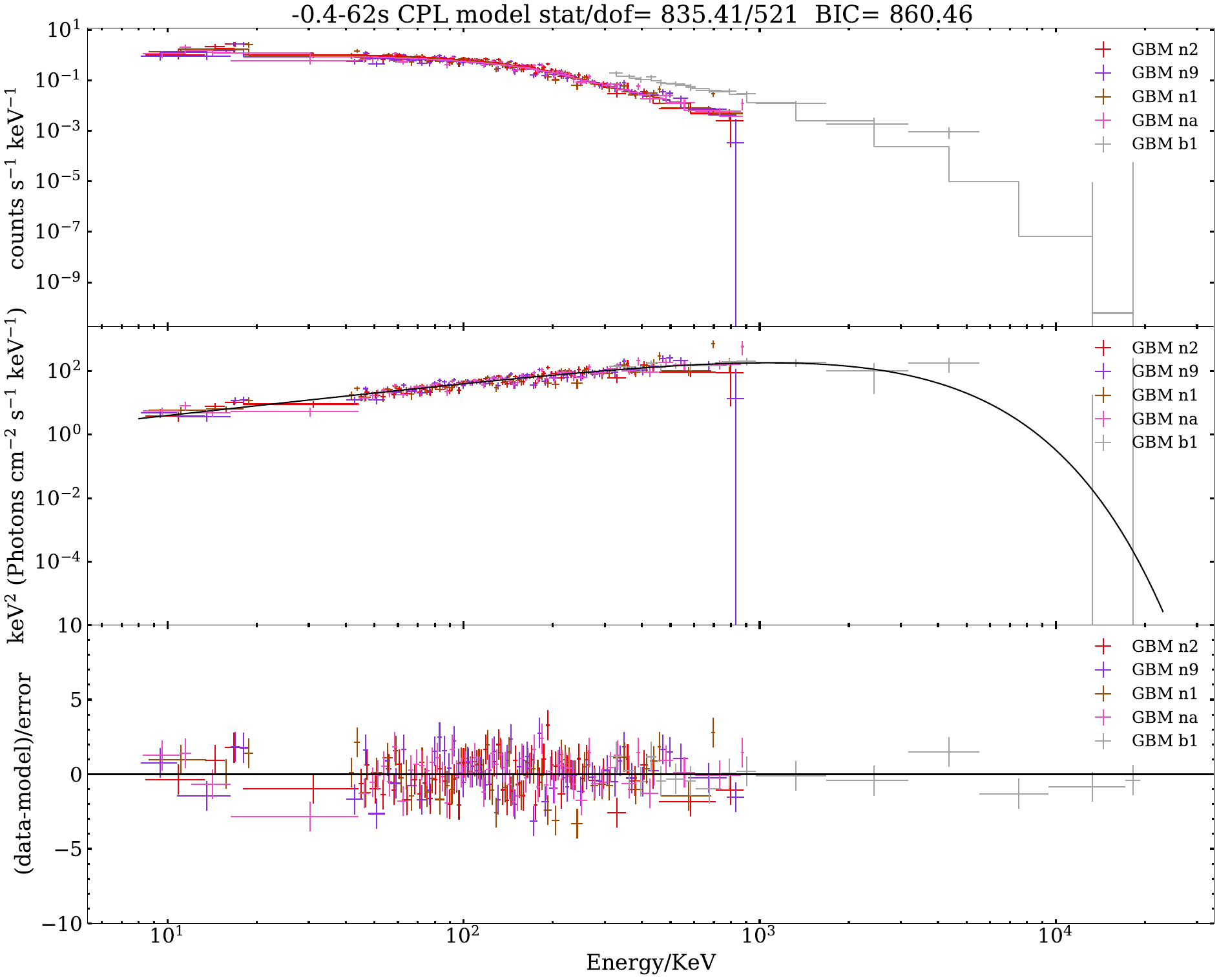}\put(0, 80){\bf a}\end{overpic} &
        \begin{overpic}[width=0.45\textwidth]{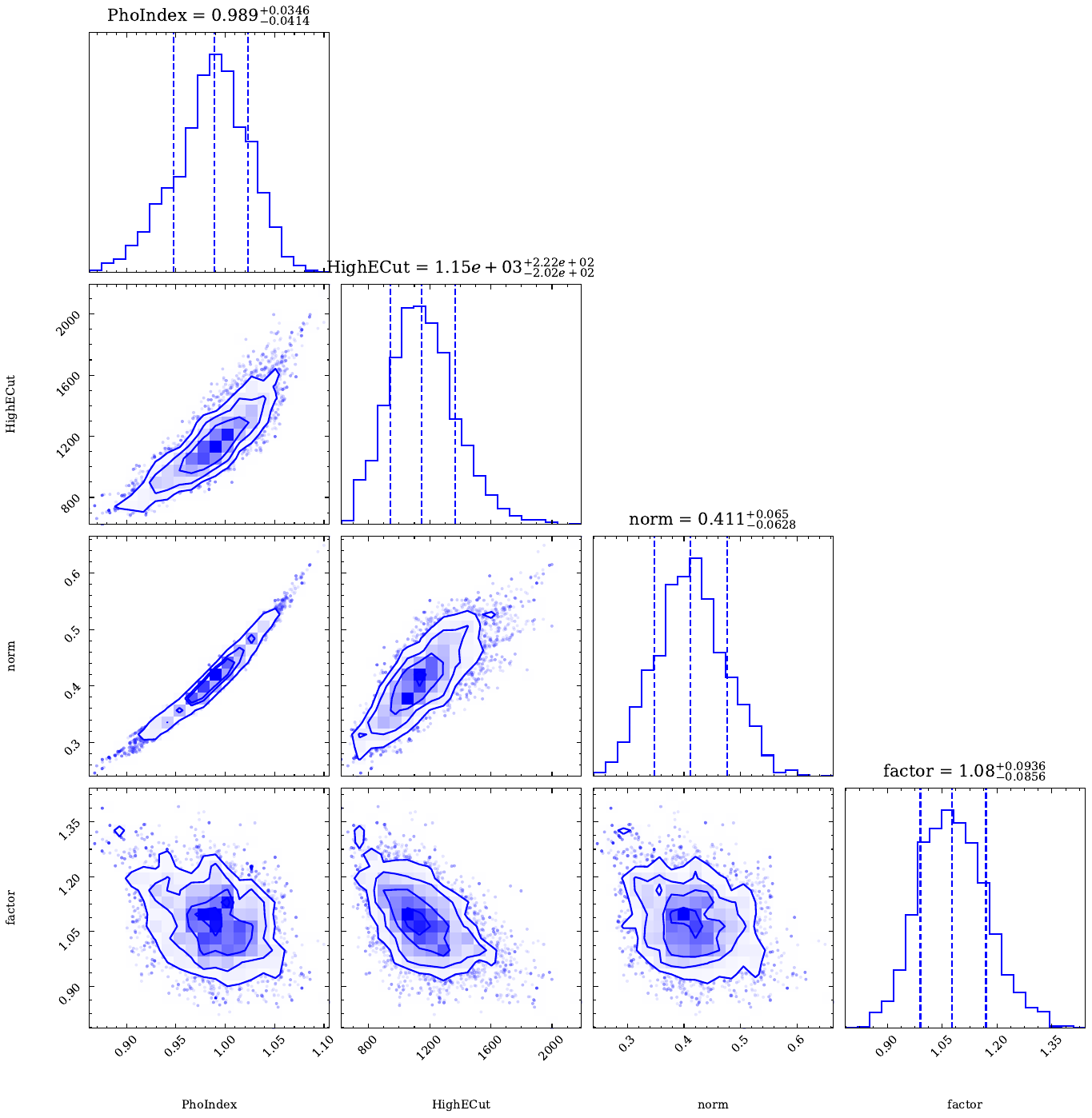}\put(0, 80){\bf b}\end{overpic} \\
\end{tabular}
\caption{\textbf{The spectral fitting result with CPL model(left panel) and corner plot of the posterior probability distributions of the parameters(right panel) of GRB 170228A.}}
\label{fig:28A_spec}
\end{figure*}

As for GRB 170228A, we conduct time-integrated spectral fitting using the Pyxspec software with Band model and cut-off power-law (CPL) to determine the best model of GRB 170228A (Fig.~\ref{fig:28A_spec}).

The spectral model used in our analysis including power-law (Eq.~\ref{equ:pl_Model}), cut-off power-law (CPL) (Eq.~\ref{equ:CPL_Model}), Band (Eq.~\ref{equ:band_Model}),  area normalized blackbody (bbodyrad) (Eq.~\ref{equ:bb_Model}), Band function with high-energy exponential cutoff (bandcut) (Eq.~\ref{equ:bandcut_Model}) and multicolor blackbody (mBB) (Eq.~\ref{equ:mbb_Model}), which are listed  as follows:
The PL model is expressed as
\begin{equation}   
N(E) = A \left(\frac{E}{E_0}\right)^{\alpha},
\label{equ:pl_Model}
\end{equation}
where $A$ is the normalization constant ($\rm photons \cdot cm^{-2} \cdot s^{-1} \cdot keV^{-1}$), $\alpha$ is the power law photon index, and $E_0$ is the pivot energy fixed at 1 keV.

The CPL model is expressed as
\begin{equation}   
N(E)=A\left(\frac{E}{E_0}\right)^{\alpha} {\rm exp}(-\frac{E}{E_{\rm c}}),
\label{equ:CPL_Model}
\end{equation}
where $A$ is the normalization constant ($\rm photons \cdot cm^{-2} \cdot s^{-1} \cdot keV^{-1}$), $\alpha$ is the power law photon index, $E_0$ is the pivot energy fixed at 1 keV, and $E_{\rm c}$ is the characteristic cutoff energy in keV. The peak energy $E_{\rm p}$ is related to the $E_{\rm c}$ through $E_{\rm p}$=$(2 + \alpha)E_{\rm c}$.

The Band model is expressed as
\begin{equation}
N(E)=\left\{
\begin{array}{l}
A(\frac{E}{100\,{\rm keV}})^{\alpha}{\rm exp}(-\frac{E}{E_{\rm c}}),\,E<(\alpha-\beta)E_{\rm c}, \\
A\big[\frac{(\alpha-\beta)E_{\rm c}}{100\,{\rm keV}}\big]^{\alpha-\beta}{\rm exp}(\beta-\alpha)(\frac{E}{100\,{\rm keV}})^{\beta}, E\geq(\alpha-\beta)E_{\rm c}, 
\end{array}\right.
\label{equ:band_Model}
\end{equation}
where $A$ is the normalization constant ($\rm photons \cdot cm^{-2} \cdot s^{-1} \cdot keV^{-1}$), $\alpha$ and $\beta$ are the low-energy and high-energy power law spectral indices, $E_{\rm c}$ is the characteristic energy in keV and the peak energy $E_{\rm p}$ is related to the $E_{\rm c}$ through $E_{\rm p}$=$(2 + \alpha)E_{\rm c}$.

The bbodyrad model is expressed as
\begin{equation}   
N(E)=\frac{1.0344 \times 10^{-3}\times AE^2}{{\rm exp}(\frac{E}{kT})-1}
\label{equ:bb_Model}
\end{equation}
where $A=R^2_{km}/D^2_{10}$ is the normalization constant, $R_{km}$ is the source radius in km and $D_{10}$ is the distance to the source in units of 10 kpc, and $kT$ is the temperature in keV.

A low energy break is observed in the prompt of some GRBs including the prompt emission of GRB 230307A \citep{Oganesyan2018bandcut,2023arXiv230705689S}, which can be described by bandcut model. The bandcut model is expressed as
\begin{equation}
N(E)=\left\{
\begin{array}{l}
A(\frac{E}{E_0})^{\alpha_1}{\rm exp}(-\frac{E}{E_1}), E\leq E_b\\
AE_b^{\alpha_1-\alpha_2}{\rm exp}(\alpha_2-\alpha_1)(\frac{E}{E_0})^{\alpha_2}{\rm exp}(-\frac{E}{E_2}), E>E_b, 
\end{array}\right.
\label{equ:bandcut_Model}
\end{equation}
where the break energy is expressed as $E_b$=$\frac{E_1 E_2}{E_2-E_1}$($\alpha_1$-$\alpha_2$), and $E_0$ is the pivot energy fixed at 1 keV.

The mbb model described a kind of quasi-thermal spectra \citep{Hou2018mbb}. And the model is expressed as
\begin{equation}
    N(E)=\frac{8.0525(m+1)K}{[(\frac{T_{\rm max}}{T_{\rm min}})^{m+1}-1]}(\frac{kT_{\rm min}}{\rm keV})^{-2}I(E)
\label{equ:mbb_Model}
\end{equation}
where
\begin{equation}
    I(E)=(\frac{E}{kT_{\rm min}})^{m-1}\int_\frac{E}{kT_{\rm max}}^\frac{E}{kT_{\rm min}}
\end{equation}
where $x=E/kT$, $m$ is the power-law index of the dependence of the luminosity distribution on the temperature, and the temperature ranges from minimum $T_{\rm min}$ to the maximum $T_{\rm max}$.

The model comparison is based on the Bayesian Information Criterion (BIC) which is defined as BIC = -2lnL + klnN, where L represents the maximum likelihood value, k denotes the number of free parameters in the model, and N signifies the number of data points. A model with a lower BIC value is preferred, particularly when the difference in BIC ($\Delta$BIC) exceeds 10. An additional condition is required for the best model that all parameters should be well–constrained, and the multiplicative factors should be approximately equal to 1.

\let\cleardoublepage\clearpage

\begin{sidewaystable}[htbp]
\tiny
\setlength{\tabcolsep}{1mm}
\begin{center}
\caption{The spectral analysis result of the precursor in GRB 230307A with different time slice}\label{tab:spec_fit}%
\begin{tabular}{cccccccccccccccccccc}
\toprule
 & & \multicolumn{5}{c}{Band model} & \multicolumn{4}{c}{CPL model} & \multicolumn{3}{c}{powerlaw model} & \multicolumn{5}{c}{mbb model} & \\
 \cline{3-19}
start time (s) & end time (s) & $\alpha$ & $\beta$ & $E_{p}$ & $log_{10}A$ & BIC & $\alpha$ & $E_{p}$ & $log_{10}A$ & BIC & $\alpha$ & $log_{10}A$ & BIC & m & kT$_{min}$ & kT$_{max}$ & $log_{10}A$ & BIC & best model \\
\hline 
\textbf{S-I} \\
-0.05 & 0.7 & -0.89$^{+0.03}_{-0.03}$ & -2.73$^{+0.07}_{-0.09}$ & 148.02$^{+4.68}_{-4.32}$ & -0.07$^{+0.02}_{-0.02}$ & 1405 & -1.05$^{+0.02}_{-0.02}$ & 177.08$^{+3.95}_{-3.87}$ & 1.93$^{+0.04}_{-0.04}$ & 1440 & -1.76$^{+0.01}_{-0.01}$ & 3.04$^{+0.01}_{-0.01}$ & 3276 & -0.66$^{+0.03}_{-0.03}$ & 5.90$^{+0.21}_{-0.19}$ & 137.58$^{+5.60}_{-5.29}$ & 3.27$^{+0.01}_{-0.01}$ & 1340 & mbb\\ 
\hline 
\textbf{S-II} \\
-0.05 & 0.09 & -0.49$^{+0.06}_{-0.07}$ & -2.60$^{+0.08}_{-0.07}$ & 155.11$^{+8.97}_{-6.39}$ & 0.33$^{+0.04}_{-0.05}$ & 1270 & -0.80$^{+0.04}_{-0.04}$ & 205.95$^{+7.91}_{-5.80}$ & 1.73$^{+0.06}_{-0.07}$ & 1322 & -1.68$^{+0.01}_{-0.01}$ & 3.14$^{+0.02}_{-0.02}$ & 2276 & -0.62$^{+0.07}_{-0.06}$ & 8.43$^{+0.61}_{-0.45}$ & 158.49$^{+13.62}_{-13.58}$ & 3.57$^{+0.01}_{-0.01}$ & 1267 & mbb\\ 
0.09 & 0.4 & -0.74$^{+0.05}_{-0.04}$ & -2.66$^{+0.06}_{-0.06}$ & 144.59$^{+3.28}_{-5.06}$ & 0.14$^{+0.03}_{-0.02}$ & 1336 & -0.98$^{+0.02}_{-0.02}$ & 182.98$^{+4.07}_{-4.10}$ & 1.96$^{+0.04}_{-0.03}$ & 1364 & -1.76$^{+0.01}_{-0.01}$ & 3.19$^{+0.01}_{-0.02}$ & 2538 & -0.74$^{+0.04}_{-0.04}$ & 6.40$^{+0.31}_{-0.28}$ & 138.68$^{+8.94}_{-7.18}$ & 3.40$^{+0.01}_{-0.01}$ & 1270 & mbb\\ 
0.4 & 0.7 & -0.82$^{+0.35}_{-0.41}$ & -2.00$^{+0.00}_{-0.00}$ & 32.36$^{+8.32}_{-5.42}$ & -0.06$^{+0.50}_{-0.40}$ & 1228 & -1.72$^{+0.06}_{-0.05}$ & 100.41$^{+13.29}_{-12.07}$ & 2.46$^{+0.08}_{-0.09}$ & 1194 & -2.00$^{+0.03}_{-0.03}$ & 2.88$^{+0.05}_{-0.05}$ & 1214 & -1.02$^{+0.10}_{-0.07}$ & 3.24$^{+0.50}_{-0.60}$ & 183.16$^{+66.47}_{-44.62}$ & 2.68$^{+0.03}_{-0.03}$ & 1195 & mbb\\ 
\hline 
\textbf{S-III} \\
-0.05 & 0.02 & -0.14$^{+0.17}_{-0.23}$ & -1.89$^{+0.06}_{-0.07}$ & 154.93$^{+21.09}_{-17.77}$ & -0.00$^{+0.11}_{-0.11}$ & 322 & -0.97$^{+0.05}_{-0.04}$ & 473.38$^{+23.54}_{-31.20}$ & 1.46$^{+0.08}_{-0.10}$ & 341 & -1.52$^{+0.02}_{-0.02}$ & 2.40$^{+0.05}_{-0.05}$ & 432 & -0.61$^{+0.07}_{-0.09}$ & 10.72$^{+1.25}_{-1.33}$ & 641.70$^{+158.92}_{-132.83}$ & 3.44$^{+0.04}_{-0.04}$ & 305 & mbb\\ 
0.02 & 0.05 & -0.47$^{+0.12}_{-0.11}$ & -3.15$^{+0.25}_{-0.29}$ & 173.18$^{+11.66}_{-11.15}$ & 0.57$^{+0.06}_{-0.05}$ & 358 & -0.64$^{+0.06}_{-0.06}$ & 199.75$^{+8.28}_{-7.12}$ & 1.75$^{+0.09}_{-0.09}$ & 362 & -1.67$^{+0.02}_{-0.02}$ & 3.40$^{+0.03}_{-0.03}$ & 875 & -0.40$^{+0.10}_{-0.10}$ & 8.49$^{+0.80}_{-0.80}$ & 124.22$^{+13.68}_{-11.42}$ & 3.83$^{+0.02}_{-0.01}$ & 342 & mbb\\ 
0.05 & 0.09 & -0.70$^{+0.04}_{-0.03}$ & -9.01$^{+0.54}_{-0.54}$ & 146.70$^{+9.05}_{-8.50}$ & 0.45$^{+0.03}_{-0.03}$ & 301 & -0.67$^{+0.05}_{-0.06}$ & 143.78$^{+5.45}_{-5.15}$ & 1.81$^{+0.10}_{-0.08}$ & 295 & -1.80$^{+0.02}_{-0.02}$ & 3.54$^{+0.04}_{-0.03}$ & 807 & -0.34$^{+0.12}_{-0.12}$ & 6.00$^{+0.84}_{-0.76}$ & 82.74$^{+7.00}_{-6.99}$ & 3.64$^{+0.01}_{-0.01}$ & 301 & mbb\\ 
0.09 & 0.12 & -0.50$^{+0.07}_{-0.10}$ & -2.78$^{+0.12}_{-0.14}$ & 154.56$^{+8.08}_{-7.55}$ & 0.49$^{+0.04}_{-0.05}$ & 337 & -0.83$^{+0.05}_{-0.06}$ & 204.55$^{+10.96}_{-10.95}$ & 1.96$^{+0.10}_{-0.09}$ & 343 & -1.70$^{+0.02}_{-0.02}$ & 3.35$^{+0.04}_{-0.04}$ & 658 & -0.62$^{+0.12}_{-0.11}$ & 8.43$^{+0.92}_{-0.96}$ & 149.14$^{+16.45}_{-17.33}$ & 3.72$^{+0.02}_{-0.02}$ & 329 & mbb\\ 
0.12 & 0.15 & -0.52$^{+0.19}_{-0.12}$ & -3.54$^{+0.57}_{-3.21}$ & 217.82$^{+20.95}_{-28.15}$ & 0.47$^{+0.10}_{-0.06}$ & 370 & -0.65$^{+0.07}_{-0.08}$ & 240.50$^{+11.78}_{-10.37}$ & 1.69$^{+0.13}_{-0.12}$ & 366 & -1.63$^{+0.01}_{-0.02}$ & 3.31$^{+0.04}_{-0.03}$ & 934 & -0.28$^{+0.08}_{-0.11}$ & 8.60$^{+1.00}_{-1.18}$ & 137.41$^{+10.29}_{-9.15}$ & 3.87$^{+0.02}_{-0.02}$ & 360 & mbb\\ 
0.15 & 0.18 & -0.06$^{+0.16}_{-0.18}$ & -2.52$^{+0.08}_{-0.14}$ & 126.66$^{+12.15}_{-8.25}$ & 0.82$^{+0.10}_{-0.11}$ & 338 & -0.72$^{+0.05}_{-0.06}$ & 194.52$^{+11.24}_{-8.83}$ & 1.84$^{+0.09}_{-0.08}$ & 358 & -1.69$^{+0.02}_{-0.02}$ & 3.40$^{+0.04}_{-0.04}$ & 777 & -0.80$^{+0.10}_{-0.12}$ & 10.44$^{+0.84}_{-0.86}$ & 163.00$^{+26.76}_{-18.14}$ & 3.78$^{+0.02}_{-0.02}$ & 317 & mbb\\ 
0.18 & 0.4 & -1.86$^{+0.04}_{-0.04}$ & -8.85$^{+0.82}_{-0.60}$ & 58.70$^{+14.29}_{-15.64}$ & -0.57$^{+0.00}_{-0.01}$ & 539 & -1.07$^{+0.04}_{-0.06}$ & 73.48$^{+2.27}_{-2.22}$ & 2.09$^{+0.08}_{-0.06}$ & 333 & -2.06$^{+0.02}_{-0.02}$ & 3.43$^{+0.03}_{-0.03}$ & 675 & -1.16$^{+0.09}_{-0.11}$ & 5.57$^{+0.34}_{-0.28}$ & 69.31$^{+8.83}_{-6.78}$ & 3.01$^{+0.01}_{-0.01}$ & 312 & mbb\\ 
\botrule
\end{tabular}
\end{center}
\end{sidewaystable}

\clearpage

\clearpage

\bibliography{apssamp}

\begin{thebibliography}{78}%
\makeatletter
\providecommand \@ifxundefined [1]{%
 \@ifx{#1\undefined}
}%
\providecommand \@ifnum [1]{%
 \ifnum #1\expandafter \@firstoftwo
 \else \expandafter \@secondoftwo
 \fi
}%
\providecommand \@ifx [1]{%
 \ifx #1\expandafter \@firstoftwo
 \else \expandafter \@secondoftwo
 \fi
}%
\providecommand \natexlab [1]{#1}%
\providecommand \enquote  [1]{``#1''}%
\providecommand \bibnamefont  [1]{#1}%
\providecommand \bibfnamefont [1]{#1}%
\providecommand \citenamefont [1]{#1}%
\providecommand \href@noop [0]{\@secondoftwo}%
\providecommand \href [0]{\begingroup \@sanitize@url \@href}%
\providecommand \@href[1]{\@@startlink{#1}\@@href}%
\providecommand \@@href[1]{\endgroup#1\@@endlink}%
\providecommand \@sanitize@url [0]{\catcode `\\12\catcode `\$12\catcode `\&12\catcode `\#12\catcode `\^12\catcode `\_12\catcode `\%12\relax}%
\providecommand \@@startlink[1]{}%
\providecommand \@@endlink[0]{}%
\providecommand \url  [0]{\begingroup\@sanitize@url \@url }%
\providecommand \@url [1]{\endgroup\@href {#1}{\urlprefix }}%
\providecommand \urlprefix  [0]{URL }%
\providecommand \Eprint [0]{\href }%
\providecommand \doibase [0]{https://doi.org/}%
\providecommand \selectlanguage [0]{\@gobble}%
\providecommand \bibinfo  [0]{\@secondoftwo}%
\providecommand \bibfield  [0]{\@secondoftwo}%
\providecommand \translation [1]{[#1]}%
\providecommand \BibitemOpen [0]{}%
\providecommand \bibitemStop [0]{}%
\providecommand \bibitemNoStop [0]{.\EOS\space}%
\providecommand \EOS [0]{\spacefactor3000\relax}%
\providecommand \BibitemShut  [1]{\csname bibitem#1\endcsname}%
\let\auto@bib@innerbib\@empty
\bibitem [{\citenamefont {{Kouveliotou}}\ \emph {et~al.}(1993)\citenamefont {{Kouveliotou}}, \citenamefont {{Meegan}}, \citenamefont {{Fishman}}, \citenamefont {{Bhat}}, \citenamefont {{Briggs}}, \citenamefont {{Koshut}}, \citenamefont {{Paciesas}},\ and\ \citenamefont {{Pendleton}}}]{1993Kouveliotou}%
  \BibitemOpen
  \bibfield  {author} {\bibinfo {author} {\bibfnamefont {C.}~\bibnamefont {{Kouveliotou}}}, \bibinfo {author} {\bibfnamefont {C.~A.}\ \bibnamefont {{Meegan}}}, \bibinfo {author} {\bibfnamefont {G.~J.}\ \bibnamefont {{Fishman}}}, \bibinfo {author} {\bibfnamefont {N.~P.}\ \bibnamefont {{Bhat}}}, \bibinfo {author} {\bibfnamefont {M.~S.}\ \bibnamefont {{Briggs}}}, \bibinfo {author} {\bibfnamefont {T.~M.}\ \bibnamefont {{Koshut}}}, \bibinfo {author} {\bibfnamefont {W.~S.}\ \bibnamefont {{Paciesas}}},\ and\ \bibinfo {author} {\bibfnamefont {G.~N.}\ \bibnamefont {{Pendleton}}},\ }\bibfield  {title} {\bibinfo {title} {{Identification of Two Classes of Gamma-Ray Bursts}},\ }\href {https://doi.org/10.1086/186969} {\bibfield  {journal} {\bibinfo  {journal} {\apjl}\ }\textbf {\bibinfo {volume} {413}},\ \bibinfo {pages} {L101} (\bibinfo {year} {1993})}\BibitemShut {NoStop}%
\bibitem [{\citenamefont {{Amati}}\ \emph {et~al.}(2002)\citenamefont {{Amati}}, \citenamefont {{Frontera}}, \citenamefont {{Tavani}}, \citenamefont {{in't Zand}}, \citenamefont {{Antonelli}}, \citenamefont {{Costa}}, \citenamefont {{Feroci}}, \citenamefont {{Guidorzi}}, \citenamefont {{Heise}}, \citenamefont {{Masetti}}, \citenamefont {{Montanari}}, \citenamefont {{Nicastro}}, \citenamefont {{Palazzi}}, \citenamefont {{Pian}}, \citenamefont {{Piro}},\ and\ \citenamefont {{Soffitta}}}]{2002A&A...390...81A}%
  \BibitemOpen
  \bibfield  {author} {\bibinfo {author} {\bibfnamefont {L.}~\bibnamefont {{Amati}}}, \bibinfo {author} {\bibfnamefont {F.}~\bibnamefont {{Frontera}}}, \bibinfo {author} {\bibfnamefont {M.}~\bibnamefont {{Tavani}}}, \bibinfo {author} {\bibfnamefont {J.~J.~M.}\ \bibnamefont {{in't Zand}}}, \bibinfo {author} {\bibfnamefont {A.}~\bibnamefont {{Antonelli}}}, \bibinfo {author} {\bibfnamefont {E.}~\bibnamefont {{Costa}}}, \bibinfo {author} {\bibfnamefont {M.}~\bibnamefont {{Feroci}}}, \bibinfo {author} {\bibfnamefont {C.}~\bibnamefont {{Guidorzi}}}, \bibinfo {author} {\bibfnamefont {J.}~\bibnamefont {{Heise}}}, \bibinfo {author} {\bibfnamefont {N.}~\bibnamefont {{Masetti}}}, \bibinfo {author} {\bibfnamefont {E.}~\bibnamefont {{Montanari}}}, \bibinfo {author} {\bibfnamefont {L.}~\bibnamefont {{Nicastro}}}, \bibinfo {author} {\bibfnamefont {E.}~\bibnamefont {{Palazzi}}}, \bibinfo {author} {\bibfnamefont {E.}~\bibnamefont {{Pian}}}, \bibinfo {author} {\bibfnamefont {L.}~\bibnamefont {{Piro}}},\ and\ \bibinfo {author}
  {\bibfnamefont {P.}~\bibnamefont {{Soffitta}}},\ }\bibfield  {title} {\bibinfo {title} {{Intrinsic spectra and energetics of BeppoSAX Gamma-Ray Bursts with known redshifts}},\ }\href {https://doi.org/10.1051/0004-6361:20020722} {\bibfield  {journal} {\bibinfo  {journal} {\aap}\ }\textbf {\bibinfo {volume} {390}},\ \bibinfo {pages} {81} (\bibinfo {year} {2002})},\ \Eprint {https://arxiv.org/abs/astro-ph/0205230} {arXiv:astro-ph/0205230 [astro-ph]} \BibitemShut {NoStop}%
\bibitem [{\citenamefont {{Yonetoku}}\ \emph {et~al.}(2004)\citenamefont {{Yonetoku}}, \citenamefont {{Murakami}}, \citenamefont {{Nakamura}}, \citenamefont {{Yamazaki}}, \citenamefont {{Inoue}},\ and\ \citenamefont {{Ioka}}}]{2004ApJ...609..935Y}%
  \BibitemOpen
  \bibfield  {author} {\bibinfo {author} {\bibfnamefont {D.}~\bibnamefont {{Yonetoku}}}, \bibinfo {author} {\bibfnamefont {T.}~\bibnamefont {{Murakami}}}, \bibinfo {author} {\bibfnamefont {T.}~\bibnamefont {{Nakamura}}}, \bibinfo {author} {\bibfnamefont {R.}~\bibnamefont {{Yamazaki}}}, \bibinfo {author} {\bibfnamefont {A.~K.}\ \bibnamefont {{Inoue}}},\ and\ \bibinfo {author} {\bibfnamefont {K.}~\bibnamefont {{Ioka}}},\ }\bibfield  {title} {\bibinfo {title} {{Gamma-Ray Burst Formation Rate Inferred from the Spectral Peak Energy-Peak Luminosity Relation}},\ }\href {https://doi.org/10.1086/421285} {\bibfield  {journal} {\bibinfo  {journal} {\apj}\ }\textbf {\bibinfo {volume} {609}},\ \bibinfo {pages} {935} (\bibinfo {year} {2004})},\ \Eprint {https://arxiv.org/abs/astro-ph/0309217} {arXiv:astro-ph/0309217 [astro-ph]} \BibitemShut {NoStop}%
\bibitem [{\citenamefont {{Golkhou}}\ \emph {et~al.}(2015)\citenamefont {{Golkhou}}, \citenamefont {{Butler}},\ and\ \citenamefont {{Littlejohns}}}]{2015ApJ...811...93G}%
  \BibitemOpen
  \bibfield  {author} {\bibinfo {author} {\bibfnamefont {V.~Z.}\ \bibnamefont {{Golkhou}}}, \bibinfo {author} {\bibfnamefont {N.~R.}\ \bibnamefont {{Butler}}},\ and\ \bibinfo {author} {\bibfnamefont {O.~M.}\ \bibnamefont {{Littlejohns}}},\ }\bibfield  {title} {\bibinfo {title} {{The Energy Dependence of GRB Minimum Variability Timescales}},\ }\href {https://doi.org/10.1088/0004-637X/811/2/93} {\bibfield  {journal} {\bibinfo  {journal} {\apj}\ }\textbf {\bibinfo {volume} {811}},\ \bibinfo {eid} {93} (\bibinfo {year} {2015})},\ \Eprint {https://arxiv.org/abs/1501.05948} {arXiv:1501.05948 [astro-ph.HE]} \BibitemShut {NoStop}%
\bibitem [{\citenamefont {{Bernardini}}\ \emph {et~al.}(2015)\citenamefont {{Bernardini}}, \citenamefont {{Ghirlanda}}, \citenamefont {{Campana}}, \citenamefont {{Covino}}, \citenamefont {{Salvaterra}}, \citenamefont {{Atteia}}, \citenamefont {{Burlon}}, \citenamefont {{Calderone}}, \citenamefont {{D'Avanzo}}, \citenamefont {{D'Elia}}, \citenamefont {{Ghisellini}}, \citenamefont {{Heussaff}}, \citenamefont {{Lazzati}}, \citenamefont {{Melandri}}, \citenamefont {{Nava}}, \citenamefont {{Vergani}},\ and\ \citenamefont {{Tagliaferri}}}]{2015MNRAS.446.1129B}%
  \BibitemOpen
  \bibfield  {author} {\bibinfo {author} {\bibfnamefont {M.~G.}\ \bibnamefont {{Bernardini}}}, \bibinfo {author} {\bibfnamefont {G.}~\bibnamefont {{Ghirlanda}}}, \bibinfo {author} {\bibfnamefont {S.}~\bibnamefont {{Campana}}}, \bibinfo {author} {\bibfnamefont {S.}~\bibnamefont {{Covino}}}, \bibinfo {author} {\bibfnamefont {R.}~\bibnamefont {{Salvaterra}}}, \bibinfo {author} {\bibfnamefont {J.~L.}\ \bibnamefont {{Atteia}}}, \bibinfo {author} {\bibfnamefont {D.}~\bibnamefont {{Burlon}}}, \bibinfo {author} {\bibfnamefont {G.}~\bibnamefont {{Calderone}}}, \bibinfo {author} {\bibfnamefont {P.}~\bibnamefont {{D'Avanzo}}}, \bibinfo {author} {\bibfnamefont {V.}~\bibnamefont {{D'Elia}}}, \bibinfo {author} {\bibfnamefont {G.}~\bibnamefont {{Ghisellini}}}, \bibinfo {author} {\bibfnamefont {V.}~\bibnamefont {{Heussaff}}}, \bibinfo {author} {\bibfnamefont {D.}~\bibnamefont {{Lazzati}}}, \bibinfo {author} {\bibfnamefont {A.}~\bibnamefont {{Melandri}}}, \bibinfo {author} {\bibfnamefont {L.}~\bibnamefont {{Nava}}}, \bibinfo
  {author} {\bibfnamefont {S.~D.}\ \bibnamefont {{Vergani}}},\ and\ \bibinfo {author} {\bibfnamefont {G.}~\bibnamefont {{Tagliaferri}}},\ }\bibfield  {title} {\bibinfo {title} {{Comparing the spectral lag of short and long gamma-ray bursts and its relation with the luminosity}},\ }\href {https://doi.org/10.1093/mnras/stu2153} {\bibfield  {journal} {\bibinfo  {journal} {\mnras}\ }\textbf {\bibinfo {volume} {446}},\ \bibinfo {pages} {1129} (\bibinfo {year} {2015})},\ \Eprint {https://arxiv.org/abs/1410.5216} {arXiv:1410.5216 [astro-ph.HE]} \BibitemShut {NoStop}%
\bibitem [{\citenamefont {{Woosley}}(1993)}]{Woosley1993}%
  \BibitemOpen
  \bibfield  {author} {\bibinfo {author} {\bibfnamefont {S.~E.}\ \bibnamefont {{Woosley}}},\ }\bibfield  {title} {\bibinfo {title} {{Gamma-Ray Bursts from Stellar Mass Accretion Disks around Black Holes}},\ }\href {https://doi.org/10.1086/172359} {\bibfield  {journal} {\bibinfo  {journal} {\apj}\ }\textbf {\bibinfo {volume} {405}},\ \bibinfo {pages} {273} (\bibinfo {year} {1993})}\BibitemShut {NoStop}%
\bibitem [{\citenamefont {{Li}}\ and\ \citenamefont {{Paczy{\'n}ski}}(1998)}]{Paczynski1998}%
  \BibitemOpen
  \bibfield  {author} {\bibinfo {author} {\bibfnamefont {L.-X.}\ \bibnamefont {{Li}}}\ and\ \bibinfo {author} {\bibfnamefont {B.}~\bibnamefont {{Paczy{\'n}ski}}},\ }\bibfield  {title} {\bibinfo {title} {{Transient Events from Neutron Star Mergers}},\ }\href {https://doi.org/10.1086/311680} {\bibfield  {journal} {\bibinfo  {journal} {\apjl}\ }\textbf {\bibinfo {volume} {507}},\ \bibinfo {pages} {L59} (\bibinfo {year} {1998})},\ \Eprint {https://arxiv.org/abs/astro-ph/9807272} {arXiv:astro-ph/9807272 [astro-ph]} \BibitemShut {NoStop}%
\bibitem [{\citenamefont {{MacFadyen}}\ and\ \citenamefont {{Woosley}}(1999)}]{MacFadyen1999}%
  \BibitemOpen
  \bibfield  {author} {\bibinfo {author} {\bibfnamefont {A.~I.}\ \bibnamefont {{MacFadyen}}}\ and\ \bibinfo {author} {\bibfnamefont {S.~E.}\ \bibnamefont {{Woosley}}},\ }\bibfield  {title} {\bibinfo {title} {{Collapsars: Gamma-Ray Bursts and Explosions in ``Failed Supernovae''}},\ }\href {https://doi.org/10.1086/307790} {\bibfield  {journal} {\bibinfo  {journal} {\apj}\ }\textbf {\bibinfo {volume} {524}},\ \bibinfo {pages} {262} (\bibinfo {year} {1999})},\ \Eprint {https://arxiv.org/abs/astro-ph/9810274} {arXiv:astro-ph/9810274 [astro-ph]} \BibitemShut {NoStop}%
\bibitem [{\citenamefont {{Eichler}}\ \emph {et~al.}(1989)\citenamefont {{Eichler}}, \citenamefont {{Livio}}, \citenamefont {{Piran}},\ and\ \citenamefont {{Schramm}}}]{Eichler1989}%
  \BibitemOpen
  \bibfield  {author} {\bibinfo {author} {\bibfnamefont {D.}~\bibnamefont {{Eichler}}}, \bibinfo {author} {\bibfnamefont {M.}~\bibnamefont {{Livio}}}, \bibinfo {author} {\bibfnamefont {T.}~\bibnamefont {{Piran}}},\ and\ \bibinfo {author} {\bibfnamefont {D.~N.}\ \bibnamefont {{Schramm}}},\ }\bibfield  {title} {\bibinfo {title} {{Nucleosynthesis, neutrino bursts and {\ensuremath{\gamma}}-rays from coalescing neutron stars}},\ }\href {https://doi.org/10.1038/340126a0} {\bibfield  {journal} {\bibinfo  {journal} {\nat}\ }\textbf {\bibinfo {volume} {340}},\ \bibinfo {pages} {126} (\bibinfo {year} {1989})}\BibitemShut {NoStop}%
\bibitem [{\citenamefont {{Narayan}}\ \emph {et~al.}(1992)\citenamefont {{Narayan}}, \citenamefont {{Paczynski}},\ and\ \citenamefont {{Piran}}}]{Narayan1992}%
  \BibitemOpen
  \bibfield  {author} {\bibinfo {author} {\bibfnamefont {R.}~\bibnamefont {{Narayan}}}, \bibinfo {author} {\bibfnamefont {B.}~\bibnamefont {{Paczynski}}},\ and\ \bibinfo {author} {\bibfnamefont {T.}~\bibnamefont {{Piran}}},\ }\bibfield  {title} {\bibinfo {title} {{Gamma-Ray Bursts as the Death Throes of Massive Binary Stars}},\ }\href {https://doi.org/10.1086/186493} {\bibfield  {journal} {\bibinfo  {journal} {\apjl}\ }\textbf {\bibinfo {volume} {395}},\ \bibinfo {pages} {L83} (\bibinfo {year} {1992})},\ \Eprint {https://arxiv.org/abs/astro-ph/9204001} {arXiv:astro-ph/9204001 [astro-ph]} \BibitemShut {NoStop}%
\bibitem [{\citenamefont {{Zhang}}(2006)}]{Zhang2006type}%
  \BibitemOpen
  \bibfield  {author} {\bibinfo {author} {\bibfnamefont {B.}~\bibnamefont {{Zhang}}},\ }\bibfield  {title} {\bibinfo {title} {{Astrophysics: A burst of new ideas}},\ }\href {https://doi.org/10.1038/4441010a} {\bibfield  {journal} {\bibinfo  {journal} {\nat}\ }\textbf {\bibinfo {volume} {444}},\ \bibinfo {pages} {1010} (\bibinfo {year} {2006})},\ \Eprint {https://arxiv.org/abs/astro-ph/0612614} {arXiv:astro-ph/0612614 [astro-ph]} \BibitemShut {NoStop}%
\bibitem [{\citenamefont {{Zhang}}(2018)}]{2018GRBbook}%
  \BibitemOpen
  \bibfield  {author} {\bibinfo {author} {\bibfnamefont {B.}~\bibnamefont {{Zhang}}},\ }\href {https://doi.org/10.1017/9781139226530} {\emph {\bibinfo {title} {{The Physics of Gamma-Ray Bursts}}}}\ (\bibinfo {year} {2018})\BibitemShut {NoStop}%
\bibitem [{\citenamefont {{Zhang}}\ \emph {et~al.}(2006)\citenamefont {{Zhang}}, \citenamefont {{Fan}}, \citenamefont {{Dyks}}, \citenamefont {{Kobayashi}}, \citenamefont {{M{\'e}sz{\'a}ros}}, \citenamefont {{Burrows}}, \citenamefont {{Nousek}},\ and\ \citenamefont {{Gehrels}}}]{Zhang2006afterglow}%
  \BibitemOpen
  \bibfield  {author} {\bibinfo {author} {\bibfnamefont {B.}~\bibnamefont {{Zhang}}}, \bibinfo {author} {\bibfnamefont {Y.~Z.}\ \bibnamefont {{Fan}}}, \bibinfo {author} {\bibfnamefont {J.}~\bibnamefont {{Dyks}}}, \bibinfo {author} {\bibfnamefont {S.}~\bibnamefont {{Kobayashi}}}, \bibinfo {author} {\bibfnamefont {P.}~\bibnamefont {{M{\'e}sz{\'a}ros}}}, \bibinfo {author} {\bibfnamefont {D.~N.}\ \bibnamefont {{Burrows}}}, \bibinfo {author} {\bibfnamefont {J.~A.}\ \bibnamefont {{Nousek}}},\ and\ \bibinfo {author} {\bibfnamefont {N.}~\bibnamefont {{Gehrels}}},\ }\bibfield  {title} {\bibinfo {title} {{Physical Processes Shaping Gamma-Ray Burst X-Ray Afterglow Light Curves: Theoretical Implications from the Swift X-Ray Telescope Observations}},\ }\href {https://doi.org/10.1086/500723} {\bibfield  {journal} {\bibinfo  {journal} {\apj}\ }\textbf {\bibinfo {volume} {642}},\ \bibinfo {pages} {354} (\bibinfo {year} {2006})},\ \Eprint {https://arxiv.org/abs/astro-ph/0508321} {arXiv:astro-ph/0508321 [astro-ph]}
  \BibitemShut {NoStop}%
\bibitem [{\citenamefont {{M{\'e}sz{\'a}ros}}\ and\ \citenamefont {{Rees}}(1997)}]{Meszaros1997afterglow}%
  \BibitemOpen
  \bibfield  {author} {\bibinfo {author} {\bibfnamefont {P.}~\bibnamefont {{M{\'e}sz{\'a}ros}}}\ and\ \bibinfo {author} {\bibfnamefont {M.~J.}\ \bibnamefont {{Rees}}},\ }\bibfield  {title} {\bibinfo {title} {{Optical and Long-Wavelength Afterglow from Gamma-Ray Bursts}},\ }\href {https://doi.org/10.1086/303625} {\bibfield  {journal} {\bibinfo  {journal} {\apj}\ }\textbf {\bibinfo {volume} {476}},\ \bibinfo {pages} {232} (\bibinfo {year} {1997})},\ \Eprint {https://arxiv.org/abs/astro-ph/9606043} {arXiv:astro-ph/9606043 [astro-ph]} \BibitemShut {NoStop}%
\bibitem [{\citenamefont {{Sari}}\ \emph {et~al.}(1998)\citenamefont {{Sari}}, \citenamefont {{Piran}},\ and\ \citenamefont {{Narayan}}}]{Sari1998afterglow}%
  \BibitemOpen
  \bibfield  {author} {\bibinfo {author} {\bibfnamefont {R.}~\bibnamefont {{Sari}}}, \bibinfo {author} {\bibfnamefont {T.}~\bibnamefont {{Piran}}},\ and\ \bibinfo {author} {\bibfnamefont {R.}~\bibnamefont {{Narayan}}},\ }\bibfield  {title} {\bibinfo {title} {{Spectra and Light Curves of Gamma-Ray Burst Afterglows}},\ }\href {https://doi.org/10.1086/311269} {\bibfield  {journal} {\bibinfo  {journal} {\apjl}\ }\textbf {\bibinfo {volume} {497}},\ \bibinfo {pages} {L17} (\bibinfo {year} {1998})},\ \Eprint {https://arxiv.org/abs/astro-ph/9712005} {arXiv:astro-ph/9712005 [astro-ph]} \BibitemShut {NoStop}%
\bibitem [{\citenamefont {{Paczynski}}\ and\ \citenamefont {{Xu}}(1994)}]{Paczynski1994IS}%
  \BibitemOpen
  \bibfield  {author} {\bibinfo {author} {\bibfnamefont {B.}~\bibnamefont {{Paczynski}}}\ and\ \bibinfo {author} {\bibfnamefont {G.}~\bibnamefont {{Xu}}},\ }\bibfield  {title} {\bibinfo {title} {{Neutrino Bursts from Gamma-Ray Bursts}},\ }\href {https://doi.org/10.1086/174178} {\bibfield  {journal} {\bibinfo  {journal} {\apj}\ }\textbf {\bibinfo {volume} {427}},\ \bibinfo {pages} {708} (\bibinfo {year} {1994})}\BibitemShut {NoStop}%
\bibitem [{\citenamefont {{Rees}}\ and\ \citenamefont {{Meszaros}}(1994)}]{Rees1994IS}%
  \BibitemOpen
  \bibfield  {author} {\bibinfo {author} {\bibfnamefont {M.~J.}\ \bibnamefont {{Rees}}}\ and\ \bibinfo {author} {\bibfnamefont {P.}~\bibnamefont {{Meszaros}}},\ }\bibfield  {title} {\bibinfo {title} {{Unsteady Outflow Models for Cosmological Gamma-Ray Bursts}},\ }\href {https://doi.org/10.1086/187446} {\bibfield  {journal} {\bibinfo  {journal} {\apjl}\ }\textbf {\bibinfo {volume} {430}},\ \bibinfo {pages} {L93} (\bibinfo {year} {1994})},\ \Eprint {https://arxiv.org/abs/astro-ph/9404038} {arXiv:astro-ph/9404038 [astro-ph]} \BibitemShut {NoStop}%
\bibitem [{\citenamefont {{Zhang}}\ and\ \citenamefont {{Yan}}(2011)}]{zhang2011ICMART}%
  \BibitemOpen
  \bibfield  {author} {\bibinfo {author} {\bibfnamefont {B.}~\bibnamefont {{Zhang}}}\ and\ \bibinfo {author} {\bibfnamefont {H.}~\bibnamefont {{Yan}}},\ }\bibfield  {title} {\bibinfo {title} {{The Internal-collision-induced Magnetic Reconnection and Turbulence (ICMART) Model of Gamma-ray Bursts}},\ }\href {https://doi.org/10.1088/0004-637X/726/2/90} {\bibfield  {journal} {\bibinfo  {journal} {\apj}\ }\textbf {\bibinfo {volume} {726}},\ \bibinfo {eid} {90} (\bibinfo {year} {2011})},\ \Eprint {https://arxiv.org/abs/1011.1197} {arXiv:1011.1197 [astro-ph.HE]} \BibitemShut {NoStop}%
\bibitem [{\citenamefont {{Zhang}}\ and\ \citenamefont {{Zhang}}(2014)}]{Zhang2014ICMART}%
  \BibitemOpen
  \bibfield  {author} {\bibinfo {author} {\bibfnamefont {B.}~\bibnamefont {{Zhang}}}\ and\ \bibinfo {author} {\bibfnamefont {B.}~\bibnamefont {{Zhang}}},\ }\bibfield  {title} {\bibinfo {title} {{Gamma-Ray Burst Prompt Emission Light Curves and Power Density Spectra in the ICMART Model}},\ }\href {https://doi.org/10.1088/0004-637X/782/2/92} {\bibfield  {journal} {\bibinfo  {journal} {\apj}\ }\textbf {\bibinfo {volume} {782}},\ \bibinfo {eid} {92} (\bibinfo {year} {2014})},\ \Eprint {https://arxiv.org/abs/1312.7701} {arXiv:1312.7701 [astro-ph.HE]} \BibitemShut {NoStop}%
\bibitem [{\citenamefont {{Pe'er}}(2008)}]{Peer2008photospheric}%
  \BibitemOpen
  \bibfield  {author} {\bibinfo {author} {\bibfnamefont {A.}~\bibnamefont {{Pe'er}}},\ }\bibfield  {title} {\bibinfo {title} {{Temporal Evolution of Thermal Emission from Relativistically Expanding Plasma}},\ }\href {https://doi.org/10.1086/588136} {\bibfield  {journal} {\bibinfo  {journal} {\apj}\ }\textbf {\bibinfo {volume} {682}},\ \bibinfo {pages} {463} (\bibinfo {year} {2008})},\ \Eprint {https://arxiv.org/abs/0802.0725} {arXiv:0802.0725 [astro-ph]} \BibitemShut {NoStop}%
\bibitem [{\citenamefont {{Meng}}\ \emph {et~al.}(2019)\citenamefont {{Meng}}, \citenamefont {{Liu}}, \citenamefont {{Wei}}, \citenamefont {{Wu}},\ and\ \citenamefont {{Zhang}}}]{Meng2019photospheric}%
  \BibitemOpen
  \bibfield  {author} {\bibinfo {author} {\bibfnamefont {Y.-Z.}\ \bibnamefont {{Meng}}}, \bibinfo {author} {\bibfnamefont {L.-D.}\ \bibnamefont {{Liu}}}, \bibinfo {author} {\bibfnamefont {J.-J.}\ \bibnamefont {{Wei}}}, \bibinfo {author} {\bibfnamefont {X.-F.}\ \bibnamefont {{Wu}}},\ and\ \bibinfo {author} {\bibfnamefont {B.-B.}\ \bibnamefont {{Zhang}}},\ }\bibfield  {title} {\bibinfo {title} {{The Time-resolved Spectra of Photospheric Emission from a Structured Jet for Gamma-Ray Bursts}},\ }\href {https://doi.org/10.3847/1538-4357/ab30c7} {\bibfield  {journal} {\bibinfo  {journal} {\apj}\ }\textbf {\bibinfo {volume} {882}},\ \bibinfo {eid} {26} (\bibinfo {year} {2019})},\ \Eprint {https://arxiv.org/abs/1904.08526} {arXiv:1904.08526 [astro-ph.HE]} \BibitemShut {NoStop}%
\bibitem [{\citenamefont {{Norris}}\ and\ \citenamefont {{Bonnell}}(2006)}]{Norris2006EE}%
  \BibitemOpen
  \bibfield  {author} {\bibinfo {author} {\bibfnamefont {J.~P.}\ \bibnamefont {{Norris}}}\ and\ \bibinfo {author} {\bibfnamefont {J.~T.}\ \bibnamefont {{Bonnell}}},\ }\bibfield  {title} {\bibinfo {title} {{Short Gamma-Ray Bursts with Extended Emission}},\ }\href {https://doi.org/10.1086/502796} {\bibfield  {journal} {\bibinfo  {journal} {\apj}\ }\textbf {\bibinfo {volume} {643}},\ \bibinfo {pages} {266} (\bibinfo {year} {2006})},\ \Eprint {https://arxiv.org/abs/astro-ph/0601190} {arXiv:astro-ph/0601190 [astro-ph]} \BibitemShut {NoStop}%
\bibitem [{\citenamefont {{Coppin}}\ \emph {et~al.}(2020)\citenamefont {{Coppin}}, \citenamefont {{de Vries}},\ and\ \citenamefont {{van Eijndhoven}}}]{GBM_precursor_PRD}%
  \BibitemOpen
  \bibfield  {author} {\bibinfo {author} {\bibfnamefont {P.}~\bibnamefont {{Coppin}}}, \bibinfo {author} {\bibfnamefont {K.~D.}\ \bibnamefont {{de Vries}}},\ and\ \bibinfo {author} {\bibfnamefont {N.}~\bibnamefont {{van Eijndhoven}}},\ }\bibfield  {title} {\bibinfo {title} {{Identification of gamma-ray burst precursors in Fermi-GBM bursts}},\ }\href {https://doi.org/10.1103/PhysRevD.102.103014} {\bibfield  {journal} {\bibinfo  {journal} {\prd}\ }\textbf {\bibinfo {volume} {102}},\ \bibinfo {eid} {103014} (\bibinfo {year} {2020})},\ \Eprint {https://arxiv.org/abs/2004.03246} {arXiv:2004.03246 [astro-ph.HE]} \BibitemShut {NoStop}%
\bibitem [{\citenamefont {{Wang}}\ \emph {et~al.}(2020)\citenamefont {{Wang}}, \citenamefont {{Peng}}, \citenamefont {{Zou}}, \citenamefont {{Zhang}},\ and\ \citenamefont {{Zhang}}}]{Wang2020_SGRB_precursor}%
  \BibitemOpen
  \bibfield  {author} {\bibinfo {author} {\bibfnamefont {J.-S.}\ \bibnamefont {{Wang}}}, \bibinfo {author} {\bibfnamefont {Z.-K.}\ \bibnamefont {{Peng}}}, \bibinfo {author} {\bibfnamefont {J.-H.}\ \bibnamefont {{Zou}}}, \bibinfo {author} {\bibfnamefont {B.-B.}\ \bibnamefont {{Zhang}}},\ and\ \bibinfo {author} {\bibfnamefont {B.}~\bibnamefont {{Zhang}}},\ }\bibfield  {title} {\bibinfo {title} {{Stringent Search for Precursor Emission in Short GRBs from Fermi/GBM Data and Physical Implications}},\ }\href {https://doi.org/10.3847/2041-8213/abbfb8} {\bibfield  {journal} {\bibinfo  {journal} {\apjl}\ }\textbf {\bibinfo {volume} {902}},\ \bibinfo {eid} {L42} (\bibinfo {year} {2020})},\ \Eprint {https://arxiv.org/abs/2010.05706} {arXiv:2010.05706 [astro-ph.HE]} \BibitemShut {NoStop}%
\bibitem [{\citenamefont {{Zhang}}\ \emph {et~al.}(2022)\citenamefont {{Zhang}}, \citenamefont {{Yi}}, \citenamefont {{Zhang}}, \citenamefont {{Xiong}},\ and\ \citenamefont {{Xiao}}}]{zhang2022superflare}%
  \BibitemOpen
  \bibfield  {author} {\bibinfo {author} {\bibfnamefont {Z.}~\bibnamefont {{Zhang}}}, \bibinfo {author} {\bibfnamefont {S.-X.}\ \bibnamefont {{Yi}}}, \bibinfo {author} {\bibfnamefont {S.-N.}\ \bibnamefont {{Zhang}}}, \bibinfo {author} {\bibfnamefont {S.-L.}\ \bibnamefont {{Xiong}}},\ and\ \bibinfo {author} {\bibfnamefont {S.}~\bibnamefont {{Xiao}}},\ }\bibfield  {title} {\bibinfo {title} {{Tidally-induced Magnetar Super Flare at the Eve of Coalescence with Its Compact Companion}},\ }\href {https://doi.org/10.3847/2041-8213/ac9b55} {\bibfield  {journal} {\bibinfo  {journal} {\apjl}\ }\textbf {\bibinfo {volume} {939}},\ \bibinfo {eid} {L25} (\bibinfo {year} {2022})},\ \Eprint {https://arxiv.org/abs/2207.12324} {arXiv:2207.12324 [astro-ph.HE]} \BibitemShut {NoStop}%
\bibitem [{\citenamefont {{Hansen}}\ and\ \citenamefont {{Lyutikov}}(2001)}]{Hansen2001pre_merger_model}%
  \BibitemOpen
  \bibfield  {author} {\bibinfo {author} {\bibfnamefont {B.~M.~S.}\ \bibnamefont {{Hansen}}}\ and\ \bibinfo {author} {\bibfnamefont {M.}~\bibnamefont {{Lyutikov}}},\ }\bibfield  {title} {\bibinfo {title} {{Radio and X-ray signatures of merging neutron stars}},\ }\href {https://doi.org/10.1046/j.1365-8711.2001.04103.x} {\bibfield  {journal} {\bibinfo  {journal} {\mnras}\ }\textbf {\bibinfo {volume} {322}},\ \bibinfo {pages} {695} (\bibinfo {year} {2001})},\ \Eprint {https://arxiv.org/abs/astro-ph/0003218} {arXiv:astro-ph/0003218 [astro-ph]} \BibitemShut {NoStop}%
\bibitem [{\citenamefont {{Tsang}}\ \emph {et~al.}(2012)\citenamefont {{Tsang}}, \citenamefont {{Read}}, \citenamefont {{Hinderer}}, \citenamefont {{Piro}},\ and\ \citenamefont {{Bondarescu}}}]{Tsang2012pre_merger_model}%
  \BibitemOpen
  \bibfield  {author} {\bibinfo {author} {\bibfnamefont {D.}~\bibnamefont {{Tsang}}}, \bibinfo {author} {\bibfnamefont {J.~S.}\ \bibnamefont {{Read}}}, \bibinfo {author} {\bibfnamefont {T.}~\bibnamefont {{Hinderer}}}, \bibinfo {author} {\bibfnamefont {A.~L.}\ \bibnamefont {{Piro}}},\ and\ \bibinfo {author} {\bibfnamefont {R.}~\bibnamefont {{Bondarescu}}},\ }\bibfield  {title} {\bibinfo {title} {{Resonant Shattering of Neutron Star Crusts}},\ }\href {https://doi.org/10.1103/PhysRevLett.108.011102} {\bibfield  {journal} {\bibinfo  {journal} {\prl}\ }\textbf {\bibinfo {volume} {108}},\ \bibinfo {eid} {011102} (\bibinfo {year} {2012})},\ \Eprint {https://arxiv.org/abs/1110.0467} {arXiv:1110.0467 [astro-ph.HE]} \BibitemShut {NoStop}%
\bibitem [{\citenamefont {{M{\'e}sz{\'a}ros}}\ and\ \citenamefont {{Rees}}(2000)}]{Meszaros2000post_merger_model}%
  \BibitemOpen
  \bibfield  {author} {\bibinfo {author} {\bibfnamefont {P.}~\bibnamefont {{M{\'e}sz{\'a}ros}}}\ and\ \bibinfo {author} {\bibfnamefont {M.~J.}\ \bibnamefont {{Rees}}},\ }\bibfield  {title} {\bibinfo {title} {{Steep Slopes and Preferred Breaks in Gamma-Ray Burst Spectra: The Role of Photospheres and Comptonization}},\ }\href {https://doi.org/10.1086/308371} {\bibfield  {journal} {\bibinfo  {journal} {\apj}\ }\textbf {\bibinfo {volume} {530}},\ \bibinfo {pages} {292} (\bibinfo {year} {2000})},\ \Eprint {https://arxiv.org/abs/astro-ph/9908126} {arXiv:astro-ph/9908126 [astro-ph]} \BibitemShut {NoStop}%
\bibitem [{\citenamefont {{Zhang}}\ \emph {et~al.}(2020)\citenamefont {{Zhang}}, \citenamefont {{Zhang}}, \citenamefont {{Li}}, \citenamefont {{Su}}, \citenamefont {{Dong}}, \citenamefont {{Chang}},\ and\ \citenamefont {{Zhang}}}]{Zhangxiaolu2020}%
  \BibitemOpen
  \bibfield  {author} {\bibinfo {author} {\bibfnamefont {X.-L.}\ \bibnamefont {{Zhang}}}, \bibinfo {author} {\bibfnamefont {C.-T.}\ \bibnamefont {{Zhang}}}, \bibinfo {author} {\bibfnamefont {X.-J.}\ \bibnamefont {{Li}}}, \bibinfo {author} {\bibfnamefont {F.-F.}\ \bibnamefont {{Su}}}, \bibinfo {author} {\bibfnamefont {X.-F.}\ \bibnamefont {{Dong}}}, \bibinfo {author} {\bibfnamefont {H.-Y.}\ \bibnamefont {{Chang}}},\ and\ \bibinfo {author} {\bibfnamefont {Z.-B.}\ \bibnamefont {{Zhang}}},\ }\bibfield  {title} {\bibinfo {title} {{Gamma-ray bursts with extended emission: classifications, energy correlations and radiation properties}},\ }\href {https://doi.org/10.1088/1674-4527/20/12/201} {\bibfield  {journal} {\bibinfo  {journal} {Research in Astronomy and Astrophysics}\ }\textbf {\bibinfo {volume} {20}},\ \bibinfo {eid} {201} (\bibinfo {year} {2020})},\ \Eprint {https://arxiv.org/abs/2010.00337} {arXiv:2010.00337 [astro-ph.HE]} \BibitemShut {NoStop}%
\bibitem [{\citenamefont {{Rowlinson}}\ \emph {et~al.}(2010)\citenamefont {{Rowlinson}}, \citenamefont {{Wiersema}}, \citenamefont {{Levan}}, \citenamefont {{Tanvir}}, \citenamefont {{O'Brien}}, \citenamefont {{Rol}}, \citenamefont {{Hjorth}}, \citenamefont {{Th{\"o}ne}}, \citenamefont {{de Ugarte Postigo}}, \citenamefont {{Fynbo}}, \citenamefont {{Jakobsson}}, \citenamefont {{Pagani}},\ and\ \citenamefont {{Stamatikos}}}]{Rowlinson2010EE}%
  \BibitemOpen
  \bibfield  {author} {\bibinfo {author} {\bibfnamefont {A.}~\bibnamefont {{Rowlinson}}}, \bibinfo {author} {\bibfnamefont {K.}~\bibnamefont {{Wiersema}}}, \bibinfo {author} {\bibfnamefont {A.~J.}\ \bibnamefont {{Levan}}}, \bibinfo {author} {\bibfnamefont {N.~R.}\ \bibnamefont {{Tanvir}}}, \bibinfo {author} {\bibfnamefont {P.~T.}\ \bibnamefont {{O'Brien}}}, \bibinfo {author} {\bibfnamefont {E.}~\bibnamefont {{Rol}}}, \bibinfo {author} {\bibfnamefont {J.}~\bibnamefont {{Hjorth}}}, \bibinfo {author} {\bibfnamefont {C.~C.}\ \bibnamefont {{Th{\"o}ne}}}, \bibinfo {author} {\bibfnamefont {A.}~\bibnamefont {{de Ugarte Postigo}}}, \bibinfo {author} {\bibfnamefont {J.~P.~U.}\ \bibnamefont {{Fynbo}}}, \bibinfo {author} {\bibfnamefont {P.}~\bibnamefont {{Jakobsson}}}, \bibinfo {author} {\bibfnamefont {C.}~\bibnamefont {{Pagani}}},\ and\ \bibinfo {author} {\bibfnamefont {M.}~\bibnamefont {{Stamatikos}}},\ }\bibfield  {title} {\bibinfo {title} {{Discovery of the afterglow and host galaxy of the low-redshift short GRB
  080905A}},\ }\href {https://doi.org/10.1111/j.1365-2966.2010.17115.x} {\bibfield  {journal} {\bibinfo  {journal} {\mnras}\ }\textbf {\bibinfo {volume} {408}},\ \bibinfo {pages} {383} (\bibinfo {year} {2010})},\ \Eprint {https://arxiv.org/abs/1006.0487} {arXiv:1006.0487 [astro-ph.HE]} \BibitemShut {NoStop}%
\bibitem [{\citenamefont {{Rowlinson}}\ \emph {et~al.}(2013)\citenamefont {{Rowlinson}}, \citenamefont {{O'Brien}}, \citenamefont {{Metzger}}, \citenamefont {{Tanvir}},\ and\ \citenamefont {{Levan}}}]{Rowlinson2013EE}%
  \BibitemOpen
  \bibfield  {author} {\bibinfo {author} {\bibfnamefont {A.}~\bibnamefont {{Rowlinson}}}, \bibinfo {author} {\bibfnamefont {P.~T.}\ \bibnamefont {{O'Brien}}}, \bibinfo {author} {\bibfnamefont {B.~D.}\ \bibnamefont {{Metzger}}}, \bibinfo {author} {\bibfnamefont {N.~R.}\ \bibnamefont {{Tanvir}}},\ and\ \bibinfo {author} {\bibfnamefont {A.~J.}\ \bibnamefont {{Levan}}},\ }\bibfield  {title} {\bibinfo {title} {{Signatures of magnetar central engines in short GRB light curves}},\ }\href {https://doi.org/10.1093/mnras/sts683} {\bibfield  {journal} {\bibinfo  {journal} {\mnras}\ }\textbf {\bibinfo {volume} {430}},\ \bibinfo {pages} {1061} (\bibinfo {year} {2013})},\ \Eprint {https://arxiv.org/abs/1301.0629} {arXiv:1301.0629 [astro-ph.HE]} \BibitemShut {NoStop}%
\bibitem [{\citenamefont {{Levan}}\ \emph {et~al.}(2024)\citenamefont {{Levan}}, \citenamefont {{Gompertz}}, \citenamefont {{Salafia}}, \citenamefont {{Bulla}}, \citenamefont {{Burns}}, \citenamefont {{Hotokezaka}}, \citenamefont {{Izzo}}, \citenamefont {{Lamb}}, \citenamefont {{Malesani}}, \citenamefont {{Oates}}, \citenamefont {{Ravasio}}, \citenamefont {{Rouco Escorial}}, \citenamefont {{Schneider}}, \citenamefont {{Sarin}}, \citenamefont {{Schulze}}, \citenamefont {{Tanvir}}, \citenamefont {{Ackley}}, \citenamefont {{Anderson}}, \citenamefont {{Brammer}}, \citenamefont {{Christensen}}, \citenamefont {{Dhillon}}, \citenamefont {{Evans}}, \citenamefont {{Fausnaugh}}, \citenamefont {{Fong}}, \citenamefont {{Fruchter}}, \citenamefont {{Fryer}}, \citenamefont {{Fynbo}}, \citenamefont {{Gaspari}}, \citenamefont {{Heintz}}, \citenamefont {{Hjorth}}, \citenamefont {{Kennea}}, \citenamefont {{Kennedy}}, \citenamefont {{Laskar}}, \citenamefont {{Leloudas}}, \citenamefont {{Mandel}}, \citenamefont
  {{Martin-Carrillo}}, \citenamefont {{Metzger}}, \citenamefont {{Nicholl}}, \citenamefont {{Nugent}}, \citenamefont {{Palmerio}}, \citenamefont {{Pugliese}}, \citenamefont {{Rastinejad}}, \citenamefont {{Rhodes}}, \citenamefont {{Rossi}}, \citenamefont {{Saccardi}}, \citenamefont {{Smartt}}, \citenamefont {{Stevance}}, \citenamefont {{Tohuvavohu}}, \citenamefont {{van der Horst}}, \citenamefont {{Vergani}}, \citenamefont {{Watson}}, \citenamefont {{Barclay}}, \citenamefont {{Bhirombhakdi}}, \citenamefont {{Breedt}}, \citenamefont {{Breeveld}}, \citenamefont {{Brown}}, \citenamefont {{Campana}}, \citenamefont {{Chrimes}}, \citenamefont {{D'Avanzo}}, \citenamefont {{D'Elia}}, \citenamefont {{De Pasquale}}, \citenamefont {{Dyer}}, \citenamefont {{Galloway}}, \citenamefont {{Garbutt}}, \citenamefont {{Green}}, \citenamefont {{Hartmann}}, \citenamefont {{Jakobsson}}, \citenamefont {{Kerry}}, \citenamefont {{Kouveliotou}}, \citenamefont {{Langeroodi}}, \citenamefont {{Le Floc'h}}, \citenamefont {{Leung}},
  \citenamefont {{Littlefair}}, \citenamefont {{Munday}}, \citenamefont {{O'Brien}}, \citenamefont {{Parsons}}, \citenamefont {{Pelisoli}}, \citenamefont {{Sahman}}, \citenamefont {{Salvaterra}}, \citenamefont {{Sbarufatti}}, \citenamefont {{Steeghs}}, \citenamefont {{Tagliaferri}}, \citenamefont {{Th{\"o}ne}}, \citenamefont {{de Ugarte Postigo}},\ and\ \citenamefont {{Kann}}}]{Levan2024KN}%
  \BibitemOpen
  \bibfield  {author} {\bibinfo {author} {\bibfnamefont {A.~J.}\ \bibnamefont {{Levan}}}, \bibinfo {author} {\bibfnamefont {B.~P.}\ \bibnamefont {{Gompertz}}}, \bibinfo {author} {\bibfnamefont {O.~S.}\ \bibnamefont {{Salafia}}}, \bibinfo {author} {\bibfnamefont {M.}~\bibnamefont {{Bulla}}}, \bibinfo {author} {\bibfnamefont {E.}~\bibnamefont {{Burns}}}, \bibinfo {author} {\bibfnamefont {K.}~\bibnamefont {{Hotokezaka}}}, \bibinfo {author} {\bibfnamefont {L.}~\bibnamefont {{Izzo}}}, \bibinfo {author} {\bibfnamefont {G.~P.}\ \bibnamefont {{Lamb}}}, \bibinfo {author} {\bibfnamefont {D.~B.}\ \bibnamefont {{Malesani}}}, \bibinfo {author} {\bibfnamefont {S.~R.}\ \bibnamefont {{Oates}}}, \bibinfo {author} {\bibfnamefont {M.~E.}\ \bibnamefont {{Ravasio}}}, \bibinfo {author} {\bibfnamefont {A.}~\bibnamefont {{Rouco Escorial}}}, \bibinfo {author} {\bibfnamefont {B.}~\bibnamefont {{Schneider}}}, \bibinfo {author} {\bibfnamefont {N.}~\bibnamefont {{Sarin}}}, \bibinfo {author} {\bibfnamefont {S.}~\bibnamefont {{Schulze}}},
  \bibinfo {author} {\bibfnamefont {N.~R.}\ \bibnamefont {{Tanvir}}}, \bibinfo {author} {\bibfnamefont {K.}~\bibnamefont {{Ackley}}}, \bibinfo {author} {\bibfnamefont {G.}~\bibnamefont {{Anderson}}}, \bibinfo {author} {\bibfnamefont {G.~B.}\ \bibnamefont {{Brammer}}}, \bibinfo {author} {\bibfnamefont {L.}~\bibnamefont {{Christensen}}}, \bibinfo {author} {\bibfnamefont {V.~S.}\ \bibnamefont {{Dhillon}}}, \bibinfo {author} {\bibfnamefont {P.~A.}\ \bibnamefont {{Evans}}}, \bibinfo {author} {\bibfnamefont {M.}~\bibnamefont {{Fausnaugh}}}, \bibinfo {author} {\bibfnamefont {W.-f.}\ \bibnamefont {{Fong}}}, \bibinfo {author} {\bibfnamefont {A.~S.}\ \bibnamefont {{Fruchter}}}, \bibinfo {author} {\bibfnamefont {C.}~\bibnamefont {{Fryer}}}, \bibinfo {author} {\bibfnamefont {J.~P.~U.}\ \bibnamefont {{Fynbo}}}, \bibinfo {author} {\bibfnamefont {N.}~\bibnamefont {{Gaspari}}}, \bibinfo {author} {\bibfnamefont {K.~E.}\ \bibnamefont {{Heintz}}}, \bibinfo {author} {\bibfnamefont {J.}~\bibnamefont {{Hjorth}}}, \bibinfo {author}
  {\bibfnamefont {J.~A.}\ \bibnamefont {{Kennea}}}, \bibinfo {author} {\bibfnamefont {M.~R.}\ \bibnamefont {{Kennedy}}}, \bibinfo {author} {\bibfnamefont {T.}~\bibnamefont {{Laskar}}}, \bibinfo {author} {\bibfnamefont {G.}~\bibnamefont {{Leloudas}}}, \bibinfo {author} {\bibfnamefont {I.}~\bibnamefont {{Mandel}}}, \bibinfo {author} {\bibfnamefont {A.}~\bibnamefont {{Martin-Carrillo}}}, \bibinfo {author} {\bibfnamefont {B.~D.}\ \bibnamefont {{Metzger}}}, \bibinfo {author} {\bibfnamefont {M.}~\bibnamefont {{Nicholl}}}, \bibinfo {author} {\bibfnamefont {A.}~\bibnamefont {{Nugent}}}, \bibinfo {author} {\bibfnamefont {J.~T.}\ \bibnamefont {{Palmerio}}}, \bibinfo {author} {\bibfnamefont {G.}~\bibnamefont {{Pugliese}}}, \bibinfo {author} {\bibfnamefont {J.}~\bibnamefont {{Rastinejad}}}, \bibinfo {author} {\bibfnamefont {L.}~\bibnamefont {{Rhodes}}}, \bibinfo {author} {\bibfnamefont {A.}~\bibnamefont {{Rossi}}}, \bibinfo {author} {\bibfnamefont {A.}~\bibnamefont {{Saccardi}}}, \bibinfo {author} {\bibfnamefont {S.~J.}\
  \bibnamefont {{Smartt}}}, \bibinfo {author} {\bibfnamefont {H.~F.}\ \bibnamefont {{Stevance}}}, \bibinfo {author} {\bibfnamefont {A.}~\bibnamefont {{Tohuvavohu}}}, \bibinfo {author} {\bibfnamefont {A.}~\bibnamefont {{van der Horst}}}, \bibinfo {author} {\bibfnamefont {S.~D.}\ \bibnamefont {{Vergani}}}, \bibinfo {author} {\bibfnamefont {D.}~\bibnamefont {{Watson}}}, \bibinfo {author} {\bibfnamefont {T.}~\bibnamefont {{Barclay}}}, \bibinfo {author} {\bibfnamefont {K.}~\bibnamefont {{Bhirombhakdi}}}, \bibinfo {author} {\bibfnamefont {E.}~\bibnamefont {{Breedt}}}, \bibinfo {author} {\bibfnamefont {A.~A.}\ \bibnamefont {{Breeveld}}}, \bibinfo {author} {\bibfnamefont {A.~J.}\ \bibnamefont {{Brown}}}, \bibinfo {author} {\bibfnamefont {S.}~\bibnamefont {{Campana}}}, \bibinfo {author} {\bibfnamefont {A.~A.}\ \bibnamefont {{Chrimes}}}, \bibinfo {author} {\bibfnamefont {P.}~\bibnamefont {{D'Avanzo}}}, \bibinfo {author} {\bibfnamefont {V.}~\bibnamefont {{D'Elia}}}, \bibinfo {author} {\bibfnamefont {M.}~\bibnamefont
  {{De Pasquale}}}, \bibinfo {author} {\bibfnamefont {M.~J.}\ \bibnamefont {{Dyer}}}, \bibinfo {author} {\bibfnamefont {D.~K.}\ \bibnamefont {{Galloway}}}, \bibinfo {author} {\bibfnamefont {J.~A.}\ \bibnamefont {{Garbutt}}}, \bibinfo {author} {\bibfnamefont {M.~J.}\ \bibnamefont {{Green}}}, \bibinfo {author} {\bibfnamefont {D.~H.}\ \bibnamefont {{Hartmann}}}, \bibinfo {author} {\bibfnamefont {P.}~\bibnamefont {{Jakobsson}}}, \bibinfo {author} {\bibfnamefont {P.}~\bibnamefont {{Kerry}}}, \bibinfo {author} {\bibfnamefont {C.}~\bibnamefont {{Kouveliotou}}}, \bibinfo {author} {\bibfnamefont {D.}~\bibnamefont {{Langeroodi}}}, \bibinfo {author} {\bibfnamefont {E.}~\bibnamefont {{Le Floc'h}}}, \bibinfo {author} {\bibfnamefont {J.~K.}\ \bibnamefont {{Leung}}}, \bibinfo {author} {\bibfnamefont {S.~P.}\ \bibnamefont {{Littlefair}}}, \bibinfo {author} {\bibfnamefont {J.}~\bibnamefont {{Munday}}}, \bibinfo {author} {\bibfnamefont {P.}~\bibnamefont {{O'Brien}}}, \bibinfo {author} {\bibfnamefont {S.~G.}\ \bibnamefont
  {{Parsons}}}, \bibinfo {author} {\bibfnamefont {I.}~\bibnamefont {{Pelisoli}}}, \bibinfo {author} {\bibfnamefont {D.~I.}\ \bibnamefont {{Sahman}}}, \bibinfo {author} {\bibfnamefont {R.}~\bibnamefont {{Salvaterra}}}, \bibinfo {author} {\bibfnamefont {B.}~\bibnamefont {{Sbarufatti}}}, \bibinfo {author} {\bibfnamefont {D.}~\bibnamefont {{Steeghs}}}, \bibinfo {author} {\bibfnamefont {G.}~\bibnamefont {{Tagliaferri}}}, \bibinfo {author} {\bibfnamefont {C.~C.}\ \bibnamefont {{Th{\"o}ne}}}, \bibinfo {author} {\bibfnamefont {A.}~\bibnamefont {{de Ugarte Postigo}}},\ and\ \bibinfo {author} {\bibfnamefont {D.~A.}\ \bibnamefont {{Kann}}},\ }\bibfield  {title} {\bibinfo {title} {{Heavy-element production in a compact object merger observed by JWST}},\ }\href {https://doi.org/10.1038/s41586-023-06759-1} {\bibfield  {journal} {\bibinfo  {journal} {\nat}\ }\textbf {\bibinfo {volume} {626}},\ \bibinfo {pages} {737} (\bibinfo {year} {2024})},\ \Eprint {https://arxiv.org/abs/2307.02098} {arXiv:2307.02098 [astro-ph.HE]}
  \BibitemShut {NoStop}%
\bibitem [{\citenamefont {{Xiao}}\ \emph {et~al.}(2022{\natexlab{a}})\citenamefont {{Xiao}}, \citenamefont {{Zhang}}, \citenamefont {{Zhu}}, \citenamefont {{Xiong}}, \citenamefont {{Gao}}, \citenamefont {{Xu}}, \citenamefont {{Zhang}}, \citenamefont {{Peng}}, \citenamefont {{Li}}, \citenamefont {{Zhang}}, \citenamefont {{Lu}}, \citenamefont {{Lin}}, \citenamefont {{Liu}}, \citenamefont {{Zhang}}, \citenamefont {{Ge}}, \citenamefont {{Tuo}}, \citenamefont {{Xue}}, \citenamefont {{Fu}}, \citenamefont {{Liu}}, \citenamefont {{Li}}, \citenamefont {{Wang}}, \citenamefont {{Zheng}}, \citenamefont {{Wang}}, \citenamefont {{Jiang}}, \citenamefont {{Li}}, \citenamefont {{Liu}}, \citenamefont {{Cao}}, \citenamefont {{Cai}}, \citenamefont {{Yi}}, \citenamefont {{Zhao}}, \citenamefont {{Xie}}, \citenamefont {{Li}}, \citenamefont {{Luo}}, \citenamefont {{Liao}}, \citenamefont {{Song}}, \citenamefont {{Zhang}}, \citenamefont {{Qu}}, \citenamefont {{Liu}}, \citenamefont {{Li}}, \citenamefont {{Xu}},\ and\ \citenamefont
  {{Li}}}]{XS_GRB211211A_QPO}%
  \BibitemOpen
  \bibfield  {author} {\bibinfo {author} {\bibfnamefont {S.}~\bibnamefont {{Xiao}}}, \bibinfo {author} {\bibfnamefont {Y.-Q.}\ \bibnamefont {{Zhang}}}, \bibinfo {author} {\bibfnamefont {Z.-P.}\ \bibnamefont {{Zhu}}}, \bibinfo {author} {\bibfnamefont {S.-L.}\ \bibnamefont {{Xiong}}}, \bibinfo {author} {\bibfnamefont {H.}~\bibnamefont {{Gao}}}, \bibinfo {author} {\bibfnamefont {D.}~\bibnamefont {{Xu}}}, \bibinfo {author} {\bibfnamefont {S.-N.}\ \bibnamefont {{Zhang}}}, \bibinfo {author} {\bibfnamefont {W.-X.}\ \bibnamefont {{Peng}}}, \bibinfo {author} {\bibfnamefont {X.-B.}\ \bibnamefont {{Li}}}, \bibinfo {author} {\bibfnamefont {P.}~\bibnamefont {{Zhang}}}, \bibinfo {author} {\bibfnamefont {F.-J.}\ \bibnamefont {{Lu}}}, \bibinfo {author} {\bibfnamefont {L.}~\bibnamefont {{Lin}}}, \bibinfo {author} {\bibfnamefont {L.-D.}\ \bibnamefont {{Liu}}}, \bibinfo {author} {\bibfnamefont {Z.}~\bibnamefont {{Zhang}}}, \bibinfo {author} {\bibfnamefont {M.-Y.}\ \bibnamefont {{Ge}}}, \bibinfo {author} {\bibfnamefont {Y.-L.}\
  \bibnamefont {{Tuo}}}, \bibinfo {author} {\bibfnamefont {W.-C.}\ \bibnamefont {{Xue}}}, \bibinfo {author} {\bibfnamefont {S.-Y.}\ \bibnamefont {{Fu}}}, \bibinfo {author} {\bibfnamefont {X.}~\bibnamefont {{Liu}}}, \bibinfo {author} {\bibfnamefont {A.}~\bibnamefont {{Li}}}, \bibinfo {author} {\bibfnamefont {T.-C.}\ \bibnamefont {{Wang}}}, \bibinfo {author} {\bibfnamefont {C.}~\bibnamefont {{Zheng}}}, \bibinfo {author} {\bibfnamefont {Y.}~\bibnamefont {{Wang}}}, \bibinfo {author} {\bibfnamefont {S.-Q.}\ \bibnamefont {{Jiang}}}, \bibinfo {author} {\bibfnamefont {J.-D.}\ \bibnamefont {{Li}}}, \bibinfo {author} {\bibfnamefont {J.-C.}\ \bibnamefont {{Liu}}}, \bibinfo {author} {\bibfnamefont {Z.-J.}\ \bibnamefont {{Cao}}}, \bibinfo {author} {\bibfnamefont {C.}~\bibnamefont {{Cai}}}, \bibinfo {author} {\bibfnamefont {Q.-B.}\ \bibnamefont {{Yi}}}, \bibinfo {author} {\bibfnamefont {Y.}~\bibnamefont {{Zhao}}}, \bibinfo {author} {\bibfnamefont {S.-L.}\ \bibnamefont {{Xie}}}, \bibinfo {author} {\bibfnamefont {C.-K.}\
  \bibnamefont {{Li}}}, \bibinfo {author} {\bibfnamefont {Q.}~\bibnamefont {{Luo}}}, \bibinfo {author} {\bibfnamefont {J.-Y.}\ \bibnamefont {{Liao}}}, \bibinfo {author} {\bibfnamefont {L.-M.}\ \bibnamefont {{Song}}}, \bibinfo {author} {\bibfnamefont {S.}~\bibnamefont {{Zhang}}}, \bibinfo {author} {\bibfnamefont {J.-L.}\ \bibnamefont {{Qu}}}, \bibinfo {author} {\bibfnamefont {C.-Z.}\ \bibnamefont {{Liu}}}, \bibinfo {author} {\bibfnamefont {X.-F.}\ \bibnamefont {{Li}}}, \bibinfo {author} {\bibfnamefont {Y.-P.}\ \bibnamefont {{Xu}}},\ and\ \bibinfo {author} {\bibfnamefont {T.-P.}\ \bibnamefont {{Li}}},\ }\bibfield  {title} {\bibinfo {title} {{The quasi-periodically oscillating precursor of a long gamma-ray burst from a binary neutron star merger}},\ }\href {https://doi.org/10.48550/arXiv.2205.02186} {\bibfield  {journal} {\bibinfo  {journal} {arXiv e-prints}\ ,\ \bibinfo {eid} {arXiv:2205.02186}} (\bibinfo {year} {2022}{\natexlab{a}})},\ \Eprint {https://arxiv.org/abs/2205.02186} {arXiv:2205.02186 [astro-ph.HE]}
  \BibitemShut {NoStop}%
\bibitem [{\citenamefont {{Yang}}\ \emph {et~al.}(2022)\citenamefont {{Yang}}, \citenamefont {{Ai}}, \citenamefont {{Zhang}}, \citenamefont {{Zhang}}, \citenamefont {{Liu}}, \citenamefont {{Wang}}, \citenamefont {{Yang}}, \citenamefont {{Yin}}, \citenamefont {{Li}},\ and\ \citenamefont {{L{\"u}}}}]{YJ_GRB211211A_nature}%
  \BibitemOpen
  \bibfield  {author} {\bibinfo {author} {\bibfnamefont {J.}~\bibnamefont {{Yang}}}, \bibinfo {author} {\bibfnamefont {S.}~\bibnamefont {{Ai}}}, \bibinfo {author} {\bibfnamefont {B.-B.}\ \bibnamefont {{Zhang}}}, \bibinfo {author} {\bibfnamefont {B.}~\bibnamefont {{Zhang}}}, \bibinfo {author} {\bibfnamefont {Z.-K.}\ \bibnamefont {{Liu}}}, \bibinfo {author} {\bibfnamefont {X.~I.}\ \bibnamefont {{Wang}}}, \bibinfo {author} {\bibfnamefont {Y.-H.}\ \bibnamefont {{Yang}}}, \bibinfo {author} {\bibfnamefont {Y.-H.}\ \bibnamefont {{Yin}}}, \bibinfo {author} {\bibfnamefont {Y.}~\bibnamefont {{Li}}},\ and\ \bibinfo {author} {\bibfnamefont {H.-J.}\ \bibnamefont {{L{\"u}}}},\ }\bibfield  {title} {\bibinfo {title} {{A long-duration gamma-ray burst with a peculiar origin}},\ }\href {https://doi.org/10.1038/s41586-022-05403-8} {\bibfield  {journal} {\bibinfo  {journal} {\nat}\ }\textbf {\bibinfo {volume} {612}},\ \bibinfo {pages} {232} (\bibinfo {year} {2022})},\ \Eprint {https://arxiv.org/abs/2204.12771} {arXiv:2204.12771
  [astro-ph.HE]} \BibitemShut {NoStop}%
\bibitem [{\citenamefont {{Sun}}\ \emph {et~al.}(2023)\citenamefont {{Sun}}, \citenamefont {{Wang}}, \citenamefont {{Yang}}, \citenamefont {{Zhang}}, \citenamefont {{Xiong}}, \citenamefont {{Yin}}, \citenamefont {{Liu}}, \citenamefont {{Li}}, \citenamefont {{Xue}}, \citenamefont {{Yan}}, \citenamefont {{Zhang}}, \citenamefont {{Tan}}, \citenamefont {{Pan}}, \citenamefont {{Liu}}, \citenamefont {{Cheng}}, \citenamefont {{Zhang}}, \citenamefont {{Hu}}, \citenamefont {{Zheng}}, \citenamefont {{An}}, \citenamefont {{Cai}}, \citenamefont {{Hu}}, \citenamefont {{Jin}}, \citenamefont {{Li}}, \citenamefont {{Li}}, \citenamefont {{Liu}}, \citenamefont {{Liu}}, \citenamefont {{Peng}}, \citenamefont {{Song}}, \citenamefont {{Sun}}, \citenamefont {{Sun}}, \citenamefont {{Wang}}, \citenamefont {{Wen}}, \citenamefont {{Xiao}}, \citenamefont {{Yi}}, \citenamefont {{Zhang}}, \citenamefont {{Zhang}}, \citenamefont {{Zhang}}, \citenamefont {{Zhang}}, \citenamefont {{Zhao}}, \citenamefont {{Zheng}}, \citenamefont {{Ling}},
  \citenamefont {{Zhang}}, \citenamefont {{Yuan}},\ and\ \citenamefont {{Zhang}}}]{2023arXiv230705689S}%
  \BibitemOpen
  \bibfield  {author} {\bibinfo {author} {\bibfnamefont {H.}~\bibnamefont {{Sun}}}, \bibinfo {author} {\bibfnamefont {C.~W.}\ \bibnamefont {{Wang}}}, \bibinfo {author} {\bibfnamefont {J.}~\bibnamefont {{Yang}}}, \bibinfo {author} {\bibfnamefont {B.~B.}\ \bibnamefont {{Zhang}}}, \bibinfo {author} {\bibfnamefont {S.~L.}\ \bibnamefont {{Xiong}}}, \bibinfo {author} {\bibfnamefont {Y.~H.~I.}\ \bibnamefont {{Yin}}}, \bibinfo {author} {\bibfnamefont {Y.}~\bibnamefont {{Liu}}}, \bibinfo {author} {\bibfnamefont {Y.}~\bibnamefont {{Li}}}, \bibinfo {author} {\bibfnamefont {W.~C.}\ \bibnamefont {{Xue}}}, \bibinfo {author} {\bibfnamefont {Z.}~\bibnamefont {{Yan}}}, \bibinfo {author} {\bibfnamefont {C.}~\bibnamefont {{Zhang}}}, \bibinfo {author} {\bibfnamefont {W.~J.}\ \bibnamefont {{Tan}}}, \bibinfo {author} {\bibfnamefont {H.~W.}\ \bibnamefont {{Pan}}}, \bibinfo {author} {\bibfnamefont {J.~C.}\ \bibnamefont {{Liu}}}, \bibinfo {author} {\bibfnamefont {H.~Q.}\ \bibnamefont {{Cheng}}}, \bibinfo {author} {\bibfnamefont
  {Y.~Q.}\ \bibnamefont {{Zhang}}}, \bibinfo {author} {\bibfnamefont {J.~W.}\ \bibnamefont {{Hu}}}, \bibinfo {author} {\bibfnamefont {C.}~\bibnamefont {{Zheng}}}, \bibinfo {author} {\bibfnamefont {Z.~H.}\ \bibnamefont {{An}}}, \bibinfo {author} {\bibfnamefont {C.}~\bibnamefont {{Cai}}}, \bibinfo {author} {\bibfnamefont {L.}~\bibnamefont {{Hu}}}, \bibinfo {author} {\bibfnamefont {C.}~\bibnamefont {{Jin}}}, \bibinfo {author} {\bibfnamefont {D.~Y.}\ \bibnamefont {{Li}}}, \bibinfo {author} {\bibfnamefont {X.~Q.}\ \bibnamefont {{Li}}}, \bibinfo {author} {\bibfnamefont {H.~Y.}\ \bibnamefont {{Liu}}}, \bibinfo {author} {\bibfnamefont {M.}~\bibnamefont {{Liu}}}, \bibinfo {author} {\bibfnamefont {W.~X.}\ \bibnamefont {{Peng}}}, \bibinfo {author} {\bibfnamefont {L.~M.}\ \bibnamefont {{Song}}}, \bibinfo {author} {\bibfnamefont {S.~L.}\ \bibnamefont {{Sun}}}, \bibinfo {author} {\bibfnamefont {X.~J.}\ \bibnamefont {{Sun}}}, \bibinfo {author} {\bibfnamefont {X.~L.}\ \bibnamefont {{Wang}}}, \bibinfo {author} {\bibfnamefont
  {X.~Y.}\ \bibnamefont {{Wen}}}, \bibinfo {author} {\bibfnamefont {S.}~\bibnamefont {{Xiao}}}, \bibinfo {author} {\bibfnamefont {S.~X.}\ \bibnamefont {{Yi}}}, \bibinfo {author} {\bibfnamefont {F.}~\bibnamefont {{Zhang}}}, \bibinfo {author} {\bibfnamefont {W.~D.}\ \bibnamefont {{Zhang}}}, \bibinfo {author} {\bibfnamefont {X.~F.}\ \bibnamefont {{Zhang}}}, \bibinfo {author} {\bibfnamefont {Y.~H.}\ \bibnamefont {{Zhang}}}, \bibinfo {author} {\bibfnamefont {D.~H.}\ \bibnamefont {{Zhao}}}, \bibinfo {author} {\bibfnamefont {S.~J.}\ \bibnamefont {{Zheng}}}, \bibinfo {author} {\bibfnamefont {Z.~X.}\ \bibnamefont {{Ling}}}, \bibinfo {author} {\bibfnamefont {S.~N.}\ \bibnamefont {{Zhang}}}, \bibinfo {author} {\bibfnamefont {W.}~\bibnamefont {{Yuan}}},\ and\ \bibinfo {author} {\bibfnamefont {B.}~\bibnamefont {{Zhang}}},\ }\bibfield  {title} {\bibinfo {title} {{Magnetar emergence in a peculiar gamma-ray burst from a compact star merger}},\ }\href {https://doi.org/10.48550/arXiv.2307.05689} {\bibfield  {journal} {\bibinfo
   {journal} {arXiv e-prints}\ ,\ \bibinfo {eid} {arXiv:2307.05689}} (\bibinfo {year} {2023})},\ \Eprint {https://arxiv.org/abs/2307.05689} {arXiv:2307.05689 [astro-ph.HE]} \BibitemShut {NoStop}%
\bibitem [{\citenamefont {{Fermi GBM Team}}(2023)}]{Fermi2023GCN}%
  \BibitemOpen
  \bibfield  {author} {\bibinfo {author} {\bibnamefont {{Fermi GBM Team}}},\ }\bibfield  {title} {\bibinfo {title} {{GRB 230307A: Fermi GBM Final Real-time Localization}},\ }\href@noop {} {\bibfield  {journal} {\bibinfo  {journal} {GRB Coordinates Network}\ }\textbf {\bibinfo {volume} {33405}},\ \bibinfo {pages} {1} (\bibinfo {year} {2023})}\BibitemShut {NoStop}%
\bibitem [{\citenamefont {{Svinkin}}\ \emph {et~al.}(2023)\citenamefont {{Svinkin}}, \citenamefont {{Frederiks}}, \citenamefont {{Ulanov}}, \citenamefont {{Tsvetkova}}, \citenamefont {{Lysenko}}, \citenamefont {{Ridnaia}}, \citenamefont {{Cline}},\ and\ \citenamefont {{Konus-Wind Team}}}]{Konus2023GCN}%
  \BibitemOpen
  \bibfield  {author} {\bibinfo {author} {\bibfnamefont {D.}~\bibnamefont {{Svinkin}}}, \bibinfo {author} {\bibfnamefont {D.}~\bibnamefont {{Frederiks}}}, \bibinfo {author} {\bibfnamefont {M.}~\bibnamefont {{Ulanov}}}, \bibinfo {author} {\bibfnamefont {A.}~\bibnamefont {{Tsvetkova}}}, \bibinfo {author} {\bibfnamefont {A.}~\bibnamefont {{Lysenko}}}, \bibinfo {author} {\bibfnamefont {A.}~\bibnamefont {{Ridnaia}}}, \bibinfo {author} {\bibfnamefont {T.}~\bibnamefont {{Cline}}},\ and\ \bibinfo {author} {\bibnamefont {{Konus-Wind Team}}},\ }\bibfield  {title} {\bibinfo {title} {{Konus-Wind detection of GRB 230307A}},\ }\href@noop {} {\bibfield  {journal} {\bibinfo  {journal} {GRB Coordinates Network}\ }\textbf {\bibinfo {volume} {33427}},\ \bibinfo {pages} {1} (\bibinfo {year} {2023})}\BibitemShut {NoStop}%
\bibitem [{\citenamefont {{Xiong}}\ \emph {et~al.}(2023)\citenamefont {{Xiong}}, \citenamefont {{Wang}}, \citenamefont {{Huang}},\ and\ \citenamefont {{Gecam Team}}}]{xiong2023GCN}%
  \BibitemOpen
  \bibfield  {author} {\bibinfo {author} {\bibfnamefont {S.}~\bibnamefont {{Xiong}}}, \bibinfo {author} {\bibfnamefont {C.}~\bibnamefont {{Wang}}}, \bibinfo {author} {\bibfnamefont {Y.}~\bibnamefont {{Huang}}},\ and\ \bibinfo {author} {\bibnamefont {{Gecam Team}}},\ }\bibfield  {title} {\bibinfo {title} {{GRB 230307A: GECAM detection of an extremely bright burst}},\ }\href@noop {} {\bibfield  {journal} {\bibinfo  {journal} {GRB Coordinates Network}\ }\textbf {\bibinfo {volume} {33406}},\ \bibinfo {pages} {1} (\bibinfo {year} {2023})}\BibitemShut {NoStop}%
\bibitem [{\citenamefont {{Yi}}\ \emph {et~al.}(2023)\citenamefont {{Yi}}, \citenamefont {{Wang}}, \citenamefont {{Zhang}}, \citenamefont {{Xiong}}, \citenamefont {{Zhang}}, \citenamefont {{Tan}}, \citenamefont {{Liu}}, \citenamefont {{Xue}}, \citenamefont {{Zhang}}, \citenamefont {{Zheng}}, \citenamefont {{Moradi}}, \citenamefont {{Wang}}, \citenamefont {{Zhang}}, \citenamefont {{An}}, \citenamefont {{Cai}}, \citenamefont {{Feng}}, \citenamefont {{Gong}}, \citenamefont {{Guo}}, \citenamefont {{Huang}}, \citenamefont {{Li}}, \citenamefont {{Li}}, \citenamefont {{Li}}, \citenamefont {{Liu}}, \citenamefont {{Liu}}, \citenamefont {{Ma}}, \citenamefont {{Peng}}, \citenamefont {{Qiao}}, \citenamefont {{Song}}, \citenamefont {{Wang}}, \citenamefont {{Wang}}, \citenamefont {{Wang}}, \citenamefont {{Wen}}, \citenamefont {{Xiao}}, \citenamefont {{Xu}}, \citenamefont {{Yang}}, \citenamefont {{Yi}}, \citenamefont {{Zhang}}, \citenamefont {{Zhang}}, \citenamefont {{Zhang}}, \citenamefont {{Zhang}}, \citenamefont
  {{Zhang}}, \citenamefont {{Zhao}}, \citenamefont {{Zhao}},\ and\ \citenamefont {{Zheng}}}]{2023arXiv231007205Y}%
  \BibitemOpen
  \bibfield  {author} {\bibinfo {author} {\bibfnamefont {S.~X.}\ \bibnamefont {{Yi}}}, \bibinfo {author} {\bibfnamefont {C.~W.}\ \bibnamefont {{Wang}}}, \bibinfo {author} {\bibfnamefont {B.}~\bibnamefont {{Zhang}}}, \bibinfo {author} {\bibfnamefont {S.~L.}\ \bibnamefont {{Xiong}}}, \bibinfo {author} {\bibfnamefont {S.~N.}\ \bibnamefont {{Zhang}}}, \bibinfo {author} {\bibfnamefont {W.~J.}\ \bibnamefont {{Tan}}}, \bibinfo {author} {\bibfnamefont {J.~C.}\ \bibnamefont {{Liu}}}, \bibinfo {author} {\bibfnamefont {W.~C.}\ \bibnamefont {{Xue}}}, \bibinfo {author} {\bibfnamefont {Y.~Q.}\ \bibnamefont {{Zhang}}}, \bibinfo {author} {\bibfnamefont {C.}~\bibnamefont {{Zheng}}}, \bibinfo {author} {\bibfnamefont {R.}~\bibnamefont {{Moradi}}}, \bibinfo {author} {\bibfnamefont {Y.}~\bibnamefont {{Wang}}}, \bibinfo {author} {\bibfnamefont {P.}~\bibnamefont {{Zhang}}}, \bibinfo {author} {\bibfnamefont {Z.~H.}\ \bibnamefont {{An}}}, \bibinfo {author} {\bibfnamefont {C.}~\bibnamefont {{Cai}}}, \bibinfo {author} {\bibfnamefont
  {P.~Y.}\ \bibnamefont {{Feng}}}, \bibinfo {author} {\bibfnamefont {K.}~\bibnamefont {{Gong}}}, \bibinfo {author} {\bibfnamefont {D.~Y.}\ \bibnamefont {{Guo}}}, \bibinfo {author} {\bibfnamefont {Y.}~\bibnamefont {{Huang}}}, \bibinfo {author} {\bibfnamefont {B.}~\bibnamefont {{Li}}}, \bibinfo {author} {\bibfnamefont {X.~B.}\ \bibnamefont {{Li}}}, \bibinfo {author} {\bibfnamefont {X.~Q.}\ \bibnamefont {{Li}}}, \bibinfo {author} {\bibfnamefont {X.~J.}\ \bibnamefont {{Liu}}}, \bibinfo {author} {\bibfnamefont {Y.~Q.}\ \bibnamefont {{Liu}}}, \bibinfo {author} {\bibfnamefont {X.}~\bibnamefont {{Ma}}}, \bibinfo {author} {\bibfnamefont {W.~X.}\ \bibnamefont {{Peng}}}, \bibinfo {author} {\bibfnamefont {R.}~\bibnamefont {{Qiao}}}, \bibinfo {author} {\bibfnamefont {L.~M.}\ \bibnamefont {{Song}}}, \bibinfo {author} {\bibfnamefont {J.}~\bibnamefont {{Wang}}}, \bibinfo {author} {\bibfnamefont {P.}~\bibnamefont {{Wang}}}, \bibinfo {author} {\bibfnamefont {Y.}~\bibnamefont {{Wang}}}, \bibinfo {author} {\bibfnamefont {X.~Y.}\
  \bibnamefont {{Wen}}}, \bibinfo {author} {\bibfnamefont {S.}~\bibnamefont {{Xiao}}}, \bibinfo {author} {\bibfnamefont {Y.~B.}\ \bibnamefont {{Xu}}}, \bibinfo {author} {\bibfnamefont {S.}~\bibnamefont {{Yang}}}, \bibinfo {author} {\bibfnamefont {Q.~B.}\ \bibnamefont {{Yi}}}, \bibinfo {author} {\bibfnamefont {D.~L.}\ \bibnamefont {{Zhang}}}, \bibinfo {author} {\bibfnamefont {F.}~\bibnamefont {{Zhang}}}, \bibinfo {author} {\bibfnamefont {H.~M.}\ \bibnamefont {{Zhang}}}, \bibinfo {author} {\bibfnamefont {J.~P.}\ \bibnamefont {{Zhang}}}, \bibinfo {author} {\bibfnamefont {Z.}~\bibnamefont {{Zhang}}}, \bibinfo {author} {\bibfnamefont {X.~Y.}\ \bibnamefont {{Zhao}}}, \bibinfo {author} {\bibfnamefont {Y.}~\bibnamefont {{Zhao}}},\ and\ \bibinfo {author} {\bibfnamefont {S.~J.}\ \bibnamefont {{Zheng}}},\ }\bibfield  {title} {\bibinfo {title} {{Evidence of mini-jet emission in a large emission zone from a magnetically-dominated gamma-ray burst jet}},\ }\href {https://doi.org/10.48550/arXiv.2310.07205} {\bibfield
  {journal} {\bibinfo  {journal} {arXiv e-prints}\ ,\ \bibinfo {eid} {arXiv:2310.07205}} (\bibinfo {year} {2023})},\ \Eprint {https://arxiv.org/abs/2310.07205} {arXiv:2310.07205 [astro-ph.HE]} \BibitemShut {NoStop}%
\bibitem [{\citenamefont {{Peng}}\ \emph {et~al.}(2024{\natexlab{a}})\citenamefont {{Peng}}, \citenamefont {{Chen}},\ and\ \citenamefont {{Mao}}}]{Peng2024compare}%
  \BibitemOpen
  \bibfield  {author} {\bibinfo {author} {\bibfnamefont {Z.-Y.}\ \bibnamefont {{Peng}}}, \bibinfo {author} {\bibfnamefont {J.-M.}\ \bibnamefont {{Chen}}},\ and\ \bibinfo {author} {\bibfnamefont {J.}~\bibnamefont {{Mao}}},\ }\bibfield  {title} {\bibinfo {title} {{A comparative analysis of two peculiar long Gamma-ray bursts: GRB 230307A and GRB 211211A}},\ }\href {https://doi.org/10.48550/arXiv.2404.17913} {\bibfield  {journal} {\bibinfo  {journal} {arXiv e-prints}\ ,\ \bibinfo {eid} {arXiv:2404.17913}} (\bibinfo {year} {2024}{\natexlab{a}})},\ \Eprint {https://arxiv.org/abs/2404.17913} {arXiv:2404.17913 [astro-ph.HE]} \BibitemShut {NoStop}%
\bibitem [{\citenamefont {{Oganesyan}}\ \emph {et~al.}(2018)\citenamefont {{Oganesyan}}, \citenamefont {{Nava}}, \citenamefont {{Ghirlanda}},\ and\ \citenamefont {{Celotti}}}]{Oganesyan2018bandcut}%
  \BibitemOpen
  \bibfield  {author} {\bibinfo {author} {\bibfnamefont {G.}~\bibnamefont {{Oganesyan}}}, \bibinfo {author} {\bibfnamefont {L.}~\bibnamefont {{Nava}}}, \bibinfo {author} {\bibfnamefont {G.}~\bibnamefont {{Ghirlanda}}},\ and\ \bibinfo {author} {\bibfnamefont {A.}~\bibnamefont {{Celotti}}},\ }\bibfield  {title} {\bibinfo {title} {{Characterization of gamma-ray burst prompt emission spectra down to soft X-rays}},\ }\href {https://doi.org/10.1051/0004-6361/201732172} {\bibfield  {journal} {\bibinfo  {journal} {\aap}\ }\textbf {\bibinfo {volume} {616}},\ \bibinfo {eid} {A138} (\bibinfo {year} {2018})},\ \Eprint {https://arxiv.org/abs/1710.09383} {arXiv:1710.09383 [astro-ph.HE]} \BibitemShut {NoStop}%
\bibitem [{\citenamefont {{Troja}}\ \emph {et~al.}(2022)\citenamefont {{Troja}}, \citenamefont {{Fryer}}, \citenamefont {{O'Connor}}, \citenamefont {{Ryan}}, \citenamefont {{Dichiara}}, \citenamefont {{Kumar}}, \citenamefont {{Ito}}, \citenamefont {{Gupta}}, \citenamefont {{Wollaeger}}, \citenamefont {{Norris}}, \citenamefont {{Kawai}}, \citenamefont {{Butler}}, \citenamefont {{Aryan}}, \citenamefont {{Misra}}, \citenamefont {{Hosokawa}}, \citenamefont {{Murata}}, \citenamefont {{Niwano}}, \citenamefont {{Pandey}}, \citenamefont {{Kutyrev}}, \citenamefont {{van Eerten}}, \citenamefont {{Chase}}, \citenamefont {{Hu}}, \citenamefont {{Caballero-Garcia}},\ and\ \citenamefont {{Castro-Tirado}}}]{Troja_KN_11A}%
  \BibitemOpen
  \bibfield  {author} {\bibinfo {author} {\bibfnamefont {E.}~\bibnamefont {{Troja}}}, \bibinfo {author} {\bibfnamefont {C.~L.}\ \bibnamefont {{Fryer}}}, \bibinfo {author} {\bibfnamefont {B.}~\bibnamefont {{O'Connor}}}, \bibinfo {author} {\bibfnamefont {G.}~\bibnamefont {{Ryan}}}, \bibinfo {author} {\bibfnamefont {S.}~\bibnamefont {{Dichiara}}}, \bibinfo {author} {\bibfnamefont {A.}~\bibnamefont {{Kumar}}}, \bibinfo {author} {\bibfnamefont {N.}~\bibnamefont {{Ito}}}, \bibinfo {author} {\bibfnamefont {R.}~\bibnamefont {{Gupta}}}, \bibinfo {author} {\bibfnamefont {R.~T.}\ \bibnamefont {{Wollaeger}}}, \bibinfo {author} {\bibfnamefont {J.~P.}\ \bibnamefont {{Norris}}}, \bibinfo {author} {\bibfnamefont {N.}~\bibnamefont {{Kawai}}}, \bibinfo {author} {\bibfnamefont {N.~R.}\ \bibnamefont {{Butler}}}, \bibinfo {author} {\bibfnamefont {A.}~\bibnamefont {{Aryan}}}, \bibinfo {author} {\bibfnamefont {K.}~\bibnamefont {{Misra}}}, \bibinfo {author} {\bibfnamefont {R.}~\bibnamefont {{Hosokawa}}}, \bibinfo {author}
  {\bibfnamefont {K.~L.}\ \bibnamefont {{Murata}}}, \bibinfo {author} {\bibfnamefont {M.}~\bibnamefont {{Niwano}}}, \bibinfo {author} {\bibfnamefont {S.~B.}\ \bibnamefont {{Pandey}}}, \bibinfo {author} {\bibfnamefont {A.}~\bibnamefont {{Kutyrev}}}, \bibinfo {author} {\bibfnamefont {H.~J.}\ \bibnamefont {{van Eerten}}}, \bibinfo {author} {\bibfnamefont {E.~A.}\ \bibnamefont {{Chase}}}, \bibinfo {author} {\bibfnamefont {Y.~D.}\ \bibnamefont {{Hu}}}, \bibinfo {author} {\bibfnamefont {M.~D.}\ \bibnamefont {{Caballero-Garcia}}},\ and\ \bibinfo {author} {\bibfnamefont {A.~J.}\ \bibnamefont {{Castro-Tirado}}},\ }\bibfield  {title} {\bibinfo {title} {{A nearby long gamma-ray burst from a merger of compact objects}},\ }\href {https://doi.org/10.1038/s41586-022-05327-3} {\bibfield  {journal} {\bibinfo  {journal} {\nat}\ }\textbf {\bibinfo {volume} {612}},\ \bibinfo {pages} {228} (\bibinfo {year} {2022})},\ \Eprint {https://arxiv.org/abs/2209.03363} {arXiv:2209.03363 [astro-ph.HE]} \BibitemShut {NoStop}%
\bibitem [{\citenamefont {{Song}}\ and\ \citenamefont {{Zhang}}(2023)}]{SXY_precursor_09A}%
  \BibitemOpen
  \bibfield  {author} {\bibinfo {author} {\bibfnamefont {X.-Y.}\ \bibnamefont {{Song}}}\ and\ \bibinfo {author} {\bibfnamefont {S.-N.}\ \bibnamefont {{Zhang}}},\ }\bibfield  {title} {\bibinfo {title} {{GRB 221009A with an Unconventional Precursor: A Typical Two-stage Collapsar Scenario?}},\ }\href {https://doi.org/10.3847/1538-4357/acfed7} {\bibfield  {journal} {\bibinfo  {journal} {\apj}\ }\textbf {\bibinfo {volume} {957}},\ \bibinfo {eid} {31} (\bibinfo {year} {2023})},\ \Eprint {https://arxiv.org/abs/2307.07104} {arXiv:2307.07104 [astro-ph.HE]} \BibitemShut {NoStop}%
\bibitem [{\citenamefont {{Zhang}}\ \emph {et~al.}(2018)\citenamefont {{Zhang}}, \citenamefont {{Zhang}}, \citenamefont {{Castro-Tirado}}, \citenamefont {{Dai}}, \citenamefont {{Tam}}, \citenamefont {{Wang}}, \citenamefont {{Hu}}, \citenamefont {{Karpov}}, \citenamefont {{Pozanenko}}, \citenamefont {{Zhang}}, \citenamefont {{Mazaeva}}, \citenamefont {{Minaev}}, \citenamefont {{Volnova}}, \citenamefont {{Oates}}, \citenamefont {{Gao}}, \citenamefont {{Wu}}, \citenamefont {{Shao}}, \citenamefont {{Tang}}, \citenamefont {{Beskin}}, \citenamefont {{Biryukov}}, \citenamefont {{Bondar}}, \citenamefont {{Ivanov}}, \citenamefont {{Katkova}}, \citenamefont {{Orekhova}}, \citenamefont {{Perkov}}, \citenamefont {{Sasyuk}}, \citenamefont {{Mankiewicz}}, \citenamefont {{{\.Z}arnecki}}, \citenamefont {{Cwiek}}, \citenamefont {{Opiela}}, \citenamefont {{Zadro{\.Z}ny}}, \citenamefont {{Aptekar}}, \citenamefont {{Frederiks}}, \citenamefont {{Svinkin}}, \citenamefont {{Kusakin}}, \citenamefont {{Inasaridze}}, \citenamefont
  {{Burhonov}}, \citenamefont {{Rumyantsev}}, \citenamefont {{Klunko}}, \citenamefont {{Moskvitin}}, \citenamefont {{Fatkhullin}}, \citenamefont {{Sokolov}}, \citenamefont {{Valeev}}, \citenamefont {{Jeong}}, \citenamefont {{Park}}, \citenamefont {{Caballero-Garc{\'\i}a}}, \citenamefont {{Cunniffe}}, \citenamefont {{Tello}}, \citenamefont {{Ferrero}}, \citenamefont {{Pandey}}, \citenamefont {{Jel{\'\i}nek}}, \citenamefont {{Peng}}, \citenamefont {{S{\'a}nchez-Ram{\'\i}rez}},\ and\ \citenamefont {{Castell{\'o}n}}}]{Zhang160625B}%
  \BibitemOpen
  \bibfield  {author} {\bibinfo {author} {\bibfnamefont {B.~B.}\ \bibnamefont {{Zhang}}}, \bibinfo {author} {\bibfnamefont {B.}~\bibnamefont {{Zhang}}}, \bibinfo {author} {\bibfnamefont {A.~J.}\ \bibnamefont {{Castro-Tirado}}}, \bibinfo {author} {\bibfnamefont {Z.~G.}\ \bibnamefont {{Dai}}}, \bibinfo {author} {\bibfnamefont {P.~H.~T.}\ \bibnamefont {{Tam}}}, \bibinfo {author} {\bibfnamefont {X.~Y.}\ \bibnamefont {{Wang}}}, \bibinfo {author} {\bibfnamefont {Y.~D.}\ \bibnamefont {{Hu}}}, \bibinfo {author} {\bibfnamefont {S.}~\bibnamefont {{Karpov}}}, \bibinfo {author} {\bibfnamefont {A.}~\bibnamefont {{Pozanenko}}}, \bibinfo {author} {\bibfnamefont {F.~W.}\ \bibnamefont {{Zhang}}}, \bibinfo {author} {\bibfnamefont {E.}~\bibnamefont {{Mazaeva}}}, \bibinfo {author} {\bibfnamefont {P.}~\bibnamefont {{Minaev}}}, \bibinfo {author} {\bibfnamefont {A.}~\bibnamefont {{Volnova}}}, \bibinfo {author} {\bibfnamefont {S.}~\bibnamefont {{Oates}}}, \bibinfo {author} {\bibfnamefont {H.}~\bibnamefont {{Gao}}}, \bibinfo {author}
  {\bibfnamefont {X.~F.}\ \bibnamefont {{Wu}}}, \bibinfo {author} {\bibfnamefont {L.}~\bibnamefont {{Shao}}}, \bibinfo {author} {\bibfnamefont {Q.~W.}\ \bibnamefont {{Tang}}}, \bibinfo {author} {\bibfnamefont {G.}~\bibnamefont {{Beskin}}}, \bibinfo {author} {\bibfnamefont {A.}~\bibnamefont {{Biryukov}}}, \bibinfo {author} {\bibfnamefont {S.}~\bibnamefont {{Bondar}}}, \bibinfo {author} {\bibfnamefont {E.}~\bibnamefont {{Ivanov}}}, \bibinfo {author} {\bibfnamefont {E.}~\bibnamefont {{Katkova}}}, \bibinfo {author} {\bibfnamefont {N.}~\bibnamefont {{Orekhova}}}, \bibinfo {author} {\bibfnamefont {A.}~\bibnamefont {{Perkov}}}, \bibinfo {author} {\bibfnamefont {V.}~\bibnamefont {{Sasyuk}}}, \bibinfo {author} {\bibfnamefont {L.}~\bibnamefont {{Mankiewicz}}}, \bibinfo {author} {\bibfnamefont {A.~F.}\ \bibnamefont {{{\.Z}arnecki}}}, \bibinfo {author} {\bibfnamefont {A.}~\bibnamefont {{Cwiek}}}, \bibinfo {author} {\bibfnamefont {R.}~\bibnamefont {{Opiela}}}, \bibinfo {author} {\bibfnamefont {A.}~\bibnamefont
  {{Zadro{\.Z}ny}}}, \bibinfo {author} {\bibfnamefont {R.}~\bibnamefont {{Aptekar}}}, \bibinfo {author} {\bibfnamefont {D.}~\bibnamefont {{Frederiks}}}, \bibinfo {author} {\bibfnamefont {D.}~\bibnamefont {{Svinkin}}}, \bibinfo {author} {\bibfnamefont {A.}~\bibnamefont {{Kusakin}}}, \bibinfo {author} {\bibfnamefont {R.}~\bibnamefont {{Inasaridze}}}, \bibinfo {author} {\bibfnamefont {O.}~\bibnamefont {{Burhonov}}}, \bibinfo {author} {\bibfnamefont {V.}~\bibnamefont {{Rumyantsev}}}, \bibinfo {author} {\bibfnamefont {E.}~\bibnamefont {{Klunko}}}, \bibinfo {author} {\bibfnamefont {A.}~\bibnamefont {{Moskvitin}}}, \bibinfo {author} {\bibfnamefont {T.}~\bibnamefont {{Fatkhullin}}}, \bibinfo {author} {\bibfnamefont {V.~V.}\ \bibnamefont {{Sokolov}}}, \bibinfo {author} {\bibfnamefont {A.~F.}\ \bibnamefont {{Valeev}}}, \bibinfo {author} {\bibfnamefont {S.}~\bibnamefont {{Jeong}}}, \bibinfo {author} {\bibfnamefont {I.~H.}\ \bibnamefont {{Park}}}, \bibinfo {author} {\bibfnamefont {M.~D.}\ \bibnamefont
  {{Caballero-Garc{\'\i}a}}}, \bibinfo {author} {\bibfnamefont {R.}~\bibnamefont {{Cunniffe}}}, \bibinfo {author} {\bibfnamefont {J.~C.}\ \bibnamefont {{Tello}}}, \bibinfo {author} {\bibfnamefont {P.}~\bibnamefont {{Ferrero}}}, \bibinfo {author} {\bibfnamefont {S.~B.}\ \bibnamefont {{Pandey}}}, \bibinfo {author} {\bibfnamefont {M.}~\bibnamefont {{Jel{\'\i}nek}}}, \bibinfo {author} {\bibfnamefont {F.~K.}\ \bibnamefont {{Peng}}}, \bibinfo {author} {\bibfnamefont {R.}~\bibnamefont {{S{\'a}nchez-Ram{\'\i}rez}}},\ and\ \bibinfo {author} {\bibfnamefont {A.}~\bibnamefont {{Castell{\'o}n}}},\ }\bibfield  {title} {\bibinfo {title} {{Transition from fireball to Poynting-flux-dominated outflow in the three-episode GRB 160625B}},\ }\href {https://doi.org/10.1038/s41550-017-0309-8} {\bibfield  {journal} {\bibinfo  {journal} {Nature Astronomy}\ }\textbf {\bibinfo {volume} {2}},\ \bibinfo {pages} {69} (\bibinfo {year} {2018})},\ \Eprint {https://arxiv.org/abs/1612.03089} {arXiv:1612.03089 [astro-ph.HE]} \BibitemShut
  {NoStop}%
\bibitem [{\citenamefont {{An}}\ \emph {et~al.}(2023)\citenamefont {{An}}, \citenamefont {{Antier}}, \citenamefont {{Bi}}, \citenamefont {{Bu}}, \citenamefont {{Cai}}, \citenamefont {{Cao}}, \citenamefont {{Camisasca}}, \citenamefont {{Chang}}, \citenamefont {{Chen}}, \citenamefont {{Chen}}, \citenamefont {{Chen}}, \citenamefont {{Chen}}, \citenamefont {{Chen}}, \citenamefont {{Chen}}, \citenamefont {{Chen}}, \citenamefont {{Coughlin}}, \citenamefont {{Cui}}, \citenamefont {{Dai}}, \citenamefont {{Hussenot-Desenonges}}, \citenamefont {{Du}}, \citenamefont {{Du}}, \citenamefont {{Du}}, \citenamefont {{Fan}}, \citenamefont {{Frontera}}, \citenamefont {{Gao}}, \citenamefont {{Gao}}, \citenamefont {{Ge}}, \citenamefont {{Gong}}, \citenamefont {{Gu}}, \citenamefont {{Guan}}, \citenamefont {{Guo}}, \citenamefont {{Guo}}, \citenamefont {{Guidorzi}}, \citenamefont {{Han}}, \citenamefont {{He}}, \citenamefont {{He}}, \citenamefont {{Hou}}, \citenamefont {{Huang}}, \citenamefont {{Huo}}, \citenamefont {{Ji}},
  \citenamefont {{Jia}}, \citenamefont {{Jiang}}, \citenamefont {{Kann}}, \citenamefont {{Klotz}}, \citenamefont {{Kong}}, \citenamefont {{Lan}}, \citenamefont {{Li}}, \citenamefont {{Li}}, \citenamefont {{Li}}, \citenamefont {{Li}}, \citenamefont {{Li}}, \citenamefont {{Li}}, \citenamefont {{Li}}, \citenamefont {{Li}}, \citenamefont {{Li}}, \citenamefont {{Li}}, \citenamefont {{Li}}, \citenamefont {{Li}}, \citenamefont {{Li}}, \citenamefont {{Liang}}, \citenamefont {{Liang}}, \citenamefont {{Liao}}, \citenamefont {{Lin}}, \citenamefont {{Liu}}, \citenamefont {{Liu}}, \citenamefont {{Liu}}, \citenamefont {{Liu}}, \citenamefont {{Liu}}, \citenamefont {{Liu}}, \citenamefont {{Liu}}, \citenamefont {{Lu}}, \citenamefont {{Lu}}, \citenamefont {{Lu}}, \citenamefont {{Luo}}, \citenamefont {{Luo}}, \citenamefont {{Ma}}, \citenamefont {{Ma}}, \citenamefont {{Ma}}, \citenamefont {{Ma}}, \citenamefont {{Maccary}}, \citenamefont {{Mao}}, \citenamefont {{Meng}}, \citenamefont {{Nie}}, \citenamefont {{Orlandini}},
  \citenamefont {{Ou}}, \citenamefont {{Peng}}, \citenamefont {{Peng}}, \citenamefont {{Qiao}}, \citenamefont {{Qu}}, \citenamefont {{Ren}}, \citenamefont {{Shi}}, \citenamefont {{Shi}}, \citenamefont {{Song}}, \citenamefont {{Song}}, \citenamefont {{Su}}, \citenamefont {{Sun}}, \citenamefont {{Sun}}, \citenamefont {{Sun}}, \citenamefont {{Tan}}, \citenamefont {{Tan}}, \citenamefont {{Tao}}, \citenamefont {{Tuo}}, \citenamefont {{Turpin}}, \citenamefont {{Wang}}, \citenamefont {{Wang}}, \citenamefont {{Wang}}, \citenamefont {{Wang}}, \citenamefont {{Wang}}, \citenamefont {{Wang}}, \citenamefont {{Wang}}, \citenamefont {{Wang}}, \citenamefont {{Wang}}, \citenamefont {{Wang}}, \citenamefont {{Wang}}, \citenamefont {{Wang}}, \citenamefont {{Wang}}, \citenamefont {{Wang}}, \citenamefont {{Wen}}, \citenamefont {{Wu}}, \citenamefont {{Wu}}, \citenamefont {{Wu}}, \citenamefont {{Xiao}}, \citenamefont {{Xiao}}, \citenamefont {{Xiao}}, \citenamefont {{Xie}}, \citenamefont {{Xiong}}, \citenamefont {{Xiong}},
  \citenamefont {{Xu}}, \citenamefont {{Xu}}, \citenamefont {{Xu}}, \citenamefont {{Xu}}, \citenamefont {{Xu}}, \citenamefont {{Xu}}, \citenamefont {{Xue}}, \citenamefont {{Yang}}, \citenamefont {{Yang}}, \citenamefont {{Yang}}, \citenamefont {{Ye}}, \citenamefont {{Yi}}, \citenamefont {{Yi}}, \citenamefont {{Yin}}, \citenamefont {{You}}, \citenamefont {{Yu}}, \citenamefont {{Yu}}, \citenamefont {{Yu}}, \citenamefont {{Zeng}}, \citenamefont {{Zhang}}, \citenamefont {{Zhang}}, \citenamefont {{Zhang}}, \citenamefont {{Zhang}}, \citenamefont {{Zhang}}, \citenamefont {{Zhang}}, \citenamefont {{Zhang}}, \citenamefont {{Zhang}}, \citenamefont {{Zhang}}, \citenamefont {{Zhang}}, \citenamefont {{Zhang}}, \citenamefont {{Zhang}}, \citenamefont {{Zhang}}, \citenamefont {{Zhang}}, \citenamefont {{Zhang}}, \citenamefont {{Zhang}}, \citenamefont {{Zhang}}, \citenamefont {{Zhang}}, \citenamefont {{Zhang}}, \citenamefont {{Zhao}}, \citenamefont {{Zhao}}, \citenamefont {{Zhao}}, \citenamefont {{Zhao}}, \citenamefont
  {{Zhao}}, \citenamefont {{Zhao}}, \citenamefont {{Zhao}}, \citenamefont {{Zhao}}, \citenamefont {{Zheng}}, \citenamefont {{Zheng}}, \citenamefont {{Zhou}}, \citenamefont {{Zhou}},\ and\ \citenamefont {{Zhu}}}]{HXMT_GECAM_221009A}%
  \BibitemOpen
  \bibfield  {author} {\bibinfo {author} {\bibfnamefont {Z.-H.}\ \bibnamefont {{An}}}, \bibinfo {author} {\bibfnamefont {S.}~\bibnamefont {{Antier}}}, \bibinfo {author} {\bibfnamefont {X.-Z.}\ \bibnamefont {{Bi}}}, \bibinfo {author} {\bibfnamefont {Q.-C.}\ \bibnamefont {{Bu}}}, \bibinfo {author} {\bibfnamefont {C.}~\bibnamefont {{Cai}}}, \bibinfo {author} {\bibfnamefont {X.-L.}\ \bibnamefont {{Cao}}}, \bibinfo {author} {\bibfnamefont {A.-E.}\ \bibnamefont {{Camisasca}}}, \bibinfo {author} {\bibfnamefont {Z.}~\bibnamefont {{Chang}}}, \bibinfo {author} {\bibfnamefont {G.}~\bibnamefont {{Chen}}}, \bibinfo {author} {\bibfnamefont {L.}~\bibnamefont {{Chen}}}, \bibinfo {author} {\bibfnamefont {T.-X.}\ \bibnamefont {{Chen}}}, \bibinfo {author} {\bibfnamefont {W.}~\bibnamefont {{Chen}}}, \bibinfo {author} {\bibfnamefont {Y.-B.}\ \bibnamefont {{Chen}}}, \bibinfo {author} {\bibfnamefont {Y.}~\bibnamefont {{Chen}}}, \bibinfo {author} {\bibfnamefont {Y.-P.}\ \bibnamefont {{Chen}}}, \bibinfo {author} {\bibfnamefont
  {M.~W.}\ \bibnamefont {{Coughlin}}}, \bibinfo {author} {\bibfnamefont {W.-W.}\ \bibnamefont {{Cui}}}, \bibinfo {author} {\bibfnamefont {Z.-G.}\ \bibnamefont {{Dai}}}, \bibinfo {author} {\bibfnamefont {T.}~\bibnamefont {{Hussenot-Desenonges}}}, \bibinfo {author} {\bibfnamefont {Y.-Q.}\ \bibnamefont {{Du}}}, \bibinfo {author} {\bibfnamefont {Y.-Y.}\ \bibnamefont {{Du}}}, \bibinfo {author} {\bibfnamefont {Y.-F.}\ \bibnamefont {{Du}}}, \bibinfo {author} {\bibfnamefont {C.-C.}\ \bibnamefont {{Fan}}}, \bibinfo {author} {\bibfnamefont {F.}~\bibnamefont {{Frontera}}}, \bibinfo {author} {\bibfnamefont {H.}~\bibnamefont {{Gao}}}, \bibinfo {author} {\bibfnamefont {M.}~\bibnamefont {{Gao}}}, \bibinfo {author} {\bibfnamefont {M.-Y.}\ \bibnamefont {{Ge}}}, \bibinfo {author} {\bibfnamefont {K.}~\bibnamefont {{Gong}}}, \bibinfo {author} {\bibfnamefont {Y.-D.}\ \bibnamefont {{Gu}}}, \bibinfo {author} {\bibfnamefont {J.}~\bibnamefont {{Guan}}}, \bibinfo {author} {\bibfnamefont {D.-Y.}\ \bibnamefont {{Guo}}}, \bibinfo
  {author} {\bibfnamefont {Z.-W.}\ \bibnamefont {{Guo}}}, \bibinfo {author} {\bibfnamefont {C.}~\bibnamefont {{Guidorzi}}}, \bibinfo {author} {\bibfnamefont {D.-W.}\ \bibnamefont {{Han}}}, \bibinfo {author} {\bibfnamefont {J.-J.}\ \bibnamefont {{He}}}, \bibinfo {author} {\bibfnamefont {J.-W.}\ \bibnamefont {{He}}}, \bibinfo {author} {\bibfnamefont {D.-J.}\ \bibnamefont {{Hou}}}, \bibinfo {author} {\bibfnamefont {Y.}~\bibnamefont {{Huang}}}, \bibinfo {author} {\bibfnamefont {J.}~\bibnamefont {{Huo}}}, \bibinfo {author} {\bibfnamefont {Z.}~\bibnamefont {{Ji}}}, \bibinfo {author} {\bibfnamefont {S.-M.}\ \bibnamefont {{Jia}}}, \bibinfo {author} {\bibfnamefont {W.-C.}\ \bibnamefont {{Jiang}}}, \bibinfo {author} {\bibfnamefont {D.~A.}\ \bibnamefont {{Kann}}}, \bibinfo {author} {\bibfnamefont {A.}~\bibnamefont {{Klotz}}}, \bibinfo {author} {\bibfnamefont {L.-D.}\ \bibnamefont {{Kong}}}, \bibinfo {author} {\bibfnamefont {L.}~\bibnamefont {{Lan}}}, \bibinfo {author} {\bibfnamefont {A.}~\bibnamefont {{Li}}}, \bibinfo
  {author} {\bibfnamefont {B.}~\bibnamefont {{Li}}}, \bibinfo {author} {\bibfnamefont {C.-Y.}\ \bibnamefont {{Li}}}, \bibinfo {author} {\bibfnamefont {C.-K.}\ \bibnamefont {{Li}}}, \bibinfo {author} {\bibfnamefont {G.}~\bibnamefont {{Li}}}, \bibinfo {author} {\bibfnamefont {M.-S.}\ \bibnamefont {{Li}}}, \bibinfo {author} {\bibfnamefont {T.-P.}\ \bibnamefont {{Li}}}, \bibinfo {author} {\bibfnamefont {W.}~\bibnamefont {{Li}}}, \bibinfo {author} {\bibfnamefont {X.-B.}\ \bibnamefont {{Li}}}, \bibinfo {author} {\bibfnamefont {X.-Q.}\ \bibnamefont {{Li}}}, \bibinfo {author} {\bibfnamefont {X.-F.}\ \bibnamefont {{Li}}}, \bibinfo {author} {\bibfnamefont {Y.-G.}\ \bibnamefont {{Li}}}, \bibinfo {author} {\bibfnamefont {Z.-W.}\ \bibnamefont {{Li}}}, \bibinfo {author} {\bibfnamefont {J.}~\bibnamefont {{Liang}}}, \bibinfo {author} {\bibfnamefont {X.-H.}\ \bibnamefont {{Liang}}}, \bibinfo {author} {\bibfnamefont {J.-Y.}\ \bibnamefont {{Liao}}}, \bibinfo {author} {\bibfnamefont {L.}~\bibnamefont {{Lin}}}, \bibinfo {author}
  {\bibfnamefont {C.-Z.}\ \bibnamefont {{Liu}}}, \bibinfo {author} {\bibfnamefont {H.-X.}\ \bibnamefont {{Liu}}}, \bibinfo {author} {\bibfnamefont {H.-W.}\ \bibnamefont {{Liu}}}, \bibinfo {author} {\bibfnamefont {J.-C.}\ \bibnamefont {{Liu}}}, \bibinfo {author} {\bibfnamefont {X.-J.}\ \bibnamefont {{Liu}}}, \bibinfo {author} {\bibfnamefont {Y.-Q.}\ \bibnamefont {{Liu}}}, \bibinfo {author} {\bibfnamefont {Y.-R.}\ \bibnamefont {{Liu}}}, \bibinfo {author} {\bibfnamefont {F.-J.}\ \bibnamefont {{Lu}}}, \bibinfo {author} {\bibfnamefont {H.}~\bibnamefont {{Lu}}}, \bibinfo {author} {\bibfnamefont {X.-F.}\ \bibnamefont {{Lu}}}, \bibinfo {author} {\bibfnamefont {Q.}~\bibnamefont {{Luo}}}, \bibinfo {author} {\bibfnamefont {T.}~\bibnamefont {{Luo}}}, \bibinfo {author} {\bibfnamefont {B.-Y.}\ \bibnamefont {{Ma}}}, \bibinfo {author} {\bibfnamefont {F.-L.}\ \bibnamefont {{Ma}}}, \bibinfo {author} {\bibfnamefont {R.-C.}\ \bibnamefont {{Ma}}}, \bibinfo {author} {\bibfnamefont {X.}~\bibnamefont {{Ma}}}, \bibinfo {author}
  {\bibfnamefont {R.}~\bibnamefont {{Maccary}}}, \bibinfo {author} {\bibfnamefont {J.-R.}\ \bibnamefont {{Mao}}}, \bibinfo {author} {\bibfnamefont {B.}~\bibnamefont {{Meng}}}, \bibinfo {author} {\bibfnamefont {J.-Y.}\ \bibnamefont {{Nie}}}, \bibinfo {author} {\bibfnamefont {M.}~\bibnamefont {{Orlandini}}}, \bibinfo {author} {\bibfnamefont {G.}~\bibnamefont {{Ou}}}, \bibinfo {author} {\bibfnamefont {J.-Q.}\ \bibnamefont {{Peng}}}, \bibinfo {author} {\bibfnamefont {W.-X.}\ \bibnamefont {{Peng}}}, \bibinfo {author} {\bibfnamefont {R.}~\bibnamefont {{Qiao}}}, \bibinfo {author} {\bibfnamefont {J.-L.}\ \bibnamefont {{Qu}}}, \bibinfo {author} {\bibfnamefont {X.-Q.}\ \bibnamefont {{Ren}}}, \bibinfo {author} {\bibfnamefont {J.-Y.}\ \bibnamefont {{Shi}}}, \bibinfo {author} {\bibfnamefont {Q.}~\bibnamefont {{Shi}}}, \bibinfo {author} {\bibfnamefont {L.-M.}\ \bibnamefont {{Song}}}, \bibinfo {author} {\bibfnamefont {X.-Y.}\ \bibnamefont {{Song}}}, \bibinfo {author} {\bibfnamefont {J.}~\bibnamefont {{Su}}}, \bibinfo
  {author} {\bibfnamefont {G.-X.}\ \bibnamefont {{Sun}}}, \bibinfo {author} {\bibfnamefont {L.}~\bibnamefont {{Sun}}}, \bibinfo {author} {\bibfnamefont {X.-L.}\ \bibnamefont {{Sun}}}, \bibinfo {author} {\bibfnamefont {W.-J.}\ \bibnamefont {{Tan}}}, \bibinfo {author} {\bibfnamefont {Y.}~\bibnamefont {{Tan}}}, \bibinfo {author} {\bibfnamefont {L.}~\bibnamefont {{Tao}}}, \bibinfo {author} {\bibfnamefont {Y.-L.}\ \bibnamefont {{Tuo}}}, \bibinfo {author} {\bibfnamefont {D.}~\bibnamefont {{Turpin}}}, \bibinfo {author} {\bibfnamefont {J.-Z.}\ \bibnamefont {{Wang}}}, \bibinfo {author} {\bibfnamefont {C.}~\bibnamefont {{Wang}}}, \bibinfo {author} {\bibfnamefont {C.-W.}\ \bibnamefont {{Wang}}}, \bibinfo {author} {\bibfnamefont {H.-J.}\ \bibnamefont {{Wang}}}, \bibinfo {author} {\bibfnamefont {H.}~\bibnamefont {{Wang}}}, \bibinfo {author} {\bibfnamefont {J.}~\bibnamefont {{Wang}}}, \bibinfo {author} {\bibfnamefont {L.-J.}\ \bibnamefont {{Wang}}}, \bibinfo {author} {\bibfnamefont {P.-J.}\ \bibnamefont {{Wang}}}, \bibinfo
  {author} {\bibfnamefont {P.}~\bibnamefont {{Wang}}}, \bibinfo {author} {\bibfnamefont {W.-S.}\ \bibnamefont {{Wang}}}, \bibinfo {author} {\bibfnamefont {X.-Y.}\ \bibnamefont {{Wang}}}, \bibinfo {author} {\bibfnamefont {X.-L.}\ \bibnamefont {{Wang}}}, \bibinfo {author} {\bibfnamefont {Y.-S.}\ \bibnamefont {{Wang}}}, \bibinfo {author} {\bibfnamefont {Y.}~\bibnamefont {{Wang}}}, \bibinfo {author} {\bibfnamefont {X.-Y.}\ \bibnamefont {{Wen}}}, \bibinfo {author} {\bibfnamefont {B.-B.}\ \bibnamefont {{Wu}}}, \bibinfo {author} {\bibfnamefont {B.-Y.}\ \bibnamefont {{Wu}}}, \bibinfo {author} {\bibfnamefont {H.}~\bibnamefont {{Wu}}}, \bibinfo {author} {\bibfnamefont {S.-H.}\ \bibnamefont {{Xiao}}}, \bibinfo {author} {\bibfnamefont {S.}~\bibnamefont {{Xiao}}}, \bibinfo {author} {\bibfnamefont {Y.-X.}\ \bibnamefont {{Xiao}}}, \bibinfo {author} {\bibfnamefont {S.-L.}\ \bibnamefont {{Xie}}}, \bibinfo {author} {\bibfnamefont {S.-L.}\ \bibnamefont {{Xiong}}}, \bibinfo {author} {\bibfnamefont {S.-L.}\ \bibnamefont
  {{Xiong}}}, \bibinfo {author} {\bibfnamefont {D.}~\bibnamefont {{Xu}}}, \bibinfo {author} {\bibfnamefont {H.}~\bibnamefont {{Xu}}}, \bibinfo {author} {\bibfnamefont {Y.-J.}\ \bibnamefont {{Xu}}}, \bibinfo {author} {\bibfnamefont {Y.-B.}\ \bibnamefont {{Xu}}}, \bibinfo {author} {\bibfnamefont {Y.-C.}\ \bibnamefont {{Xu}}}, \bibinfo {author} {\bibfnamefont {Y.-P.}\ \bibnamefont {{Xu}}}, \bibinfo {author} {\bibfnamefont {W.-C.}\ \bibnamefont {{Xue}}}, \bibinfo {author} {\bibfnamefont {S.}~\bibnamefont {{Yang}}}, \bibinfo {author} {\bibfnamefont {Y.-J.}\ \bibnamefont {{Yang}}}, \bibinfo {author} {\bibfnamefont {Z.-X.}\ \bibnamefont {{Yang}}}, \bibinfo {author} {\bibfnamefont {W.-T.}\ \bibnamefont {{Ye}}}, \bibinfo {author} {\bibfnamefont {Q.-B.}\ \bibnamefont {{Yi}}}, \bibinfo {author} {\bibfnamefont {S.-X.}\ \bibnamefont {{Yi}}}, \bibinfo {author} {\bibfnamefont {Q.-Q.}\ \bibnamefont {{Yin}}}, \bibinfo {author} {\bibfnamefont {Y.}~\bibnamefont {{You}}}, \bibinfo {author} {\bibfnamefont {Y.-W.}\ \bibnamefont
  {{Yu}}}, \bibinfo {author} {\bibfnamefont {W.}~\bibnamefont {{Yu}}}, \bibinfo {author} {\bibfnamefont {W.-H.}\ \bibnamefont {{Yu}}}, \bibinfo {author} {\bibfnamefont {M.}~\bibnamefont {{Zeng}}}, \bibinfo {author} {\bibfnamefont {B.}~\bibnamefont {{Zhang}}}, \bibinfo {author} {\bibfnamefont {B.-B.}\ \bibnamefont {{Zhang}}}, \bibinfo {author} {\bibfnamefont {D.-L.}\ \bibnamefont {{Zhang}}}, \bibinfo {author} {\bibfnamefont {F.}~\bibnamefont {{Zhang}}}, \bibinfo {author} {\bibfnamefont {H.-M.}\ \bibnamefont {{Zhang}}}, \bibinfo {author} {\bibfnamefont {J.}~\bibnamefont {{Zhang}}}, \bibinfo {author} {\bibfnamefont {L.}~\bibnamefont {{Zhang}}}, \bibinfo {author} {\bibfnamefont {P.}~\bibnamefont {{Zhang}}}, \bibinfo {author} {\bibfnamefont {P.}~\bibnamefont {{Zhang}}}, \bibinfo {author} {\bibfnamefont {S.}~\bibnamefont {{Zhang}}}, \bibinfo {author} {\bibfnamefont {S.-N.}\ \bibnamefont {{Zhang}}}, \bibinfo {author} {\bibfnamefont {W.-C.}\ \bibnamefont {{Zhang}}}, \bibinfo {author} {\bibfnamefont {X.-F.}\
  \bibnamefont {{Zhang}}}, \bibinfo {author} {\bibfnamefont {X.-L.}\ \bibnamefont {{Zhang}}}, \bibinfo {author} {\bibfnamefont {Y.-Q.}\ \bibnamefont {{Zhang}}}, \bibinfo {author} {\bibfnamefont {Y.-T.}\ \bibnamefont {{Zhang}}}, \bibinfo {author} {\bibfnamefont {Y.-F.}\ \bibnamefont {{Zhang}}}, \bibinfo {author} {\bibfnamefont {Y.-H.}\ \bibnamefont {{Zhang}}}, \bibinfo {author} {\bibfnamefont {Z.}~\bibnamefont {{Zhang}}}, \bibinfo {author} {\bibfnamefont {G.-Y.}\ \bibnamefont {{Zhao}}}, \bibinfo {author} {\bibfnamefont {H.-S.}\ \bibnamefont {{Zhao}}}, \bibinfo {author} {\bibfnamefont {H.-Y.}\ \bibnamefont {{Zhao}}}, \bibinfo {author} {\bibfnamefont {Q.-X.}\ \bibnamefont {{Zhao}}}, \bibinfo {author} {\bibfnamefont {S.-J.}\ \bibnamefont {{Zhao}}}, \bibinfo {author} {\bibfnamefont {X.-Y.}\ \bibnamefont {{Zhao}}}, \bibinfo {author} {\bibfnamefont {X.-F.}\ \bibnamefont {{Zhao}}}, \bibinfo {author} {\bibfnamefont {Y.}~\bibnamefont {{Zhao}}}, \bibinfo {author} {\bibfnamefont {C.}~\bibnamefont {{Zheng}}}, \bibinfo
  {author} {\bibfnamefont {S.-J.}\ \bibnamefont {{Zheng}}}, \bibinfo {author} {\bibfnamefont {D.-K.}\ \bibnamefont {{Zhou}}}, \bibinfo {author} {\bibfnamefont {X.}~\bibnamefont {{Zhou}}},\ and\ \bibinfo {author} {\bibfnamefont {X.-C.}\ \bibnamefont {{Zhu}}},\ }\bibfield  {title} {\bibinfo {title} {{Insight-HXMT and GECAM-C observations of the brightest-of-all-time GRB 221009A}},\ }\href {https://doi.org/10.48550/arXiv.2303.01203} {\bibfield  {journal} {\bibinfo  {journal} {arXiv e-prints}\ ,\ \bibinfo {eid} {arXiv:2303.01203}} (\bibinfo {year} {2023})},\ \Eprint {https://arxiv.org/abs/2303.01203} {arXiv:2303.01203 [astro-ph.HE]} \BibitemShut {NoStop}%
\bibitem [{\citenamefont {{Dichiara}}\ \emph {et~al.}(2023)\citenamefont {{Dichiara}}, \citenamefont {{Tsang}}, \citenamefont {{Troja}}, \citenamefont {{Neill}}, \citenamefont {{Norris}},\ and\ \citenamefont {{Yang}}}]{2023Dichiara}%
  \BibitemOpen
  \bibfield  {author} {\bibinfo {author} {\bibfnamefont {S.}~\bibnamefont {{Dichiara}}}, \bibinfo {author} {\bibfnamefont {D.}~\bibnamefont {{Tsang}}}, \bibinfo {author} {\bibfnamefont {E.}~\bibnamefont {{Troja}}}, \bibinfo {author} {\bibfnamefont {D.}~\bibnamefont {{Neill}}}, \bibinfo {author} {\bibfnamefont {J.~P.}\ \bibnamefont {{Norris}}},\ and\ \bibinfo {author} {\bibfnamefont {Y.~H.}\ \bibnamefont {{Yang}}},\ }\bibfield  {title} {\bibinfo {title} {{A Luminous Precursor in the Extremely Bright GRB 230307A}},\ }\href {https://doi.org/10.3847/2041-8213/acf21d} {\bibfield  {journal} {\bibinfo  {journal} {\apjl}\ }\textbf {\bibinfo {volume} {954}},\ \bibinfo {eid} {L29} (\bibinfo {year} {2023})},\ \Eprint {https://arxiv.org/abs/2307.02996} {arXiv:2307.02996 [astro-ph.HE]} \BibitemShut {NoStop}%
\bibitem [{\citenamefont {{Troja}}\ \emph {et~al.}(2010)\citenamefont {{Troja}}, \citenamefont {{Rosswog}},\ and\ \citenamefont {{Gehrels}}}]{2010ApJ...723.1711T}%
  \BibitemOpen
  \bibfield  {author} {\bibinfo {author} {\bibfnamefont {E.}~\bibnamefont {{Troja}}}, \bibinfo {author} {\bibfnamefont {S.}~\bibnamefont {{Rosswog}}},\ and\ \bibinfo {author} {\bibfnamefont {N.}~\bibnamefont {{Gehrels}}},\ }\bibfield  {title} {\bibinfo {title} {{Precursors of Short Gamma-ray Bursts}},\ }\href {https://doi.org/10.1088/0004-637X/723/2/1711} {\bibfield  {journal} {\bibinfo  {journal} {\apj}\ }\textbf {\bibinfo {volume} {723}},\ \bibinfo {pages} {1711} (\bibinfo {year} {2010})},\ \Eprint {https://arxiv.org/abs/1009.1385} {arXiv:1009.1385 [astro-ph.HE]} \BibitemShut {NoStop}%
\bibitem [{\citenamefont {Neill}\ \emph {et~al.}(2022)\citenamefont {Neill}, \citenamefont {Tsang}, \citenamefont {van Eerten}, \citenamefont {Ryan},\ and\ \citenamefont {Newton}}]{10.1093/mnras/stac1645}%
  \BibitemOpen
  \bibfield  {author} {\bibinfo {author} {\bibfnamefont {D.}~\bibnamefont {Neill}}, \bibinfo {author} {\bibfnamefont {D.}~\bibnamefont {Tsang}}, \bibinfo {author} {\bibfnamefont {H.}~\bibnamefont {van Eerten}}, \bibinfo {author} {\bibfnamefont {G.}~\bibnamefont {Ryan}},\ and\ \bibinfo {author} {\bibfnamefont {W.~G.}\ \bibnamefont {Newton}},\ }\bibfield  {title} {\bibinfo {title} {{Resonant shattering flares in black hole-neutron star and binary neutron star mergers}},\ }\href {https://doi.org/10.1093/mnras/stac1645} {\bibfield  {journal} {\bibinfo  {journal} {Monthly Notices of the Royal Astronomical Society}\ }\textbf {\bibinfo {volume} {514}},\ \bibinfo {pages} {5385} (\bibinfo {year} {2022})},\ \Eprint {https://arxiv.org/abs/https://academic.oup.com/mnras/article-pdf/514/4/5385/44741776/stac1645.pdf} {https://academic.oup.com/mnras/article-pdf/514/4/5385/44741776/stac1645.pdf} \BibitemShut {NoStop}%
\bibitem [{\citenamefont {{Lander}}(2013)}]{Lander2013NS}%
  \BibitemOpen
  \bibfield  {author} {\bibinfo {author} {\bibfnamefont {S.~K.}\ \bibnamefont {{Lander}}},\ }\bibfield  {title} {\bibinfo {title} {{Magnetic Fields in Superconducting Neutron Stars}},\ }\href {https://doi.org/10.1103/PhysRevLett.110.071101} {\bibfield  {journal} {\bibinfo  {journal} {\prl}\ }\textbf {\bibinfo {volume} {110}},\ \bibinfo {eid} {071101} (\bibinfo {year} {2013})},\ \Eprint {https://arxiv.org/abs/1211.3912} {arXiv:1211.3912 [astro-ph.SR]} \BibitemShut {NoStop}%
\bibitem [{\citenamefont {{Henriksson}}\ and\ \citenamefont {{Wasserman}}(2013)}]{Henriksson2013NS}%
  \BibitemOpen
  \bibfield  {author} {\bibinfo {author} {\bibfnamefont {K.~T.}\ \bibnamefont {{Henriksson}}}\ and\ \bibinfo {author} {\bibfnamefont {I.}~\bibnamefont {{Wasserman}}},\ }\bibfield  {title} {\bibinfo {title} {{Poloidal magnetic fields in superconducting neutron stars}},\ }\href {https://doi.org/10.1093/mnras/stt338} {\bibfield  {journal} {\bibinfo  {journal} {\mnras}\ }\textbf {\bibinfo {volume} {431}},\ \bibinfo {pages} {2986} (\bibinfo {year} {2013})},\ \Eprint {https://arxiv.org/abs/1212.5842} {arXiv:1212.5842 [astro-ph.HE]} \BibitemShut {NoStop}%
\bibitem [{\citenamefont {{Goldreich}}\ and\ \citenamefont {{Reisenegger}}(1992)}]{Goldreich1992Magnetic}%
  \BibitemOpen
  \bibfield  {author} {\bibinfo {author} {\bibfnamefont {P.}~\bibnamefont {{Goldreich}}}\ and\ \bibinfo {author} {\bibfnamefont {A.}~\bibnamefont {{Reisenegger}}},\ }\bibfield  {title} {\bibinfo {title} {{Magnetic Field Decay in Isolated Neutron Stars}},\ }\href {https://doi.org/10.1086/171646} {\bibfield  {journal} {\bibinfo  {journal} {\apj}\ }\textbf {\bibinfo {volume} {395}},\ \bibinfo {pages} {250} (\bibinfo {year} {1992})}\BibitemShut {NoStop}%
\bibitem [{\citenamefont {{Hollerbach}}\ and\ \citenamefont {{R{\"u}diger}}(2002)}]{Hollerbach2002Halldriftohmic}%
  \BibitemOpen
  \bibfield  {author} {\bibinfo {author} {\bibfnamefont {R.}~\bibnamefont {{Hollerbach}}}\ and\ \bibinfo {author} {\bibfnamefont {G.}~\bibnamefont {{R{\"u}diger}}},\ }\bibfield  {title} {\bibinfo {title} {{The influence of Hall drift on the magnetic fields of neutron stars}},\ }\href {https://doi.org/10.1046/j.1365-8711.2002.05905.x} {\bibfield  {journal} {\bibinfo  {journal} {\mnras}\ }\textbf {\bibinfo {volume} {337}},\ \bibinfo {pages} {216} (\bibinfo {year} {2002})},\ \Eprint {https://arxiv.org/abs/astro-ph/0208312} {arXiv:astro-ph/0208312 [astro-ph]} \BibitemShut {NoStop}%
\bibitem [{\citenamefont {{Igoshev}}\ \emph {et~al.}(2023)\citenamefont {{Igoshev}}, \citenamefont {{Hollerbach}},\ and\ \citenamefont {{Wood}}}]{Igoshev2023Ambipolar}%
  \BibitemOpen
  \bibfield  {author} {\bibinfo {author} {\bibfnamefont {A.~P.}\ \bibnamefont {{Igoshev}}}, \bibinfo {author} {\bibfnamefont {R.}~\bibnamefont {{Hollerbach}}},\ and\ \bibinfo {author} {\bibfnamefont {T.}~\bibnamefont {{Wood}}},\ }\bibfield  {title} {\bibinfo {title} {{Three-dimensional magnetothermal evolution of off-centred dipole magnetic field configurations in neutron stars}},\ }\href {https://doi.org/10.1093/mnras/stad2404} {\bibfield  {journal} {\bibinfo  {journal} {\mnras}\ }\textbf {\bibinfo {volume} {525}},\ \bibinfo {pages} {3354} (\bibinfo {year} {2023})},\ \Eprint {https://arxiv.org/abs/2308.09132} {arXiv:2308.09132 [astro-ph.HE]} \BibitemShut {NoStop}%
\bibitem [{\citenamefont {{Giacomazzo}}\ \emph {et~al.}(2015)\citenamefont {{Giacomazzo}}, \citenamefont {{Zrake}}, \citenamefont {{Duffell}}, \citenamefont {{MacFadyen}},\ and\ \citenamefont {{Perna}}}]{Giacomazzo2015amplify}%
  \BibitemOpen
  \bibfield  {author} {\bibinfo {author} {\bibfnamefont {B.}~\bibnamefont {{Giacomazzo}}}, \bibinfo {author} {\bibfnamefont {J.}~\bibnamefont {{Zrake}}}, \bibinfo {author} {\bibfnamefont {P.~C.}\ \bibnamefont {{Duffell}}}, \bibinfo {author} {\bibfnamefont {A.~I.}\ \bibnamefont {{MacFadyen}}},\ and\ \bibinfo {author} {\bibfnamefont {R.}~\bibnamefont {{Perna}}},\ }\bibfield  {title} {\bibinfo {title} {{Producing Magnetar Magnetic Fields in the Merger of Binary Neutron Stars}},\ }\href {https://doi.org/10.1088/0004-637X/809/1/39} {\bibfield  {journal} {\bibinfo  {journal} {\apj}\ }\textbf {\bibinfo {volume} {809}},\ \bibinfo {eid} {39} (\bibinfo {year} {2015})},\ \Eprint {https://arxiv.org/abs/1410.0013} {arXiv:1410.0013 [astro-ph.HE]} \BibitemShut {NoStop}%
\bibitem [{\citenamefont {{Wang}}\ \emph {et~al.}(2024{\natexlab{a}})\citenamefont {{Wang}}, \citenamefont {{Moradi}},\ and\ \citenamefont {{Li}}}]{2024arXiv240409251W}%
  \BibitemOpen
  \bibfield  {author} {\bibinfo {author} {\bibfnamefont {Y.}~\bibnamefont {{Wang}}}, \bibinfo {author} {\bibfnamefont {R.}~\bibnamefont {{Moradi}}},\ and\ \bibinfo {author} {\bibfnamefont {L.}~\bibnamefont {{Li}}},\ }\bibfield  {title} {\bibinfo {title} {{Multipolar Electromagnetic Emission of Newborn Magnetar}},\ }\href {https://doi.org/10.48550/arXiv.2404.09251} {\bibfield  {journal} {\bibinfo  {journal} {arXiv e-prints}\ ,\ \bibinfo {eid} {arXiv:2404.09251}} (\bibinfo {year} {2024}{\natexlab{a}})},\ \Eprint {https://arxiv.org/abs/2404.09251} {arXiv:2404.09251 [astro-ph.HE]} \BibitemShut {NoStop}%
\bibitem [{\citenamefont {Daigne}\ and\ \citenamefont {Mochkovitch}(2002)}]{10.1046/j.1365-8711.2002.05875.x}%
  \BibitemOpen
  \bibfield  {author} {\bibinfo {author} {\bibfnamefont {F.}~\bibnamefont {Daigne}}\ and\ \bibinfo {author} {\bibfnamefont {R.}~\bibnamefont {Mochkovitch}},\ }\bibfield  {title} {\bibinfo {title} {{The expected thermal precursors of gamma-ray bursts in the internal shock model}},\ }\href {https://doi.org/10.1046/j.1365-8711.2002.05875.x} {\bibfield  {journal} {\bibinfo  {journal} {Monthly Notices of the Royal Astronomical Society}\ }\textbf {\bibinfo {volume} {336}},\ \bibinfo {pages} {1271} (\bibinfo {year} {2002})},\ \Eprint {https://arxiv.org/abs/https://academic.oup.com/mnras/article-pdf/336/4/1271/3044413/336-4-1271.pdf} {https://academic.oup.com/mnras/article-pdf/336/4/1271/3044413/336-4-1271.pdf} \BibitemShut {NoStop}%
\bibitem [{\citenamefont {{Li}}\ and\ \citenamefont {{Yu}}(2016)}]{2016ApJ...819..120L}%
  \BibitemOpen
  \bibfield  {author} {\bibinfo {author} {\bibfnamefont {S.-Z.}\ \bibnamefont {{Li}}}\ and\ \bibinfo {author} {\bibfnamefont {Y.-W.}\ \bibnamefont {{Yu}}},\ }\bibfield  {title} {\bibinfo {title} {{Shock Breakout Driven by the Remnant of a Neutron Star Binary Merger: An X-Ray Precursor of Mergernova Emission}},\ }\href {https://doi.org/10.3847/0004-637X/819/2/120} {\bibfield  {journal} {\bibinfo  {journal} {\apj}\ }\textbf {\bibinfo {volume} {819}},\ \bibinfo {eid} {120} (\bibinfo {year} {2016})},\ \Eprint {https://arxiv.org/abs/1511.01229} {arXiv:1511.01229 [astro-ph.HE]} \BibitemShut {NoStop}%
\bibitem [{\citenamefont {Nakar}(2020)}]{NAKAR20201}%
  \BibitemOpen
  \bibfield  {author} {\bibinfo {author} {\bibfnamefont {E.}~\bibnamefont {Nakar}},\ }\bibfield  {title} {\bibinfo {title} {The electromagnetic counterparts of compact binary mergers},\ }\href {https://doi.org/https://doi.org/10.1016/j.physrep.2020.08.008} {\bibfield  {journal} {\bibinfo  {journal} {Physics Reports}\ }\textbf {\bibinfo {volume} {886}},\ \bibinfo {pages} {1} (\bibinfo {year} {2020})},\ \bibinfo {note} {the electromagnetic counterparts of compact binary mergers}\BibitemShut {NoStop}%
\bibitem [{\citenamefont {{Lazzati}}(2005)}]{Lazzati2005Precursor}%
  \BibitemOpen
  \bibfield  {author} {\bibinfo {author} {\bibfnamefont {D.}~\bibnamefont {{Lazzati}}},\ }\bibfield  {title} {\bibinfo {title} {{Precursor activity in bright, long BATSE gamma-ray bursts}},\ }\href {https://doi.org/10.1111/j.1365-2966.2005.08687.x} {\bibfield  {journal} {\bibinfo  {journal} {\mnras}\ }\textbf {\bibinfo {volume} {357}},\ \bibinfo {pages} {722} (\bibinfo {year} {2005})},\ \Eprint {https://arxiv.org/abs/astro-ph/0411753} {arXiv:astro-ph/0411753 [astro-ph]} \BibitemShut {NoStop}%
\bibitem [{\citenamefont {{Du}}\ \emph {et~al.}(2024)\citenamefont {{Du}}, \citenamefont {{L{\"u}}}, \citenamefont {{Liu}},\ and\ \citenamefont {{Liang}}}]{lv2024ICMART}%
  \BibitemOpen
  \bibfield  {author} {\bibinfo {author} {\bibfnamefont {Z.-W.}\ \bibnamefont {{Du}}}, \bibinfo {author} {\bibfnamefont {H.}~\bibnamefont {{L{\"u}}}}, \bibinfo {author} {\bibfnamefont {X.}~\bibnamefont {{Liu}}},\ and\ \bibinfo {author} {\bibfnamefont {E.}~\bibnamefont {{Liang}}},\ }\bibfield  {title} {\bibinfo {title} {{The jet composition of GRB 230307A: Poynting-flux-dominated outflow?}},\ }\href {https://doi.org/10.1093/mnrasl/slad203} {\bibfield  {journal} {\bibinfo  {journal} {\mnras}\ }\textbf {\bibinfo {volume} {529}},\ \bibinfo {pages} {L67} (\bibinfo {year} {2024})},\ \Eprint {https://arxiv.org/abs/2401.05002} {arXiv:2401.05002 [astro-ph.HE]} \BibitemShut {NoStop}%
\bibitem [{\citenamefont {{Peng}}\ \emph {et~al.}(2024{\natexlab{b}})\citenamefont {{Peng}}, \citenamefont {{Chen}},\ and\ \citenamefont {{Mao}}}]{Peng2024Comparative}%
  \BibitemOpen
  \bibfield  {author} {\bibinfo {author} {\bibfnamefont {Z.-Y.}\ \bibnamefont {{Peng}}}, \bibinfo {author} {\bibfnamefont {J.-M.}\ \bibnamefont {{Chen}}},\ and\ \bibinfo {author} {\bibfnamefont {J.}~\bibnamefont {{Mao}}},\ }\bibfield  {title} {\bibinfo {title} {{A Comparative Analysis of Two Peculiar Gamma-Ray Bursts: GRB 230307A and GRB 211211A}},\ }\href {https://doi.org/10.3847/1538-4357/ad45fc} {\bibfield  {journal} {\bibinfo  {journal} {\apj}\ }\textbf {\bibinfo {volume} {969}},\ \bibinfo {eid} {26} (\bibinfo {year} {2024}{\natexlab{b}})}\BibitemShut {NoStop}%
\bibitem [{\citenamefont {{Metzger}}\ \emph {et~al.}(2008)\citenamefont {{Metzger}}, \citenamefont {{Quataert}},\ and\ \citenamefont {{Thompson}}}]{Metzger08}%
  \BibitemOpen
  \bibfield  {author} {\bibinfo {author} {\bibfnamefont {B.~D.}\ \bibnamefont {{Metzger}}}, \bibinfo {author} {\bibfnamefont {E.}~\bibnamefont {{Quataert}}},\ and\ \bibinfo {author} {\bibfnamefont {T.~A.}\ \bibnamefont {{Thompson}}},\ }\bibfield  {title} {\bibinfo {title} {{Short-duration gamma-ray bursts with extended emission from protomagnetar spin-down}},\ }\href {https://doi.org/10.1111/j.1365-2966.2008.12923.x} {\bibfield  {journal} {\bibinfo  {journal} {\mnras}\ }\textbf {\bibinfo {volume} {385}},\ \bibinfo {pages} {1455} (\bibinfo {year} {2008})},\ \Eprint {https://arxiv.org/abs/0712.1233} {arXiv:0712.1233 [astro-ph]} \BibitemShut {NoStop}%
\bibitem [{\citenamefont {Bucciantini}\ \emph {et~al.}(2011)\citenamefont {Bucciantini}, \citenamefont {Metzger}, \citenamefont {Thompson},\ and\ \citenamefont {Quataert}}]{10.1111/j.1365-2966.2011.19810.x}%
  \BibitemOpen
  \bibfield  {author} {\bibinfo {author} {\bibfnamefont {N.}~\bibnamefont {Bucciantini}}, \bibinfo {author} {\bibfnamefont {B.~D.}\ \bibnamefont {Metzger}}, \bibinfo {author} {\bibfnamefont {T.~A.}\ \bibnamefont {Thompson}},\ and\ \bibinfo {author} {\bibfnamefont {E.}~\bibnamefont {Quataert}},\ }\bibfield  {title} {\bibinfo {title} {{Short gamma-ray bursts with extended emission from magnetar birth: jet formation and collimation}},\ }\href {https://doi.org/10.1111/j.1365-2966.2011.19810.x} {\bibfield  {journal} {\bibinfo  {journal} {Monthly Notices of the Royal Astronomical Society}\ }\textbf {\bibinfo {volume} {419}},\ \bibinfo {pages} {1537} (\bibinfo {year} {2011})},\ \Eprint {https://arxiv.org/abs/https://academic.oup.com/mnras/article-pdf/419/2/1537/3125386/mnras0419-1537.pdf} {https://academic.oup.com/mnras/article-pdf/419/2/1537/3125386/mnras0419-1537.pdf} \BibitemShut {NoStop}%
\bibitem [{\citenamefont {{Jordana-Mitjans}}\ \emph {et~al.}(2022)\citenamefont {{Jordana-Mitjans}}, \citenamefont {{Mundell}}, \citenamefont {{Guidorzi}}, \citenamefont {{Smith}}, \citenamefont {{Ram{\'\i}rez-Ruiz}}, \citenamefont {{Metzger}}, \citenamefont {{Kobayashi}}, \citenamefont {{Gomboc}}, \citenamefont {{Steele}}, \citenamefont {{Shrestha}}, \citenamefont {{Marongiu}}, \citenamefont {{Rossi}},\ and\ \citenamefont {{Rothberg}}}]{2022ApJ...939..106J}%
  \BibitemOpen
  \bibfield  {author} {\bibinfo {author} {\bibfnamefont {N.}~\bibnamefont {{Jordana-Mitjans}}}, \bibinfo {author} {\bibfnamefont {C.~G.}\ \bibnamefont {{Mundell}}}, \bibinfo {author} {\bibfnamefont {C.}~\bibnamefont {{Guidorzi}}}, \bibinfo {author} {\bibfnamefont {R.~J.}\ \bibnamefont {{Smith}}}, \bibinfo {author} {\bibfnamefont {E.}~\bibnamefont {{Ram{\'\i}rez-Ruiz}}}, \bibinfo {author} {\bibfnamefont {B.~D.}\ \bibnamefont {{Metzger}}}, \bibinfo {author} {\bibfnamefont {S.}~\bibnamefont {{Kobayashi}}}, \bibinfo {author} {\bibfnamefont {A.}~\bibnamefont {{Gomboc}}}, \bibinfo {author} {\bibfnamefont {I.~A.}\ \bibnamefont {{Steele}}}, \bibinfo {author} {\bibfnamefont {M.}~\bibnamefont {{Shrestha}}}, \bibinfo {author} {\bibfnamefont {M.}~\bibnamefont {{Marongiu}}}, \bibinfo {author} {\bibfnamefont {A.}~\bibnamefont {{Rossi}}},\ and\ \bibinfo {author} {\bibfnamefont {B.}~\bibnamefont {{Rothberg}}},\ }\bibfield  {title} {\bibinfo {title} {{A Short Gamma-Ray Burst from a Protomagnetar Remnant}},\ }\href
  {https://doi.org/10.3847/1538-4357/ac972b} {\bibfield  {journal} {\bibinfo  {journal} {\apj}\ }\textbf {\bibinfo {volume} {939}},\ \bibinfo {eid} {106} (\bibinfo {year} {2022})},\ \Eprint {https://arxiv.org/abs/2211.05810} {arXiv:2211.05810 [astro-ph.HE]} \BibitemShut {NoStop}%
\bibitem [{\citenamefont {{Gompertz}}\ \emph {et~al.}(2013)\citenamefont {{Gompertz}}, \citenamefont {{O'Brien}}, \citenamefont {{Wynn}},\ and\ \citenamefont {{Rowlinson}}}]{2013MNRAS.431.1745G}%
  \BibitemOpen
  \bibfield  {author} {\bibinfo {author} {\bibfnamefont {B.~P.}\ \bibnamefont {{Gompertz}}}, \bibinfo {author} {\bibfnamefont {P.~T.}\ \bibnamefont {{O'Brien}}}, \bibinfo {author} {\bibfnamefont {G.~A.}\ \bibnamefont {{Wynn}}},\ and\ \bibinfo {author} {\bibfnamefont {A.}~\bibnamefont {{Rowlinson}}},\ }\bibfield  {title} {\bibinfo {title} {{Can magnetar spin-down power extended emission in some short GRBs?}},\ }\href {https://doi.org/10.1093/mnras/stt293} {\bibfield  {journal} {\bibinfo  {journal} {\mnras}\ }\textbf {\bibinfo {volume} {431}},\ \bibinfo {pages} {1745} (\bibinfo {year} {2013})},\ \Eprint {https://arxiv.org/abs/1302.3643} {arXiv:1302.3643 [astro-ph.HE]} \BibitemShut {NoStop}%
\bibitem [{\citenamefont {{Gibson}}\ \emph {et~al.}(2017)\citenamefont {{Gibson}}, \citenamefont {{Wynn}}, \citenamefont {{Gompertz}},\ and\ \citenamefont {{O'Brien}}}]{Gibson17}%
  \BibitemOpen
  \bibfield  {author} {\bibinfo {author} {\bibfnamefont {S.~L.}\ \bibnamefont {{Gibson}}}, \bibinfo {author} {\bibfnamefont {G.~A.}\ \bibnamefont {{Wynn}}}, \bibinfo {author} {\bibfnamefont {B.~P.}\ \bibnamefont {{Gompertz}}},\ and\ \bibinfo {author} {\bibfnamefont {P.~T.}\ \bibnamefont {{O'Brien}}},\ }\bibfield  {title} {\bibinfo {title} {{Fallback accretion on to a newborn magnetar: short GRBs with extended emission}},\ }\href {https://doi.org/10.1093/mnras/stx1531} {\bibfield  {journal} {\bibinfo  {journal} {\mnras}\ }\textbf {\bibinfo {volume} {470}},\ \bibinfo {pages} {4925} (\bibinfo {year} {2017})},\ \Eprint {https://arxiv.org/abs/1706.04802} {arXiv:1706.04802 [astro-ph.HE]} \BibitemShut {NoStop}%
\bibitem [{\citenamefont {{Meng}}\ \emph {et~al.}(2024)\citenamefont {{Meng}}, \citenamefont {{Wang}},\ and\ \citenamefont {{Liu}}}]{Meng2024NSBH}%
  \BibitemOpen
  \bibfield  {author} {\bibinfo {author} {\bibfnamefont {Y.-Z.}\ \bibnamefont {{Meng}}}, \bibinfo {author} {\bibfnamefont {X.~I.}\ \bibnamefont {{Wang}}},\ and\ \bibinfo {author} {\bibfnamefont {Z.-K.}\ \bibnamefont {{Liu}}},\ }\bibfield  {title} {\bibinfo {title} {{Significant Cocoon Emission and Photosphere Duration Stretching in GRB 211211A: A Burst from a Neutron Star‑Black Hole Merger}},\ }\href {https://doi.org/10.3847/1538-4357/ad1bd7} {\bibfield  {journal} {\bibinfo  {journal} {\apj}\ }\textbf {\bibinfo {volume} {963}},\ \bibinfo {eid} {112} (\bibinfo {year} {2024})},\ \Eprint {https://arxiv.org/abs/2304.00893} {arXiv:2304.00893 [astro-ph.HE]} \BibitemShut {NoStop}%
\bibitem [{\citenamefont {{Kaspi}}\ and\ \citenamefont {{Beloborodov}}(2017)}]{Kaspi2017Magnetars}%
  \BibitemOpen
  \bibfield  {author} {\bibinfo {author} {\bibfnamefont {V.~M.}\ \bibnamefont {{Kaspi}}}\ and\ \bibinfo {author} {\bibfnamefont {A.~M.}\ \bibnamefont {{Beloborodov}}},\ }\bibfield  {title} {\bibinfo {title} {{Magnetars}},\ }\href {https://doi.org/10.1146/annurev-astro-081915-023329} {\bibfield  {journal} {\bibinfo  {journal} {\araa}\ }\textbf {\bibinfo {volume} {55}},\ \bibinfo {pages} {261} (\bibinfo {year} {2017})},\ \Eprint {https://arxiv.org/abs/1703.00068} {arXiv:1703.00068 [astro-ph.HE]} \BibitemShut {NoStop}%
\bibitem [{\citenamefont {{Xiao}}\ \emph {et~al.}(2022{\natexlab{b}})\citenamefont {{Xiao}}, \citenamefont {{Liu}}, \citenamefont {{Peng}}, \citenamefont {{An}}, \citenamefont {{Xiong}}, \citenamefont {{Tuo}}, \citenamefont {{Gong}}, \citenamefont {{Zhang}}, \citenamefont {{Zhang}}, \citenamefont {{Zheng}}, \citenamefont {{Li}}, \citenamefont {{Gao}}, \citenamefont {{Guo}}, \citenamefont {{Li}}, \citenamefont {{Liang}}, \citenamefont {{Liu}}, \citenamefont {{Qiao}}, \citenamefont {{Sun}}, \citenamefont {{Wang}}, \citenamefont {{Wen}}, \citenamefont {{Xu}}, \citenamefont {{Yang}}, \citenamefont {{Zhang}}, \citenamefont {{Zhang}}, \citenamefont {{Zhang}}, \citenamefont {{Zhao}}, \citenamefont {{Qi}}, \citenamefont {{Han}}, \citenamefont {{Li}}, \citenamefont {{Huang}}, \citenamefont {{Song}}, \citenamefont {{Cai}}, \citenamefont {{Yi}}, \citenamefont {{Zhao}}, \citenamefont {{Song}}, \citenamefont {{Huang}}, \citenamefont {{Ge}}, \citenamefont {{Ma}}, \citenamefont {{Li}}, \citenamefont {{Li}}, \citenamefont
  {{Wang}}, \citenamefont {{Wang}}, \citenamefont {{Zhang}}, \citenamefont {{Zhang}}, \citenamefont {{Zhang}}, \citenamefont {{Zhao}}, \citenamefont {{Guo}}, \citenamefont {{Chen}}, \citenamefont {{Xie}},\ and\ \citenamefont {{Zhang}}}]{xiao_GECAMB_time_calibration}%
  \BibitemOpen
  \bibfield  {author} {\bibinfo {author} {\bibfnamefont {S.}~\bibnamefont {{Xiao}}}, \bibinfo {author} {\bibfnamefont {Y.~Q.}\ \bibnamefont {{Liu}}}, \bibinfo {author} {\bibfnamefont {W.~X.}\ \bibnamefont {{Peng}}}, \bibinfo {author} {\bibfnamefont {Z.~H.}\ \bibnamefont {{An}}}, \bibinfo {author} {\bibfnamefont {S.~L.}\ \bibnamefont {{Xiong}}}, \bibinfo {author} {\bibfnamefont {Y.~L.}\ \bibnamefont {{Tuo}}}, \bibinfo {author} {\bibfnamefont {K.}~\bibnamefont {{Gong}}}, \bibinfo {author} {\bibfnamefont {P.}~\bibnamefont {{Zhang}}}, \bibinfo {author} {\bibfnamefont {K.}~\bibnamefont {{Zhang}}}, \bibinfo {author} {\bibfnamefont {S.~J.}\ \bibnamefont {{Zheng}}}, \bibinfo {author} {\bibfnamefont {C.~Y.}\ \bibnamefont {{Li}}}, \bibinfo {author} {\bibfnamefont {M.}~\bibnamefont {{Gao}}}, \bibinfo {author} {\bibfnamefont {D.~Y.}\ \bibnamefont {{Guo}}}, \bibinfo {author} {\bibfnamefont {X.~Q.}\ \bibnamefont {{Li}}}, \bibinfo {author} {\bibfnamefont {X.~H.}\ \bibnamefont {{Liang}}}, \bibinfo {author} {\bibfnamefont
  {X.~J.}\ \bibnamefont {{Liu}}}, \bibinfo {author} {\bibfnamefont {R.}~\bibnamefont {{Qiao}}}, \bibinfo {author} {\bibfnamefont {X.~L.}\ \bibnamefont {{Sun}}}, \bibinfo {author} {\bibfnamefont {J.~Z.}\ \bibnamefont {{Wang}}}, \bibinfo {author} {\bibfnamefont {X.~Y.}\ \bibnamefont {{Wen}}}, \bibinfo {author} {\bibfnamefont {Y.~B.}\ \bibnamefont {{Xu}}}, \bibinfo {author} {\bibfnamefont {S.}~\bibnamefont {{Yang}}}, \bibinfo {author} {\bibfnamefont {D.~L.}\ \bibnamefont {{Zhang}}}, \bibinfo {author} {\bibfnamefont {F.}~\bibnamefont {{Zhang}}}, \bibinfo {author} {\bibfnamefont {F.}~\bibnamefont {{Zhang}}}, \bibinfo {author} {\bibfnamefont {X.~Y.}\ \bibnamefont {{Zhao}}}, \bibinfo {author} {\bibfnamefont {J.~L.}\ \bibnamefont {{Qi}}}, \bibinfo {author} {\bibfnamefont {X.~B.}\ \bibnamefont {{Han}}}, \bibinfo {author} {\bibfnamefont {Z.~D.}\ \bibnamefont {{Li}}}, \bibinfo {author} {\bibfnamefont {J.}~\bibnamefont {{Huang}}}, \bibinfo {author} {\bibfnamefont {L.~M.}\ \bibnamefont {{Song}}}, \bibinfo {author}
  {\bibfnamefont {C.}~\bibnamefont {{Cai}}}, \bibinfo {author} {\bibfnamefont {Q.~B.}\ \bibnamefont {{Yi}}}, \bibinfo {author} {\bibfnamefont {Y.}~\bibnamefont {{Zhao}}}, \bibinfo {author} {\bibfnamefont {X.~Y.}\ \bibnamefont {{Song}}}, \bibinfo {author} {\bibfnamefont {Y.}~\bibnamefont {{Huang}}}, \bibinfo {author} {\bibfnamefont {M.~Y.}\ \bibnamefont {{Ge}}}, \bibinfo {author} {\bibfnamefont {X.}~\bibnamefont {{Ma}}}, \bibinfo {author} {\bibfnamefont {X.~B.}\ \bibnamefont {{Li}}}, \bibinfo {author} {\bibfnamefont {B.}~\bibnamefont {{Li}}}, \bibinfo {author} {\bibfnamefont {P.}~\bibnamefont {{Wang}}}, \bibinfo {author} {\bibfnamefont {J.}~\bibnamefont {{Wang}}}, \bibinfo {author} {\bibfnamefont {Y.~Q.}\ \bibnamefont {{Zhang}}}, \bibinfo {author} {\bibfnamefont {Z.}~\bibnamefont {{Zhang}}}, \bibinfo {author} {\bibfnamefont {X.~L.}\ \bibnamefont {{Zhang}}}, \bibinfo {author} {\bibfnamefont {H.~Y.}\ \bibnamefont {{Zhao}}}, \bibinfo {author} {\bibfnamefont {Z.~W.}\ \bibnamefont {{Guo}}}, \bibinfo {author}
  {\bibfnamefont {C.}~\bibnamefont {{Chen}}}, \bibinfo {author} {\bibfnamefont {S.~L.}\ \bibnamefont {{Xie}}},\ and\ \bibinfo {author} {\bibfnamefont {S.~N.}\ \bibnamefont {{Zhang}}},\ }\bibfield  {title} {\bibinfo {title} {{On-ground and on-orbit time calibrations of GECAM}},\ }\href {https://doi.org/10.1093/mnras/stac085} {\bibfield  {journal} {\bibinfo  {journal} {\mnras}\ }\textbf {\bibinfo {volume} {511}},\ \bibinfo {pages} {964} (\bibinfo {year} {2022}{\natexlab{b}})}\BibitemShut {NoStop}%
\bibitem [{\citenamefont {{Zhang}}\ \emph {et~al.}(2023)\citenamefont {{Zhang}}, \citenamefont {{Zheng}}, \citenamefont {{Liu}}, \citenamefont {{An}}, \citenamefont {{Wang}}, \citenamefont {{Wen}}, \citenamefont {{Li}}, \citenamefont {{Sun}}, \citenamefont {{Gong}}, \citenamefont {{Liu}}, \citenamefont {{Liu}}, \citenamefont {{Yang}}, \citenamefont {{Peng}}, \citenamefont {{Qiao}}, \citenamefont {{Guo}}, \citenamefont {{Feng}}, \citenamefont {{Zhang}}, \citenamefont {{Xue}}, \citenamefont {{Tan}}, \citenamefont {{Cai}}, \citenamefont {{Xiao}}, \citenamefont {{Yi}}, \citenamefont {{Xu}}, \citenamefont {{Gao}}, \citenamefont {{Wang}}, \citenamefont {{Hou}}, \citenamefont {{Huang}}, \citenamefont {{Zhao}}, \citenamefont {{Ma}}, \citenamefont {{Wang}}, \citenamefont {{Wang}}, \citenamefont {{Li}}, \citenamefont {{Zhang}}, \citenamefont {{Zhang}}, \citenamefont {{Li}}, \citenamefont {{Wang}}, \citenamefont {{Liang}}, \citenamefont {{Wang}}, \citenamefont {{Li}}, \citenamefont {{Ye}}, \citenamefont {{Zheng}},
  \citenamefont {{Song}}, \citenamefont {{Zhang}}, \citenamefont {{Chen}},\ and\ \citenamefont {{Xiong}}}]{Zhang2023HEBS}%
  \BibitemOpen
  \bibfield  {author} {\bibinfo {author} {\bibfnamefont {D.}~\bibnamefont {{Zhang}}}, \bibinfo {author} {\bibfnamefont {C.}~\bibnamefont {{Zheng}}}, \bibinfo {author} {\bibfnamefont {J.}~\bibnamefont {{Liu}}}, \bibinfo {author} {\bibfnamefont {Z.}~\bibnamefont {{An}}}, \bibinfo {author} {\bibfnamefont {C.}~\bibnamefont {{Wang}}}, \bibinfo {author} {\bibfnamefont {X.}~\bibnamefont {{Wen}}}, \bibinfo {author} {\bibfnamefont {X.}~\bibnamefont {{Li}}}, \bibinfo {author} {\bibfnamefont {X.}~\bibnamefont {{Sun}}}, \bibinfo {author} {\bibfnamefont {K.}~\bibnamefont {{Gong}}}, \bibinfo {author} {\bibfnamefont {Y.}~\bibnamefont {{Liu}}}, \bibinfo {author} {\bibfnamefont {X.}~\bibnamefont {{Liu}}}, \bibinfo {author} {\bibfnamefont {S.}~\bibnamefont {{Yang}}}, \bibinfo {author} {\bibfnamefont {W.}~\bibnamefont {{Peng}}}, \bibinfo {author} {\bibfnamefont {R.}~\bibnamefont {{Qiao}}}, \bibinfo {author} {\bibfnamefont {D.}~\bibnamefont {{Guo}}}, \bibinfo {author} {\bibfnamefont {P.}~\bibnamefont {{Feng}}}, \bibinfo {author}
  {\bibfnamefont {Y.}~\bibnamefont {{Zhang}}}, \bibinfo {author} {\bibfnamefont {W.}~\bibnamefont {{Xue}}}, \bibinfo {author} {\bibfnamefont {W.}~\bibnamefont {{Tan}}}, \bibinfo {author} {\bibfnamefont {C.}~\bibnamefont {{Cai}}}, \bibinfo {author} {\bibfnamefont {S.}~\bibnamefont {{Xiao}}}, \bibinfo {author} {\bibfnamefont {Q.}~\bibnamefont {{Yi}}}, \bibinfo {author} {\bibfnamefont {Y.}~\bibnamefont {{Xu}}}, \bibinfo {author} {\bibfnamefont {M.}~\bibnamefont {{Gao}}}, \bibinfo {author} {\bibfnamefont {J.}~\bibnamefont {{Wang}}}, \bibinfo {author} {\bibfnamefont {D.}~\bibnamefont {{Hou}}}, \bibinfo {author} {\bibfnamefont {Y.}~\bibnamefont {{Huang}}}, \bibinfo {author} {\bibfnamefont {X.}~\bibnamefont {{Zhao}}}, \bibinfo {author} {\bibfnamefont {X.}~\bibnamefont {{Ma}}}, \bibinfo {author} {\bibfnamefont {P.}~\bibnamefont {{Wang}}}, \bibinfo {author} {\bibfnamefont {J.}~\bibnamefont {{Wang}}}, \bibinfo {author} {\bibfnamefont {X.}~\bibnamefont {{Li}}}, \bibinfo {author} {\bibfnamefont {P.}~\bibnamefont
  {{Zhang}}}, \bibinfo {author} {\bibfnamefont {Z.}~\bibnamefont {{Zhang}}}, \bibinfo {author} {\bibfnamefont {Y.}~\bibnamefont {{Li}}}, \bibinfo {author} {\bibfnamefont {H.}~\bibnamefont {{Wang}}}, \bibinfo {author} {\bibfnamefont {X.}~\bibnamefont {{Liang}}}, \bibinfo {author} {\bibfnamefont {Y.}~\bibnamefont {{Wang}}}, \bibinfo {author} {\bibfnamefont {B.}~\bibnamefont {{Li}}}, \bibinfo {author} {\bibfnamefont {J.}~\bibnamefont {{Ye}}}, \bibinfo {author} {\bibfnamefont {S.}~\bibnamefont {{Zheng}}}, \bibinfo {author} {\bibfnamefont {L.}~\bibnamefont {{Song}}}, \bibinfo {author} {\bibfnamefont {F.}~\bibnamefont {{Zhang}}}, \bibinfo {author} {\bibfnamefont {G.}~\bibnamefont {{Chen}}},\ and\ \bibinfo {author} {\bibfnamefont {S.}~\bibnamefont {{Xiong}}},\ }\bibfield  {title} {\bibinfo {title} {{The performance of SiPM-based gamma-ray detector (GRD) of GECAM-C}},\ }\href {https://doi.org/10.1016/j.nima.2023.168586} {\bibfield  {journal} {\bibinfo  {journal} {Nuclear Instruments and Methods in Physics Research
  A}\ }\textbf {\bibinfo {volume} {1056}},\ \bibinfo {eid} {168586} (\bibinfo {year} {2023})},\ \Eprint {https://arxiv.org/abs/2303.00537} {arXiv:2303.00537 [astro-ph.IM]} \BibitemShut {NoStop}%
\bibitem [{\citenamefont {{Wang}}\ \emph {et~al.}(2024{\natexlab{b}})\citenamefont {{Wang}}, \citenamefont {{Zhang}}, \citenamefont {{Zheng}}, \citenamefont {{Xiong}}, \citenamefont {{An}}, \citenamefont {{Peng}}, \citenamefont {{Zhao}}, \citenamefont {{Zhao}}, \citenamefont {{Zheng}}, \citenamefont {{Feng}}, \citenamefont {{Gong}}, \citenamefont {{Guo}}, \citenamefont {{Li}}, \citenamefont {{Liu}}, \citenamefont {{Liu}}, \citenamefont {{Tan}}, \citenamefont {{Wang}}, \citenamefont {{Xue}}, \citenamefont {{Yang}}, \citenamefont {{Zhang}}, \citenamefont {{Zhang}},\ and\ \citenamefont {{Zhang}}}]{Wang2024GTM}%
  \BibitemOpen
  \bibfield  {author} {\bibinfo {author} {\bibfnamefont {C.}~\bibnamefont {{Wang}}}, \bibinfo {author} {\bibfnamefont {J.}~\bibnamefont {{Zhang}}}, \bibinfo {author} {\bibfnamefont {S.}~\bibnamefont {{Zheng}}}, \bibinfo {author} {\bibfnamefont {S.}~\bibnamefont {{Xiong}}}, \bibinfo {author} {\bibfnamefont {Z.}~\bibnamefont {{An}}}, \bibinfo {author} {\bibfnamefont {W.}~\bibnamefont {{Peng}}}, \bibinfo {author} {\bibfnamefont {H.}~\bibnamefont {{Zhao}}}, \bibinfo {author} {\bibfnamefont {X.}~\bibnamefont {{Zhao}}}, \bibinfo {author} {\bibfnamefont {C.}~\bibnamefont {{Zheng}}}, \bibinfo {author} {\bibfnamefont {P.}~\bibnamefont {{Feng}}}, \bibinfo {author} {\bibfnamefont {K.}~\bibnamefont {{Gong}}}, \bibinfo {author} {\bibfnamefont {D.}~\bibnamefont {{Guo}}}, \bibinfo {author} {\bibfnamefont {X.}~\bibnamefont {{Li}}}, \bibinfo {author} {\bibfnamefont {J.}~\bibnamefont {{Liu}}}, \bibinfo {author} {\bibfnamefont {Y.}~\bibnamefont {{Liu}}}, \bibinfo {author} {\bibfnamefont {W.}~\bibnamefont {{Tan}}}, \bibinfo
  {author} {\bibfnamefont {Y.}~\bibnamefont {{Wang}}}, \bibinfo {author} {\bibfnamefont {W.}~\bibnamefont {{Xue}}}, \bibinfo {author} {\bibfnamefont {S.}~\bibnamefont {{Yang}}}, \bibinfo {author} {\bibfnamefont {D.}~\bibnamefont {{Zhang}}}, \bibinfo {author} {\bibfnamefont {F.}~\bibnamefont {{Zhang}}},\ and\ \bibinfo {author} {\bibfnamefont {Y.}~\bibnamefont {{Zhang}}},\ }\bibfield  {title} {\bibinfo {title} {{Simulation of the in-flight background and performance of DRO/GTM}},\ }\href {https://doi.org/10.1007/s10686-024-09946-8} {\bibfield  {journal} {\bibinfo  {journal} {Experimental Astronomy}\ }\textbf {\bibinfo {volume} {57}},\ \bibinfo {eid} {26} (\bibinfo {year} {2024}{\natexlab{b}})}\BibitemShut {NoStop}%
\bibitem [{\citenamefont {Meegan}\ \emph {et~al.}(2009)\citenamefont {Meegan}, \citenamefont {Lichti}, \citenamefont {Bhat}, \citenamefont {Bissaldi}, \citenamefont {Briggs}, \citenamefont {Connaughton}, \citenamefont {Diehl}, \citenamefont {Fishman}, \citenamefont {Greiner},\ and\ \citenamefont {Hoover}}]{GBM_overview}%
  \BibitemOpen
  \bibfield  {author} {\bibinfo {author} {\bibfnamefont {C.}~\bibnamefont {Meegan}}, \bibinfo {author} {\bibfnamefont {G.}~\bibnamefont {Lichti}}, \bibinfo {author} {\bibfnamefont {P.~N.}\ \bibnamefont {Bhat}}, \bibinfo {author} {\bibfnamefont {E.}~\bibnamefont {Bissaldi}}, \bibinfo {author} {\bibfnamefont {M.~S.}\ \bibnamefont {Briggs}}, \bibinfo {author} {\bibfnamefont {V.}~\bibnamefont {Connaughton}}, \bibinfo {author} {\bibfnamefont {R.}~\bibnamefont {Diehl}}, \bibinfo {author} {\bibfnamefont {G.}~\bibnamefont {Fishman}}, \bibinfo {author} {\bibfnamefont {J.}~\bibnamefont {Greiner}},\ and\ \bibinfo {author} {\bibfnamefont {A.~S.}\ \bibnamefont {Hoover}},\ }\bibfield  {title} {\bibinfo {title} {The fermi gamma-ray burst monitor},\ }\href@noop {} {\bibfield  {journal} {\bibinfo  {journal} {The Astrophysical Journal}\ }\textbf {\bibinfo {volume} {702}},\ \bibinfo {pages} {791} (\bibinfo {year} {2009})}\BibitemShut {NoStop}%
\bibitem [{\citenamefont {Bissaldi}\ \emph {et~al.}(2009)\citenamefont {Bissaldi}, \citenamefont {Kienlin}, \citenamefont {Lichti}, \citenamefont {Steinle}, \citenamefont {Bhat}, \citenamefont {Briggs}, \citenamefont {Fishman}, \citenamefont {Hoover}, \citenamefont {Kippen},\ and\ \citenamefont {Krumrey}}]{GBM_calibration}%
  \BibitemOpen
  \bibfield  {author} {\bibinfo {author} {\bibfnamefont {E.}~\bibnamefont {Bissaldi}}, \bibinfo {author} {\bibfnamefont {A.~V.}\ \bibnamefont {Kienlin}}, \bibinfo {author} {\bibfnamefont {G.}~\bibnamefont {Lichti}}, \bibinfo {author} {\bibfnamefont {H.}~\bibnamefont {Steinle}}, \bibinfo {author} {\bibfnamefont {P.~N.}\ \bibnamefont {Bhat}}, \bibinfo {author} {\bibfnamefont {M.~S.}\ \bibnamefont {Briggs}}, \bibinfo {author} {\bibfnamefont {G.~J.}\ \bibnamefont {Fishman}}, \bibinfo {author} {\bibfnamefont {A.~S.}\ \bibnamefont {Hoover}}, \bibinfo {author} {\bibfnamefont {R.~M.}\ \bibnamefont {Kippen}},\ and\ \bibinfo {author} {\bibfnamefont {M.~a.}\ \bibnamefont {Krumrey}},\ }\bibfield  {title} {\bibinfo {title} {Ground-based calibration and characterization of the fermi gamma-ray burst monitor detectors},\ }\href@noop {} {\bibfield  {journal} {\bibinfo  {journal} {Experimental Astronomy}\ }\textbf {\bibinfo {volume} {24}},\ \bibinfo {pages} {47} (\bibinfo {year} {2009})}\BibitemShut {NoStop}%
\bibitem [{\citenamefont {{Scargle}}\ \emph {et~al.}(2013)\citenamefont {{Scargle}}, \citenamefont {{Norris}}, \citenamefont {{Jackson}},\ and\ \citenamefont {{Chiang}}}]{2013ApJ...764..167S}%
  \BibitemOpen
  \bibfield  {author} {\bibinfo {author} {\bibfnamefont {J.~D.}\ \bibnamefont {{Scargle}}}, \bibinfo {author} {\bibfnamefont {J.~P.}\ \bibnamefont {{Norris}}}, \bibinfo {author} {\bibfnamefont {B.}~\bibnamefont {{Jackson}}},\ and\ \bibinfo {author} {\bibfnamefont {J.}~\bibnamefont {{Chiang}}},\ }\bibfield  {title} {\bibinfo {title} {{Studies in Astronomical Time Series Analysis. VI. Bayesian Block Representations}},\ }\href {https://doi.org/10.1088/0004-637X/764/2/167} {\bibfield  {journal} {\bibinfo  {journal} {\apj}\ }\textbf {\bibinfo {volume} {764}},\ \bibinfo {eid} {167} (\bibinfo {year} {2013})},\ \Eprint {https://arxiv.org/abs/1207.5578} {arXiv:1207.5578 [astro-ph.IM]} \BibitemShut {NoStop}%
\bibitem [{\citenamefont {{Band}}(1997)}]{1997ApJ...486..928B}%
  \BibitemOpen
  \bibfield  {author} {\bibinfo {author} {\bibfnamefont {D.~L.}\ \bibnamefont {{Band}}},\ }\bibfield  {title} {\bibinfo {title} {{Gamma-Ray Burst Spectral Evolution through Cross-Correlations of Discriminator Light Curves}},\ }\href {https://doi.org/10.1086/304566} {\bibfield  {journal} {\bibinfo  {journal} {\apj}\ }\textbf {\bibinfo {volume} {486}},\ \bibinfo {pages} {928} (\bibinfo {year} {1997})},\ \Eprint {https://arxiv.org/abs/astro-ph/9704206} {arXiv:astro-ph/9704206 [astro-ph]} \BibitemShut {NoStop}%
\bibitem [{\citenamefont {{Guidorzi}}\ \emph {et~al.}(2016)\citenamefont {{Guidorzi}}, \citenamefont {{Dichiara}},\ and\ \citenamefont {{Amati}}}]{GuidorziPSD}%
  \BibitemOpen
  \bibfield  {author} {\bibinfo {author} {\bibfnamefont {C.}~\bibnamefont {{Guidorzi}}}, \bibinfo {author} {\bibfnamefont {S.}~\bibnamefont {{Dichiara}}},\ and\ \bibinfo {author} {\bibfnamefont {L.}~\bibnamefont {{Amati}}},\ }\bibfield  {title} {\bibinfo {title} {{Individual power density spectra of Swift gamma-ray bursts}},\ }\href {https://doi.org/10.1051/0004-6361/201527642} {\bibfield  {journal} {\bibinfo  {journal} {\aap}\ }\textbf {\bibinfo {volume} {589}},\ \bibinfo {eid} {A98} (\bibinfo {year} {2016})},\ \Eprint {https://arxiv.org/abs/1603.06890} {arXiv:1603.06890 [astro-ph.HE]} \BibitemShut {NoStop}%
\bibitem [{\citenamefont {{Gordon}}\ and\ \citenamefont {{Arnaud}}(2021)}]{pyxspec}%
  \BibitemOpen
  \bibfield  {author} {\bibinfo {author} {\bibfnamefont {C.}~\bibnamefont {{Gordon}}}\ and\ \bibinfo {author} {\bibfnamefont {K.}~\bibnamefont {{Arnaud}}},\ }\href@noop {} {\bibinfo {title} {{PyXspec: Python interface to XSPEC spectral-fitting program}}},\ \bibinfo {howpublished} {Astrophysics Source Code Library, record ascl:2101.014} (\bibinfo {year} {2021}),\ \Eprint {https://arxiv.org/abs/2101.014} {ascl:2101.014} \BibitemShut {NoStop}%
\bibitem [{\citenamefont {{Hou}}\ \emph {et~al.}(2018)\citenamefont {{Hou}}, \citenamefont {{Zhang}}, \citenamefont {{Meng}}, \citenamefont {{Wu}}, \citenamefont {{Liang}}, \citenamefont {{L{\"u}}}, \citenamefont {{Liu}}, \citenamefont {{Liang}}, \citenamefont {{Lin}}, \citenamefont {{Lu}}, \citenamefont {{Huang}},\ and\ \citenamefont {{Zhang}}}]{Hou2018mbb}%
  \BibitemOpen
  \bibfield  {author} {\bibinfo {author} {\bibfnamefont {S.-J.}\ \bibnamefont {{Hou}}}, \bibinfo {author} {\bibfnamefont {B.-B.}\ \bibnamefont {{Zhang}}}, \bibinfo {author} {\bibfnamefont {Y.-Z.}\ \bibnamefont {{Meng}}}, \bibinfo {author} {\bibfnamefont {X.-F.}\ \bibnamefont {{Wu}}}, \bibinfo {author} {\bibfnamefont {E.-W.}\ \bibnamefont {{Liang}}}, \bibinfo {author} {\bibfnamefont {H.-J.}\ \bibnamefont {{L{\"u}}}}, \bibinfo {author} {\bibfnamefont {T.}~\bibnamefont {{Liu}}}, \bibinfo {author} {\bibfnamefont {Y.-F.}\ \bibnamefont {{Liang}}}, \bibinfo {author} {\bibfnamefont {L.}~\bibnamefont {{Lin}}}, \bibinfo {author} {\bibfnamefont {R.-j.}\ \bibnamefont {{Lu}}}, \bibinfo {author} {\bibfnamefont {J.-S.}\ \bibnamefont {{Huang}}},\ and\ \bibinfo {author} {\bibfnamefont {B.}~\bibnamefont {{Zhang}}},\ }\bibfield  {title} {\bibinfo {title} {{Multicolor Blackbody Emission in GRB 081221}},\ }\href {https://doi.org/10.3847/1538-4357/aadc07} {\bibfield  {journal} {\bibinfo  {journal} {\apj}\ }\textbf {\bibinfo
  {volume} {866}},\ \bibinfo {eid} {13} (\bibinfo {year} {2018})},\ \Eprint {https://arxiv.org/abs/1808.03942} {arXiv:1808.03942 [astro-ph.HE]} \BibitemShut {NoStop}%
\end{thebibliography}%

\end{document}